\documentclass[prb,twocolumn,showpacs,floatfix]{revtex4}
\usepackage{amsmath}
\usepackage{graphicx}
\usepackage{epsfig}
\usepackage{dcolumn}% Align table columns on decimal point
\usepackage{bm}% bold math

\newcommand{\pbcs}{|p\,{\rule[.1cm]{1.25mm}{0.25mm}}{\rm BCS}\rangle}

\newcommand{\pbcss}{p\,{\rule[.1cm]{1.25mm}{0.25mm}}{\rm BCS}}

\newcommand{\up}{\uparrow}
\newcommand{\dn}{\downarrow}
\newcommand{\eps}{\varepsilon}
\newcommand{\cc}{\tilde{C}}
\newcommand{\ccd}{\tilde{C}^\dag}
\newcommand{\tl}{\tilde}

\newcommand{\np}{N^{(\rm ph)}}

\begin{document}
%\draft
%@twocolumnfalse\endcsname
%%%%%%%%%%%%%%%%%%
%\widetext

\title{ Two spin liquid phases in the spatially anisotropic 
triangular Heisenberg model } 
\author{Seiji Yunoki and Sandro Sorella}
\date{\today}
\affiliation{Istituto Nazionale di Fisica della Materia (INFM)-Democritos, 
National Simulation Centre, and Scuola Internazionale Superiore di Studi 
Avanzati (SISSA), I-34014 Trieste, Italy}

\begin{abstract} 
The quantum spin-1/2 antiferromagnetic Heisenberg model on a 
two dimensional triangular 
lattice geometry with spatial anisotropy is relevant to describe 
materials like ${\rm Cs_2 Cu Cl_4}$ and organic compounds like 
{$\kappa$-(ET)$_2$Cu$_2$(CN)$_3$}. The strength of the spatial anisotropy can 
increase quantum fluctuations and can destabilize the magnetically ordered 
state leading to non conventional spin liquid phases. 
In order to understand these intriguing phenomena, quantum Monte Carlo 
methods are used to study this model system as a function 
of the anisotropic strength, represented by the ratio $J'/J$ between the 
intra-chain nearest neighbor coupling $J$ and the inter-chain one 
$J'$. We have found evidence of two spin liquid regions. The first one is 
stable for small values of the coupling $J'/J \alt 0.65$, and appears 
gapless and fractionalized, whereas the second one 
is a more conventional spin liquid with a small spin gap and is energetically 
favored in the  region $0.65\alt J'/J \alt 0.8$. 
We have also shown that in both spin liquid phases there is no evidence 
of broken translation symmetry with dimer or spin-Peirls order or 
any broken spatial reflection symmetry of the lattice. 
The various phases are in good agreement with the experimental findings, thus 
supporting the existence of spin liquid phases in two dimensional quantum 
spin-1/2 systems. 
\end{abstract}
%]
%
%  71.10.-w 	Theories and models of many-electron systems
%  71.10.Pm 	Fermions in reduced dimensions (anyons, 
%                   composite fermions, Luttinger liquid, etc.)
%
%  75. 	Magnetic properties and materials
%  75.10.-b 	General theory and models of magnetic ordering
%  75.10.Jm 	Quantized spin models 
%  75.30.Kz 	Magnetic phase boundaries (including magnetic transitions, 
%                                           metamagnetism, etc.)
%  75.40.Cx 	Static properties (order parameter, static susceptibility, 
%                                  heat capacities, critical exponents, etc.)
%  75.40.Gb 	Dynamic properties (dynamic susceptibility, spin waves, 
%                                   spin diffusion, dynamic scaling, etc.)
%  75.40.Mg 	Numerical simulation studies
%  75.90.+w 	Other topics in magnetic properties and materials 
%                               (restricted to new topics in section 75)
%
%%%%%%%%%%%%%%%%%%
%\narrowtext

\maketitle

%%%%%%%%%%%%%%%%%%%%%%%%%%%%%%%
%%%%%%%%%%%%%%%%%%%%%%%%%%%%%%%
\section{Introduction}
%%%%%%%%%%%%%%%%%%%%%%%%%%%%%%%
%%%%%%%%%%%%%%%%%%%%%%%%%%%%%%%

Since the pioneering work by Anderson and Fazekas~\cite{pwa} the 
spin-1/2 antiferromagnetic Heisenberg model on the triangular 
lattice has been considered one of the most promising candidates for 
a spin liquid phase in a frustrated antiferromagnet. 
However several numerical studies~\cite{lhiullier,bernu,caprio,kenzie_2} 
have all consistently confirmed that in the isotropic triangular lattice 
the classical magnetically ordered state appears stable. Nevertheless 
the ordered moment is found considerably  smaller than the classical 
value,~\cite{caprio,kenzie_2} suggesting that the model is very close 
to a quantum critical point,~\cite{nelson} namely to a phase where 
the long range antiferromagnetic order is completely destroyed. 
This picture is supported by the  recently established  result that in 
the quantum dimer model on the triangular lattice geometry~\cite{sondhi} 
a spin liquid phase is stable,  a result that is particularly important 
because for instance the same model on the square lattice displays only 
non spin liquid  phases with broken translation 
symmetry.~\cite{kivelson,fradkin,becca} 

\begin{figure}[hbt]
\includegraphics[width=4.cm,angle=0]{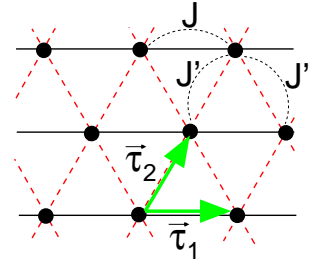}
\begin{center}
\caption{
Antiferromagnetic Heisenberg models on the triangular lattice studied. 
A spin-1/2 is located at each dot. 
${\vec\tau}_1=(1,0)$ and 
${\vec\tau}_2=({\frac{1}{2}},{\frac{\sqrt{3}}{2}})$ 
denote the primitive translational vectors.
A lattice constant is set to be one.
}
\label{lattice}
\end{center}
\end{figure}

Recent experiments on two different materials~\cite{coldea,kanoda} have 
renewed the interest in the spin-1/2 antiferromagnetic Heisenberg model 
on the triangular lattice described by the following Hamiltonian:~\cite{note5} 
\begin{equation} \label{model}
{\hat H}= J \sum\limits_{\langle {\bf i},{\bf j}\rangle}  
\vec S_{\bf i} \cdot \vec S_{\bf j}   + J^\prime 
 \sum_{\langle\langle {\bf i},{\bf j}\rangle\rangle}  
\vec S_{\bf i} \cdot \vec S_{\bf j},
\end{equation} 
where $\vec S_{\bf i}$ is a spin 1/2 located at site ${\bf i}$ on the 
triangular lattice, $\langle {\bf i},{\bf j}\rangle$ 
($\langle\langle {\bf i},{\bf j}\rangle\rangle$) indicates nearest neighbor 
sites along the chain (between different chains), and 
the corresponding antiferromagnetic couplings 
are denoted by $J$ and $J'$ (see Fig.~\ref{lattice}). 

Clearly the anisotropy increases the quantum fluctuations in this model 
as for $J'=0$ the Hamiltonian ${\hat H}$ reduces to a system of uncoupled 
one dimensional (1D) chains, implying 
spin fractionalization and no antiferromagnetic order.
 In this limit  the spin one excitations should form a broad two-spinon 
continuum of states as predicted theoretically,~\cite{faddeev} and indeed 
several  experiments by inelastic neutron scattering have revealed this non 
trivial spin dynamics.~\cite{exp1d}

In a series of experiments by Coldea, {\it et al.},~\cite{coldea} 
the low energy spin dynamics of ${\rm Cs_2 Cu Cl_4}$ have been studied 
systematically and found to be described essentially by the model 
Hamiltonian (\ref{model}) with anisotropic exchange interaction 
$J/J'\simeq 3.0$. In their experiments,~\cite{coldea} there are two facts 
supporting the existence of a spin liquid phase in this system: 
i) a spin spiral phase appears at temperatures below $T_N\simeq 0.6$ K, 
which is about the same order of magnitude as the inter-plane 
coupling $J''/J\simeq0.045$ 
in the third spatial direction, 
and therefore sizably smaller than the two dimensional (2D) couplings 
$J$ ($\simeq0.37$ meV) and $J'$. 
In fact, true long range order in 2D is possible only at zero temperature and 
a finite $T_N$ is due to the inter plane coupling $J''$ which allows to 
cut-off the logarithmically divergent quantum fluctuations in 2D. 
A finite $T_N$ is therefore expected to be of order $\simeq J'/\log(J'/J'')$, 
which appears slightly off from the observed $T_N\simeq0.6$ K. 
ii) At temperatures larger than $T_N$ (or by applying an external magnetic 
field), the inelastic neutron scattering experiments~\cite{coldea} found that 
the line shape of the spectrum consists of a broad continuum, 
which is in contrast to the expected behavior of a magnetically ordered 
state, but is instead similar to the broad two-spinon continuum expected in 
the 1D 
systems mentioned above. 
Therefore these experiments suggest that spin fractionalization can be 
realized also in this 2D system for temperatures higher than the 
corresponding 3D transition temperature $T_N$.

Another spectacular experiments on spin-1/2 quasi 2D triangular systems have 
been recently reported by Shimizu {\it et al}.~\cite{kanoda} 
Here two organic materials {$\kappa$-(ET)$_2$Cu$_2$(CN)$_3$} and 
{$\kappa$-(ET)$_2$Cu$_2$[N(CN)$_2$]Cl} 
(where ET denotes bis(ethylenedithio)-tetrathiafulvalence) 
are synthesized, corresponding essentially 
to two different values of $J'/J\simeq$ 0.8--0.9 and $J'/J\simeq1.8$, 
respectively.~\cite{kanoda} 
While the latter material shows commensurate spin ordering at $T_N=27$ K, 
for the former material no magnetic order is observed even down to 
the milli-Kelvin region ($\sim32$ mK) regardless the fact that the estimated 
value of $J$ is about $250K$. This observation strongly indicates possible 
realization of a spin liquid state for 
{$\kappa$-(ET)$_2$Cu$_2$(CN)$_3$}.~\cite{kanoda}

Motivated by these experiments, we shall consider here the spin-1/2 
antiferromagnetic Heisenberg model on the triangular 
lattice with spatial anisotropy described by Eq.~(\ref{model}). 
In particular, we report a detailed and systematic 
quantum Monte Carlo (QMC) study of the ground state as well as the low-lying 
excitations of this model system as a function of $J'/J$. 
Using both quantum variational Monte Carlo (VMC) method and Green function 
Monte Carlo method with an improved extension of the 
 fixed node approximation (named effective 
Hamiltonian approach in this paper), it is found that there exist two 
spin liquid regions in the phase diagram of this model by varying the ratio 
$J'/J$ for $J'/J\alt0.8$. 
The first one is stable for small values of the coupling 
$J'/J\alt 0.65$, and appears to show gapless, fractionalized fermionic 
excitations.  
The second one is energetically favored in the region 
$0.65\alt J'/J \alt 0.8$, and is a more conventional spin liquid 
with a small 
spin gap in the excitation spectrum, the same type of spin liquid phase 
realized in the quantum dimer model in the isotropic triangular 
geometry.~\cite{sondhi} 
It is argued that the two experimental observations 
of spin liquid like behaviors for ${\rm Cs_2 Cu Cl_4}$ and 
{$\kappa$-(ET)$_2$Cu$_2$(CN)$_3$} mentioned above should 
correspond to these two 
different spin liquid phases of the model, respectively.

The paper is organized as follows. 
In Sec.~\ref{method}, we first introduce the variational wave functions 
considered (Sec.~\ref{vwf}), and describe our original optimization method to 
obtain the minimum energy variational wave function containing a large 
number of variational parameters (Sec.~\ref{srmethod}). 
In order to systematically correct this variational ansatz, 
an effective Hamiltonian approach is introduced in Sec.~\ref{fnemethod} 
along the line of the well established diffusion Monte Carlo 
technique,~\cite{dmc} allowing, for continuous systems, to achieve the best 
variational wave function with the same phases of the chosen  
variational state. As explained in Sec.~\ref{secdyn}, within this 
approach, it is also possible to calculate the low-lying excitation spectrum, 
which can be compared directly with dynamical experimental measurements. 
In Sec.~\ref{secresults}, 
all our numerical results are reported for the spin-1/2 
antiferromagnetic Heisenberg model on the triangular lattice as a function 
of $J'/J$, including the decoupled chains case ($J'=0$). 
Finally our conclusions and remarks are presented in Sec.~\ref{conclusion}. 
The paper is also supplemented by several important appendices for the 
detailed explanation of the methods and the wave functions used, which 
should be very useful for reproducing our results or extending 
our approach to other 
model systems. A part of this work have been reported briefly as a short 
communication.~\cite{letter}

%%%%%%%%%%%%%%%%%%%%%%%%%%%%%%%
%%%%%%%%%%%%%%%%%%%%%%%%%%%%%%%
\section{Numerical method} \label{method}
%%%%%%%%%%%%%%%%%%%%%%%%%%%%%%%
%%%%%%%%%%%%%%%%%%%%%%%%%%%%%%%

\subsection{Variational wave functions: projected BCS states} \label{vwf}

The variational wave function considered in this study is the so-called 
projected BCS state defined by 
\begin{equation} \label{wf}
 \pbcs ={\cal P}_{\rm G} \exp\left[ \sum_{i<j} f_{i,j}  \left(c^{\dag}_{i,\uparrow} 
 c^{\dag}_{j,\downarrow}+ c^{\dag}_{j,\uparrow}
 c^{\dag}_{i,\downarrow} \right) \right] |0\rangle, 
\end{equation}
where $c^{\dag}_{i,\sigma}$ ($c_{i,\sigma}$) is an electron creation 
(annihilation) operator at site $i$ with spin 
$\sigma$($=\uparrow,\downarrow$), $|0\rangle$ is the vacuum state with no 
electrons, the function $f_{i,j}$  is the so-called pairing function that 
contains all variational freedom of the $\pbcs$, and is determined 
by the minimum energy condition,  whereas  
${\cal P}_{\rm G}$ is the usual 
Gutzwiller projection operator onto the subspace 
of singly occupied sites, implying that the total number of electrons $N$ is 
equal to the number of sites $L$. 
The  pairing function $f_{i,j}$ of this 
projected BCS state can be parameterized using 
a BCS Hamiltonian:
\begin{equation} \label{bcsh}
{\hat H}_{\rm BCS} = 
\sum\limits_{i,j} \left[t_{i,j}  \left( \sum_\sigma c^{\dag}_{i,\sigma}   
 c_{j,\sigma}  \right) + \left( \Delta_{i,j}   c^{\dag}_{i,\uparrow}
 c^{\dag}_{j,\downarrow} +~ {\rm h.c.} \right) \right]
\end{equation}
Here $t_{i,j}$ and $\Delta_{i,j}$ as well as the chemical potential 
$t_{i,i}=-\mu$ (assumed uniform) can be considered variational 
parameters, which implicitly determine the pairing function $f_{i,j}$ 
corresponding to the ground state (GS)  of 
${\hat H}_{\rm BCS}$.
Here $i,j$ ($=1,2,\cdots,L$) label the sites of the lattice 
(see Fig.~\ref{lattice}), {\it i.e.},  
${\vec r}_i=i_1{\vec\tau_1}+i_2{\vec\tau_2}$, in the lexicographic order,  
so that the condition $i<j$ in Eq.~(\ref{wf}) is meaningful 
in any spatial dimension.

When  $t_{i,j}$ and $\Delta_{i,j}$ depend 
only on ${\vec\l}={\vec r}_i-{\vec r}_j$, {\it i.e.},  
$t_{i,j}=t_{\vec\l}$ and 
$\Delta_{i,j}=\Delta_{\vec\l}$, respectively, 
${\hat H}_{\rm BCS}$ can be described more compactly by 
\begin{equation} \label{bcshk}
{\hat H}_{\rm BCS} = \sum_{{\bf k},\sigma}(\eps_{\bf k}-\mu)
c^\dag_{{\bf k},\sigma}c_{{\bf k},\sigma} 
+ \sum_{\bf k}(\Delta_{\bf k}c^{\dag}_{{\bf k},\uparrow} 
 c^{\dag}_{-{\bf k},\downarrow}+{\rm h.c}). 
\end{equation}
Here 
\begin{equation}
c^\dag_{{\bf k},\sigma}
={\frac{1}{\sqrt{L}}}\sum_j {\rm e}^{-i{\vec k}\cdot{\vec r_j}} c^{\dag}_{j,\sigma},
\end{equation}
\begin{equation}
\eps_{\bf k}
 =\sum_{\vec\l}{\rm e}^{-i{\vec k}\cdot{\vec\l}} t_{\vec\l}, 
\end{equation}
and 
\begin{equation} \label{defdeltak}
\Delta_{\bf k}
=\sum_{\vec\l}{\rm e}^{-i{\vec k}\cdot{\vec\l}} \Delta_{\vec\l}\;. 
\end{equation} 
For a singlet pairing, 
$\Delta_{i,j}=\Delta_{j,i}$, and thus 
$\Delta_{\bf k}=\Delta_{-{\bf k}}$. 
In this case the projected BCS wave function defined by Eq.~(\ref{wf}) reads 
\begin{equation}\label{wfq}
\pbcs ={\cal P}_{\rm G} |{\rm BCS}\rangle, 
\end{equation}
with $|{\rm BCS}\rangle$ being the ground state of ${\hat H}_{\rm BCS}$ given 
by Eq.~(\ref{bcshk}): 
\begin{equation}
|{\rm BCS}\rangle = \exp\left[ \sum_{\bf k} f_{\bf k} 
c^{\dag}_{{\bf k},\uparrow} c^{\dag}_{{\bf k},\downarrow}
\right] |0\rangle, 
\label{wfbcs}
\end{equation}
where $f_{\bf k}=v_{\bf k}/u_{\bf k}=\Delta_{\bf k}/\left(\xi_{\bf k}+E_{\bf k}\right)$, 
$u_{\bf k}=\sqrt{{\frac{1}{2}}\left[1+{\frac{\xi_{\bf k}}{E_{\bf k}}}\right]}$, 
%\begin{equation}
% \left\{
% \begin{array}{c}
%u_{\bf k}^2={\frac{1}{2}}\left[1+{\frac{\xi_{\bf k}}{E_{\bf k}}}\right]\\ 
%v_{\bf k}^2={\frac{1}{2}}\left[1-{\frac{\xi_{\bf k}}{E_{\bf k}}}\right],
%\end{array}
%\right.
%\end{equation}
and 
\begin{equation}\label{bcsspectrum}
E_{\bf k}=\sqrt{\xi_{\bf k}^2+\Delta_{\bf k}^2}  
\end{equation}
with $\xi_{\bf k}=\eps_{\bf k}-\mu$.

At the variational level, both $\xi_{\bf k}$  and $\Delta_{\bf k}$ have to be 
parameterized in order to minimize the variational energy of the $\pbcs$ wave 
function.
The most relevant parameters for lowering the energy are the short range 
terms, and we have chosen to expand $\eps_{\vec k} = \sum_{\nu=1}^3 
2 t_{{\vec\tau}_\nu} \cos\left({\vec k}\cdot {\vec\tau}_{\nu}\right)$, where 
$t_{{\vec\tau}_\nu}$ are variational parameters and ${\vec \tau}_{\nu}$ are 
the nearest neighbor vectors (${\vec \tau}_3={\vec \tau}_2-{\vec \tau}_1$). 
Analogously, the gap function $\Delta_{\vec l}$ is truncated up to the 
third nearest distance along the chain (${\vec \tau}_1$) 
direction. This is also because, as will be discussed in Sec.~\ref{sub:1d}, 
for the 1D spin-1/2 antiferromagnetic Heisenberg model ${\hat H}_{\rm 1D}$, 
inclusion of the parameter $\Delta_{3\vec\tau_1}$ is known to be crucial for 
this type of projected BCS states to represent almost exactly the ground 
state of ${\hat H}_{\rm 1D}$.~\cite{sorella} 
For the present 2D system described by Eq.~(\ref{model}), which preserve 
$C_{2v}$ symmetry for $J'\ne J$, the projected BCS state $\pbcs$ thus 
contains ten independent variational parameters in $\Delta_{\bf k}$ 
(see, {\it e.g.}, Tab.~\ref{tablecoldea}) and the chemical potential 
$\mu$ which, as opposed to the 1D case, may differ from zero in the 
triangular lattice case. As will be discussed in Sec.~\ref{sub:coldea}, 
we found that the variational parameters $t_{{\vec\tau}_2}$ and 
$t_{{\vec\tau}_3}$ are 
irrelevant and the energetically favorable symmetry of $\Delta_{\bf k}$ is 
$A_1$.~\cite{letter}

As shown in Sec.~\ref{secresults}, the projected BCS state described 
above is a very good variational state for $J'/J\alt$ 0.7. 
However, close to the isotropic limit $J'/J \simeq 1.0$, the translation 
invariant ansatz state, also previously attempted in the presence of hole 
doping,~\cite{dagotto,palee} is not very accurate when compared with the 
exact diagonalization results possible on the $6\times6$ 
cluster.~\cite{liliana} 
In this region of $J'/J$, as shown in Sec.~\ref{secresults:sondhi}, we have 
found that it is more convenient to consider a BCS Hamiltonian defined on 
a $(2 \times 1)$ unit cell [cf. Eq.~(\ref{pairbroken})] for a much better 
variational wave function. As shown in App.~\ref{appendix:sondhi}, 
by using the projected BCS state thereby constructed, 
it is then possible to represent the well known short range resonant 
valence bond (RVB) wave function, with a particular choice of the 
variational parameters. 
The RVB state is a good variational ansatz for the isotropic triangular 
lattice~\cite{bernu} and represents a very convenient initial guess 
for defining, within the present $\pbcs$ framework, an accurate 
variational state in the nearly isotropic triangular lattice.

\subsection{ Minimization method} \label{srmethod}
 
In order to evaluate the optimal variational parameters that minimize  the  
energy expectation value,
\begin{equation} 
E(\Psi)={\frac{\langle \Psi |H | \Psi \rangle} {\langle \Psi | \Psi \rangle}}, 
\end{equation}
we follow the method, recently introduced for calculations of electronic 
structure,~\cite{casula} which will be described in some detail in the 
following.
According to this method, in order to reach the minimum energy  
$E(\Psi)$ in a stable and efficient way, the logarithmic derivative  
$O_k(x)$ of the wave function 
$\Psi_{\{\alpha_k\}}(x)=\langle x|\Psi_{\{\alpha_k\}}\rangle$ 
with respect to each variational parameter  $ \alpha_k= t_{i,j} 
$ and/or  $\Delta_{i,j}$ ($k=1,2,\dots,p$) has to be evaluated on a 
given $N$-electron configuration 
$x=\{ \mathbf{r}_1, \ldots, \mathbf{r}_N \}$ of the projected 
Hilbert space with one electron per site, namely,  
\begin{equation} \label{opx}
O_k(x) = {\frac{\partial}{\partial\alpha_k}} \ln \Psi_{\{\alpha_k\}} (x). 
\end{equation} 
In fact, for infinitesimal changes of these variational parameters,  
$\alpha_k\to \alpha_k'=\alpha_k+\delta\alpha_k$, 
the corresponding change of  the wave function reads 
\begin{equation} \label{opx2}
\Psi_{\{\alpha_k'\}}(x)=\Psi_{\{\alpha_k\}}(x)
\left[1+\sum_{k=1}^p O_k(x)\cdot\delta\alpha_k+
{\cal O}(\delta\alpha_k^2)\right].
\end{equation}
In order to simplify the notations, here we introduce the operator 
${\hat O}_k$ corresponding to $O_k(x)$ which is defined by 
\begin{equation}
\langle x|{\hat O}_k |x'\rangle = O_k(x)\cdot \delta_{xx'}. 
\end{equation}
Using this operator form, the above equation is more compactly written as 
\begin{equation}\label{derwf}
|\Psi_{\{\alpha_k'\}}\rangle=\left[1+\sum_{k=1}^p 
\delta\alpha_k {\hat O}_k  \right] |\Psi_{\{\alpha_k\}}\rangle
\end{equation}
valid up to ${\cal O}(\delta\alpha_k^2)$.

For the minimization method described below, it is important to evaluate 
numerically the value $O_k(x)$ corresponding to a given real space 
configuration of electrons $x$ satisfying the constraint of no doubly 
occupied sites. To this purpose, we have to recall that the 
variational parameters  $\Delta_{i,j}$ and $t_{i,j}$ are explicitly 
defined in ${\hat H}_{\rm BCS}$ [Eq.~(\ref{bcsh})], but only implicitly 
in the wave function itself, defined as the ground state of 
this unprojected BCS Hamiltonian $|{\rm BCS}\rangle$ [Eq.~(\ref{wf})]. 
Thus, in order to evaluate the logarithmic derivatives of the wave function 
$|\Psi_{\{\alpha_k\}}\rangle$ with respect to a variational parameter 
$\alpha_k$, we apply simple perturbation theory to ${\hat H}_{\rm BCS}$ 
and calculate the perturbed state 
$|\Psi_{\{\alpha_k+ \delta \alpha_k \}}\rangle$, within linear order 
in $\delta \alpha_k$. It is then possible 
to compute  $\langle x| \Psi_{\{\alpha_k+ \delta \alpha_k \}} \rangle$ 
using simple algebra and within the same accuracy ${\cal O}(\delta\alpha_k^2)$.
After recasting the calculation by using appropriate matrix-matrix 
operations, the evaluation of $O_k(x)$ is possible in an efficient way, using 
only $\simeq (2L)^2$ operations for each variational parameter $\alpha_k$. 
More details of the method are found in App.~\ref{appmethod}.

The minimization method used here is similar to the standard and well known 
steepest descent method, where the expectation value of the  energy
$E(\Psi_{\{\alpha_k\}})$ is optimized by iteratively changing 
the parameters $\alpha_k$ ($k=1,\cdots,p$) according to 
the corresponding derivatives  of the energy (generalized forces):
\begin{widetext}
\begin{equation} \label{forces}
f_k = - {\partial E(\Psi_{\{\alpha_k\}}) \over \partial  \alpha_k}
= - { \langle \Psi_{\{\alpha_k\}} |\left[ {\hat O}_k {\hat H} + {\hat H} {\hat O}_k \right] | \Psi_{\{\alpha_k\}} \rangle
 \over \langle \Psi_{\{\alpha_k\}} | \Psi_{\{\alpha_k\}} \rangle }
+ 2 E(\Psi_{\{\alpha_k\}}) {   \langle \Psi_{\{\alpha_k\}} |  {\hat O}_k |\Psi_{\{\alpha_k\}}  \rangle 
\over \langle \Psi_{\{\alpha_k\}} | \Psi_{\{\alpha_k\}} \rangle },
\end{equation}
\end{widetext}
namely, 
\begin{equation} \label{itersd}
 \alpha_k \to \alpha_k'=\alpha_k +  f_k\cdot\delta t.
\end{equation}
where $\delta t$ in the standard steepest descent method is 
determined at each iteration by minimizing the energy expectation 
value.~\cite{num} 
The above expressions for the forces $f_k$ can be computed statistically 
by appropriately averaging  the local energy 
$$e_L(x) = 
{ \langle \Psi_{\{\alpha_k\}} |  {\hat H} |x \rangle  \over 
\langle  \Psi_{\{\alpha_k\}} |x \rangle } $$ 
and $O_k(x)$ over a set 
of configurations $ \{ x_l \}$, $l=1,2,\cdots, M$ 
distributed according to the square of the wave function, 
$\left|\langle x| \Psi_{\{\alpha_k\}} \rangle\right|^2$, generated 
with a standard variational Monte Carlo (VMC) scheme, namely,
\begin{eqnarray}
f_k  &=& - {\frac{2}{M}}  \sum_{l=1}^M  O_k (x_l) e_L(x_l)  + 2 \bar O_k  \bar  e_L  \label{defforza} \\
\bar O_k&=& {\frac{1}{M}} \sum_{l=1}^M O_k (x_l)  \label{defobark} \\
\bar e_L &=&  {\frac{1}{M}} \sum_{l=1}^M e_L (x_l)  \label{defbarel} 
\end{eqnarray}
and thus $f_k$ can be computed efficiently once   $O_k(x)$ and 
$e_L(x)$ are evaluated  for any sampled configuration. Here we assumed that 
all quantities are real. However an extension to the complex case is 
straightforward.  

Now, for simplicity, we assume that $\delta t$ is 
positive and small 
enough. Indeed the variation of the total energy 
\begin{eqnarray}
\Delta E & =& E(\Psi_{\{\alpha_k'\}})-E(\Psi_{\{\alpha_k\}}) \nonumber \\
&=&-\sum_{k=1}^p f_k\cdot\delta\alpha_k + {\cal O}(\delta\alpha_k^2)
\end{eqnarray}
at each step is easily shown to be negative for small enough 
$\delta t$ because in this limit 
\begin{equation}
\Delta E = - \delta t \sum_{k=1}^p f_k^2 + {\cal O}(\delta t^2).
\end{equation}
Thus the steepest descent method certainly converges to the minimum of 
the energy when all the forces vanish. 

Let us now generalize the steepest descent method by slightly 
modifying the basic iteration (\ref{itersd}) with  
a suitably chosen positive definite matrix $s$:
\begin{equation} \label{iterforce}
\alpha_k \to \alpha_k'=\alpha_k +  \delta t   \sum_{\l=1}^p s^{-1}_{k,l} f_{l}.
\end{equation}
Again,  using the analogy with the steepest descent method,
convergence to the energy minimum is reached when
the value of $\delta t$ is sufficiently small and
is kept positive constant for each iteration.
In fact, similarly to the steepest descent method,  
 the energy variation corresponding to 
 a small change of the parameters is:
\begin{equation}
 \Delta E = -\delta t  \sum_{k=1}^p\sum_{\l=1}^p  s^{-1}_{k,\l} f_k f_{\l}+{\cal O}({\delta t}^2).
\end{equation}
and is always 
negative for small enough $\delta t$, unless the minimum condition of 
$f_k=0$ is reached and the variational parameters no longer change 
because  $\delta \alpha_k=0$. 
It should be noted here that the steepest descent method 
is a special case with the matrix $s=1$ (unit matrix). 

A more convenient  choice for the matrix $s_{j,k}$ is given by\cite{casula} 
\begin{widetext}
\begin{equation}  \label{defsr}
s_{j,k}= 
 \frac{\langle \Psi_{\{\alpha_k\}} | {\hat O}_j {\hat O}_k |\Psi_{\{\alpha_k\}} \rangle} {\langle \Psi_{\{\alpha_k\}} |\Psi_{\{\alpha_k\}}\rangle }
-\frac{\langle \Psi_{\{\alpha_k\}} | {\hat O}_j  |\Psi_{\{\alpha_k\}} \rangle} {\langle \Psi_{\{\alpha_k\}} |\Psi_{\{\alpha_k\}} \rangle} 
 \frac{\langle \Psi_{\{\alpha_k\}} |  {\hat O}_k |\Psi_{\{\alpha_k\}} \rangle} {\langle \Psi_{\{\alpha_k\}} |\Psi_{\{\alpha_k\}} \rangle}.
\end{equation}
\end{widetext}
This matrix can be efficiently evaluated statistically, and 
similarly to the forces $f_k$,  can be obtained within a VMC scheme once 
$O_k(x)$ are computed for a set of configurations $\{ x_l \}$, 
$l=1,2,\cdots, M$ distributed according to the wave function squared 
$\left|\langle x| \Psi_{\{\alpha_k\}} \rangle\right|^2$, namely,
\begin{equation} \label{defoverlap}
s_{j,k} = {\frac{1}{M}} \sum_{l=1}^M \left[  O_j (x_l) -\bar O_j\right]  
\cdot \left[ O_k (x_l) - \bar O_k\right] 
\end{equation}
where $\bar O_k$ and $ \bar O_j$ are defined in Eq.~(\ref{defobark}).
For whatsoever choice   of the  $M$ configurations $\{ x_l \}$, this 
matrix remain positive definite regardless of the statistical noise 
(at most has vanishing eigenvalues), because it is explicitly written 
in Eq.~(\ref{defoverlap}) as an overlap matrix in ${\cal R}^M$ 
between the $p$ vectors $O_k (x_l)-\bar O_k$ $(k=1,\cdots, p)$. 
In this paper, this generalized method defined by 
Eq.~(\ref{iterforce}) with the matrix ${\bar s}$ given by Eq.~(\ref{defsr}) 
will be called Stochastic Reconfiguration (SR) optimization method. 
It should be noted here that other convenient types of positive definite 
matrix have been recently proposed.~\cite{umrigarhess,sorhess} 
However for the present purpose the improvement does not appear important.  

Let us next discuss why the choice Eq.~(\ref{defsr}) for the matrix $s$ 
in the SR scheme [Eq.~(\ref{iterforce})] is particularly simple and 
convenient compared to the simplest steepest descent method. 
For a stable iterative method for the energy optimization, 
such as the SR method and the steepest descent one, 
a basic ingredient is that at each iteration the new set of parameters
$\{\alpha_k^\prime\}$ are determined close enough to the previous set 
$\{\alpha_k\}$ in terms of a prescribed distance. 
The fundamental difference between the SR minimization and the standard
steepest descent method is simply related to the definition of this 
distance ${\bf\Delta}_\alpha$. 

Within the SR scheme, ${\bf\Delta}_\alpha$ is chosen to be the square 
distance between the two normalized wave functions corresponding to the 
two different sets of variational parameters 
$\{ \alpha_k^\prime \}$ and $\{ \alpha_k \}$, {\it i.e.}, 
\begin{equation} \label{defdistsr}
{\bf\Delta}_{\alpha}^{(\rm SR)}  = 2- 2 {  \langle \Psi_{\{\alpha_k\}} | 
\Psi_{\{\alpha_k^\prime\}} \rangle 
\over  \sqrt{ \langle \Psi_{\{\alpha_k\}} |
\Psi_{\{\alpha_k \}} \rangle  \langle \Psi_{\{\alpha_k^\prime \}} |
\Psi_{\{\alpha_k^\prime\}} \rangle  } }. 
\end{equation}
The reason to normalize the two wave functions before computing their distance 
is obvious because, within a VMC scheme, the normalization of the wave 
function is irrelevant for quantum mechanical averages. 
The basic advantage of the SR method 
is the possibility to work directly  with the wave function distance 
${\bf \Delta}_{\alpha}^{(\rm SR)}$. In fact, this quantity can be explicitly 
written in terms of the matrix $s$ [Eq,~(\ref{defsr})] by substituting 
Eq.~(\ref{derwf}) in its definition Eq.~(\ref{defdistsr}), yielding  
\begin{equation} \label{expdelta}
{\bf\Delta}_\alpha^{(\rm SR)}=  
  \sum_{k=1}^p\sum_{\l=1}^p  s_{k,\l}  
(\alpha^\prime _k-\alpha_k)( \alpha^\prime_{\l}-\alpha_{\l})
+O(|\alpha_k -\alpha_k^\prime|^3).
\end{equation} 
Therefore the most convenient change of the variational parameters is to 
minimize the functional
$${\cal F}(\{\alpha_k'-\alpha_k\})=\Delta E +\bar \Lambda  {\bf \Delta}_\alpha^{(\rm SR)}.$$
Here $\Delta E$ is the linear change in the energy 
$\Delta E = -\sum_{k} f_k (\alpha^\prime_k-\alpha_k)$, and for 
${\bf \Delta}_\alpha^{(\rm SR)}$ the leading term of the expansion in 
small $\alpha_k^\prime-\alpha_k$ given in Eq.~(\ref{expdelta}) can be used.  
$\bar \Lambda$ is a Lagrange multiplier that allows for a stable minimization 
with small change ${\bf \Delta}_\alpha^{(\rm SR)}$ of the wave function 
$|\Psi_{\{\alpha_k\}} \rangle$. Then the stationary condition 
$\delta{\cal F}(\{\alpha_k'-\alpha_k\})/\delta{(\alpha_k'-\alpha_k)}=0$ 
naturally lead to the SR iteration scheme described by 
Eq.~(\ref{iterforce}) with $\delta t = 1/(2{\bar\Lambda})$.

In a similar manner, it is also possible to obtain the standard  
steepest descent method. In this case 
 the Cartesian metric defined 
in the $p$-dimensional space of the variational parameters 
is implicitly assumed to distinguish the two sets of variational parameters, 
{\it i.e.}, the distance here is defined to be 
$$ {\bf \Delta}_{\alpha}^{(\rm SD)}=\sum_{k=1}^p  (\alpha^\prime_k  - \alpha_k)^2.$$ 
The same argument used above to minimize 
$\Delta E + {\bar\Lambda} {\bf \Delta}_{\alpha}^{(\rm SD)}$ with respect to 
the variational parameter change $\alpha_k-\alpha_k'$ 
will then lead to the standard steepest descent algorithm described by 
Eq.~(\ref{itersd}).

The advantage of the SR method compared with the steepest descent one is now 
transparent. Sometimes a small change of the variational parameters 
corresponds to a large change of the wave function, and conversely a large 
change of the variational parameters can imply only a small change of the 
wave function. The SR method takes into account this effect through 
a better definition of the distance ${\bf\Delta}_{\alpha}^{(\rm SR)}$ 
[Eq.~(\ref{defdistsr})].

Here a single SR iteration of the SR minimization scheme is 
summarized as follows: i) a set of variational parameters 
$\{\alpha_k\}=\{\alpha_k^{(i)}\}$ is given after the $i$-th iteration, 
ii) the generalized force $f_k$ [Eqs.~(\ref{forces}) and (\ref{defforza})] 
and the matrix $s$ [Eqs.~(\ref{defsr}) and (\ref{defoverlap})] are 
calculated statistically 
using a small variational Monte Carlo simulation (bin) containing typically 
a few thousand samples ($M=1000 \div 10000$) 
distributed according to the wave function squared
$|\Psi_{\{\alpha_k\}}(x)|^2$, and 
 iii) a new set of the 
variational parameters $\{\alpha_k^{(i+1)}\}$ is determined from 
Eq.~(\ref{iterforce}) 
with a suitable choice of $\delta t$. After a few  hundred (or sometimes 
thousand) iterations needed for equilibration, the iteration described above 
is further repeated to statistically determine the optimized variational 
parameters until the desired statistical accuracy is reached. 
This method allows for a very accurate determination of the optimized 
variational parameters with very small statistical uncertainty.
In the present study, all the variational parameters are optimized using 
this SR minimization method.

\begin{figure}[hbt]
\includegraphics[width=8.0cm,angle=-0]{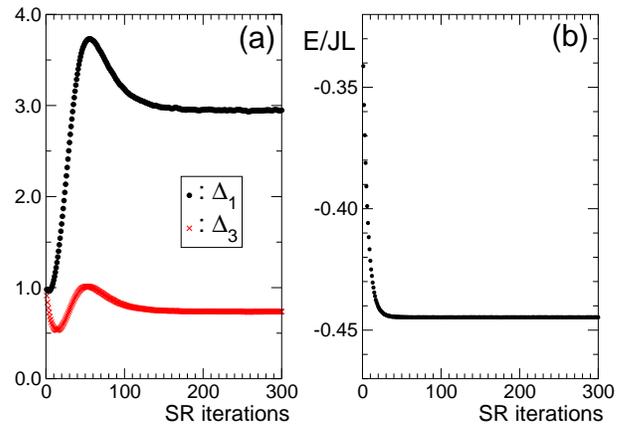}
\begin{center}
\caption{
Monte Carlo evolution of (a) variational parameters 
($\Delta_1$ and $\Delta_3$) and (b) energy as a function of Stochastic 
Reconfiguration (SR) iterations 
for the 1D spin-1/2 antiferromagnetic Heisenberg model ${\hat H}_{\rm 1D}$ 
on a $L=22$ ring. Here the SR method with $\delta t=0.2/J$ 
[Eq.~(\ref{iterforce})] is used. 
In (a), $\Delta_1$ and $\Delta_3$ are the two variational parameters 
of the wave function (for details, see in \ref{sub:1d}), which are both 
initialized to the value $\Delta_1=\Delta_3=1$. In (b), at each iteration, 
the energy is computed by a small VMC simulation for the wave function with 
the fixed variational parameters given at the corresponding iteration in (a). 
}
\label{plotmin}
\end{center}
\end{figure}

A sample case study for the SR minimization scheme is presented in 
Fig.~\ref{plotmin} for the 1D antiferromagnetic Heisenberg model on a 
$L=22$ ring. As seen in Fig.~\ref{plotmin}, after the first few hundred 
iterations needed for equilibration, the variational parameters 
fluctuate around the stable mean values. 
It is interesting to notice in Fig.~\ref{plotmin} that the variational 
parameters may continue 
to change substantially, even after the energy appears to reach its 
equilibrium value only after the first $\simeq 50$ iterations. 
The reason is that a very tiny energy gain, not visible in this plot, 
is implicitly reached at equilibrium, by satisfying very accurately 
the Euler condition of minimum energy, {\it i.e.}, $f_k=0$ for 
$k=1,2,\cdots,p$. 
A stochastic minimization method like the steepest descent method or the 
SR one, which is based not only on the energy but also on its 
derivatives ($f_k$), is 
therefore much more efficient, especially for statistical 
methods where energy differences for slight changes of the variational 
parameters are often very noisy.

Having determined the optimized variational parameters $\{\alpha_k^*\}$ as 
described above, a single standard VMC simulation is performed to calculate 
various physical quantities for $| \Psi_{\{\alpha_k^*\}} \rangle$ using a bin 
length much larger than the one used in the SR minimization procedure. 
Nevertheless, the mentioned  Euler conditions ($f_k=0$ for all $k$) 
are usually satisfied within a few standard deviations simply because the 
statistically averaged variational parameters are much more accurate than 
the ones obtained at the final SR iteration.~\cite{casulamol}

\subsection{ Effective Hamiltonian approach} \label{fnemethod}
As is well known, within the variational approach, 
it is sometimes very difficult to describe long range properties accurately. 
This is simply because these long range properties are rather insensitive 
to the energy, which is instead determined mainly by short range correlation 
functions. 
A clear example of this limitation is found when this method is applied, 
for instance, to study the long range antiferromagnetic order
for the spin-1/2 antiferromagnetic Heisenberg model on the square lattice. 
The long range antiferromagnetic order 
is now considered a well accepted and established property of this 
model.~\cite{reger,sandvick,troyer} 
Nevertheless it has been shown in Ref.~\onlinecite{doucot} that two wave 
functions with completely different long-distance properties, with or without 
antiferromagnetic long range order, provide similar and very 
accurate (within $0.1\%$ accuracy) energy per site in the thermodynamic 
limit. 

Here we shall consider a possible way to overcome this 
limitation of the variational method by introducing what will be named 
``effective Hamiltonian approach'' in the following. The main idea of this 
method is to approximate, as accurately as possible,  the exact 
Hamiltonian  ${\hat H}$ by an 
effective Hamiltonian ${\hat H}^{\rm eff}$, 
for which the exact ground state  can be numerically sampled   
 by Green function quantum Monte Carlo schemes.
The important point is that, within this construction,
some of the important short range properties of the Hamiltonian ${\hat H}$ 
are preserved, and therefore a much better control of correlation functions 
is achieved compared to a mere variational approach.

The short range properties of the Hamiltonian ${\hat H}$ can be conveniently 
defined in the configuration basis $\left\{ x \right\}$. 
Each element of this basis is defined by the positions and spins 
of the $N$ electrons. 
In this basis the matrix elements of the  Hamiltonian ${\hat H}$  are then 
indicated by $ \langle x|{\hat H}|x^\prime\rangle=H_{x,x^\prime}$. 
In this paper, we define a short range Hamiltonian as 
a Hamiltonian ${\hat H}$ for which the {\em off-diagonal} matrix elements 
$ H_{x,x^\prime}$ are non zero only for configurations $x$ and $x^\prime$  
connected one to the other   by local short-range ``moves'' of electrons 
({\it e.g}, hoppings or spin-flips). More precisely,
\begin{equation}\label{srcondition}
\left\{ \begin{array}{ll}
H_{x^\prime,x} \ne 0 & ~~{\rm for}~~|x-x^\prime| \le  {\rm R}\\[0.1cm]
H_{x^\prime,x} = 0 & ~~{\rm otherwise}
\end{array}  \right. 
\end{equation}
where $R$ determines the  range of the off-diagonal matrix elements. 
In this definition, not only conventional Hubbard, Heisenberg, and $t$-$J$ 
models can be thought of short range Hamiltonians (with ${\rm R} =1$), but 
also models with any interaction ${\hat V}$ which is diagonal 
in the configuration basis $\{ x \}$ ($V_{x,x'}\propto V_x \delta_{x,x'}$), 
{\it e.g.}, the long range Coulomb interactions, can be considered  
short range Hamiltonians. 
Therefore, within this definition, 
most of the physically relevant Hamiltonians are 
short range, essentially because quantum fluctuations are important only 
through the short range off-diagonal matrix elements, in the absence of 
which the Hamiltonian is simply classical. 

At present, short-range spin Hamiltonians which can be solved numerically 
 by quantum Monte Carlo methods are the ones for which the quantity 
\begin{equation} \label{nosignproblem}
s_{x^\prime,x} = \psi_G(x^\prime) H_{x^\prime,x} \psi_G(x) <0
\end{equation}
is strictly negative for all non zero matrix elements $H_{x^\prime,x}$.
Whenever a wave function  $\psi_G(x)$ 
satisfying this condition exists, a unitary 
transformation $H_{x,x^\prime} \to {\rm Sgn} \left[\psi_G(x)\right] 
{\rm Sgn} \left[\psi_G(x^\prime)\right]  H_{x,x^\prime} $ 
transforms the Hamiltonian 
to a standard bosonic one (namely with all non-positive  
off-diagonal matrix-elements) that can be solved numerically with quantum 
Monte Carlo techniques, without facing the so-called  ''sign 
problem''.~\cite{note8} 
For instance, for the antiferromagnetic Heisenberg model 
in any bipartite lattice, a  particularly simple variational wave function  
$\langle x|\psi_G\rangle=\psi_G(x)$ satisfying the Marshall sign
rule, $\psi_G(x) \propto  (-1)^{N_{\rm A}}$, allows to satisfy 
the condition $s_{x,x^\prime} < 0$, where $N_{\rm A}$ is the 
number of electrons with down spin on one of the sublattices for the given 
configuration $\{x\}$ (see App.~\ref{app:marshall}). 
With this wave function $\psi_G(x)$, it is therefore possible to solve 
numerically the spin-1/2 antiferromagnetic Heisenberg model in 1D, on a 
two-leg ladder, and on the 2D square lattice. In these cases, it is also 
clear that with the same Marshall sign of the wave function $\psi_G(x)$, 
different low energy properties can be obtained, {\it i.e.}, a gapless spin 
liquid ground state for the single chain,  a gapped spin liquid for the 
two-leg ladder, and a quantum antiferromagnet for the 2D limit.

However, as is well known, only for very particular models the Marshall sign 
rule and Eq.~(\ref{nosignproblem}) are 
satisfied,~\cite{santoro,troyer,sandvick} and in general 
Eq.~(\ref{nosignproblem}) is violated regardless of the wave function 
$|\psi_G\rangle$ used. 
Even when a wave function $|\psi_G\rangle$ with the optimal signs, 
{\it i.e.}, the exact GS signs, is used for generic frustrated Hamiltonians, 
there still exist off-diagonal matrix 
elements with $s_{x,x^\prime} >0$ (notorious sign problem). Namely, due to 
frustration, they do not decrease the energy expectation value.

In order to overcome this difficulty and treat more generic models, 
we will define below an effective Hamiltonian ${\hat H}^{\rm eff}$, 
which is closely related to ${\hat H}$, by using 
an optimal wave function $\psi_G(x)$, that we name in the following the 
''guiding function''. This guiding wave function is required to 
be non zero for all configurations $\{x\}$. Even if this requirement is 
not satisfied, all the forthcoming analysis remain valid in the 
subspace of configurations $\{x\}$ for which $\psi_G(x)\ne 0$.
Once the guiding wave function is provided and thus ${\hat H}^{\rm eff}$ is 
defined, the effective model system ${\hat H}^{\rm eff}$ is solved exactly 
using the standard Green function quantum Monte Carlo 
method.~\cite{calandra} 
As will be shown later, the low energy properties of 
${\hat H}^{\rm eff}$ are 
weakly dependent on the low energy properties, {\it i.e.}, long range 
behavior, of $\psi_G(x)$.

The GS wave function of  ${\hat H}$ is approximated 
by the GS of an  effective Hamiltonian ${\hat H}^{\rm eff}$. 
This approximate variational state is very good in energy because most of 
the matrix elements of ${\hat H}$ are treated exactly in ${\hat H}^{\rm eff}$, 
whereas the remaining ones are removed and traced to the 
diagonal terms of ${\hat H}^{\rm eff}$.
As it will be shown later on,  this enforces the constraint that the GS 
of ${\hat H}^{\rm eff}$  has the same non trivial signs  of $\psi_G(x)$. 
Therefore, if $\psi_G(x)$ is chosen to have the same signs of the exact 
ground state of ${\hat H}$ for  most configurations, this approach 
becomes essentially exact.~\cite{note4}

Within the effective Hamiltonian approach, the problem to construct an 
accurate approximate wave function for the GS of ${\hat H}$ is therefore 
related to how to know phases of the GS.
This, we believe, is a much simpler task, because a good variational wave 
function of a short range Hamiltonian should provide also good phases. 
In fact, the variational energy of the guiding function 
$|\psi_G\rangle$, 
$$E(\psi_G)={ \sum_{x,x^\prime} {\rm Sgn}\left[s_{x,x^\prime}\right]
\left|\psi_G(x)  \psi_G(x^\prime) H_{x,x^\prime}\right| 
\over \sum_x |\psi_G(x)|^2 },$$
depends strongly on the signs of $\psi_G(x)$ via the short range 
terms appearing in  $s_{x,x^\prime}$. 
This assumption that a variational wave function with accurate energy 
should have  very good signs  can be checked directly on small size clusters.
Moreover, for the nearest neighbor antiferromagnetic Heisenberg model on the 
square lattice, it is known that all good variational wave functions 
satisfy the Marshall sign rule, although they may display different large 
distance behaviors.~\cite{doucot}  
This fact clearly confirms this assumption because all these 
wave functions  differ only in their amplitudes but not in their phases.

From these considerations, in this paper, we will chose as a guiding 
function the projected BCS wave function $\pbcs$ described in the previous 
subsection with the variational parameters optimized for each $J'/J$ by a 
careful energy minimization using the SR scheme (Sec.~\ref{srmethod}). 
In most cases, due to the quality of the guiding function used, no relevant 
corrections are found in the various large distance correlations calculated 
for $\pbcs$ when these are compared with the ground state correlations of 
${\hat H}^{\rm eff}$. In some special cases, the use of the effective 
Hamiltonian approach is instead of crucial importance.

\subsubsection{ Definition of the effective Hamiltonian}

As mentioned previously, the effective Hamiltonian ${\hat H}^{\rm eff}$ is 
defined in terms of the matrix elements of ${\hat H}$, which are chosen 
to generate a dynamic as close as possible to the exact one.
An obvious  condition to require is that if $|\psi_G\rangle$ is exact, then
the ground state of ${\hat H}^{\rm eff}$ and its eigenvalue have to coincide 
with the ones of ${\hat H}$. 
In order to fulfill this condition, the so-called lattice fixed node (FN) 
was proposed\cite{ceperley}. 
In the standard lattice fixed node approach, all the matrix elements which 
satisfy Eq.~(\ref{nosignproblem}) are unchanged, whereas the remaining 
off-diagonal 
matrix elements are dealt semiclassically and traced to the diagonal term, 
defining the standard FN Hamiltonian ${\hat H}^{\rm eff}={\hat H}^{\rm FN}$. 
The FN Hamiltonian ${\hat H}^{\rm FN}$ is obtained  
by modifying its  diagonal elements in a way that the local energies  
corresponding to the FN Hamiltonian ${\hat H}^{\rm FN}$ and the exact one  
${\hat H}$ coincide for all configuration $x$, namely, 
\begin{equation} \label{eqlocal}
{ \langle \psi_G |  {\hat H} |x \rangle  \over 
\langle  \psi_G |x \rangle }
={ \langle \psi_G |  {\hat H}^{\rm FN} |x \rangle  \over 
\langle  \psi_G |x \rangle },
\end{equation}
and therefore the FN Hamiltonian is defined by 
\begin{equation}
H_{x',x}^{\rm FN}= \left\{ \begin{array}{lrl}
  ~~~~~H_{x^\prime,x}  ~~~~~~~~~{\rm if}~x^\prime \ne x   &{\rm and }~~  s_{x^\prime,x} \le& 0   \\[0.1cm]
   ~~~~~~~0   \,~~~~~~~~~~~~{\rm  if}~x^\prime \ne x   & {\rm and }~~  s_{x^\prime,x}  >&0 \\[0.1cm]
  H_{x,x}+{\cal V}^{\rm sf}(x)  ~~~{\rm if}~x^\prime = x   &   

\end{array}  \right. 
\label{fnham}
\end{equation}
where
$$
{\cal V}^{\rm sf}(x)=\sum_{\{x^\prime(\ne x),s_{x^\prime,x}>0\}}\psi_G(x^\prime) H_{x^\prime,x} /\psi_G(x).
$$
The FN approach was inspired from the similar fixed node method on  
continuous systems,~\cite{dmc} and indeed is a well established 
approach~\cite{ceperley} which provides 
also variational upper bounds of the ground state energy, {\it i.e.}, 
$E_0^{\rm FN}\le E(\psi_G)$ where $E_0^{\rm FN}$ is the ground state energy 
of ${\hat H}^{\rm FN}$.

As we will show below, for lattice systems there is a better way to choose 
this effective Hamiltonian~\cite{effectiveyunoki}, which not only provides 
better variational energies, but also allows for a better accuracy  of the 
low energy long distance properties of the ground state. For this end, 
it is important to notice the following key difference between a lattice 
system and a continuous one: in the lattice system the configurations 
$\{ x \}$  which do not satisfy the condition (\ref{nosignproblem}) 
may be a relevant fraction of the total number of configurations, 
whereas in the continuous case such configurations represent just an 
irrelevant  ''nodal surface'' of the phase space.
Therefore dropping all the off-diagonal matrix elements with 
$s_{x,x^\prime} > 0$ as in the standard FN method 
seems to be a too drastic approximation for the lattice case. 
This approximation can be indeed improved for lattice systems 
because, contrary to the continuous case, the FN Hamiltonian does not 
provide the best variational state with the same 
signs of the chosen guiding function $\psi_G(x)$.

The main consequence of neglecting all the off-diagonal matrix elements 
with $s_{x,x^\prime} > 0$ is a bias of the dynamic, as electrons cannot 
move freely in some of the configurations. In order to compensate 
this bias in the diffusion of the electrons, we  
introduce a renormalization constant $K \le 1$, which reduces 
the off-diagonal ``hopping'' in the allowed configurations with 
$s_{x,x^\prime}<0$: 
\begin{equation}
H_{x',x}^{\rm eff}= \left\{ \begin{array}{lrl}
  K  H_{x^\prime,x} ~~~~~~~~~~~{\rm if}~x^\prime \ne x   &{\rm and }~~  s_{x^\prime,x} \le& 0   \\[0.1cm]
   ~~0   ~~~~~~~~~~~~~~~~~{\rm  if}~x^\prime \ne x   & {\rm and }~~  s_{x^\prime,x}  >&0 \\[0.1cm]
  H_{x,x}+{\cal V}^{\rm sf}(x)  ~~~{\rm if}~x^\prime = x   &
\end{array}  \right. 
\label{gammaham}
\end{equation}
whereas in order to satisfy the condition (\ref{eqlocal}) 
${\cal V}^{\rm sf}$ is modified as follows: 
\begin{eqnarray*}
{\cal V}^{\rm sf}(x)&=& (1-K)
 \sum\limits_{ \{ x^\prime (\ne x),s_{x^\prime,x}<0 \}} 
\psi_G(x^\prime) H_{x^\prime,x} /\psi_G(x)   \\
&+&\sum\limits_{\{ x^\prime(\ne x),s_{x^\prime,x}>0 \}}  
\psi_G(x^\prime) H_{x^\prime,x} /\psi_G(x).
\end{eqnarray*}

\subsubsection{ Optimal choice for the constant $K$ }

Whenever there is no sign problem and $s_{x,x^\prime} \le 0$, 
$K=1$ is obviously the best choice for which  ${\hat H}={\hat H}^{\rm eff}$.
Also this choice $K=1$  coincides with the standard lattice FN 
approach [Eq.~(\ref{fnham})] (which will be denoted by the acronymous FN). 
Conversely, when the sign of the off-diagonal term in ${\hat H}$ are 
frustrated ($s_{x,x^\prime} \ge 0$), a better choice of the constant $K$ 
can be obtained by using a 
relation which has been well known for continuous systems, and 
used to correct efficiently the error due to the finite time slice 
discretization in the diffusion Monte Carlo (DMC) 
calculations.~\cite{reynolds}

Let us first discuss this relation used in DMC before considering the 
lattice case. In the DMC, the small imaginary time ($\Delta \tau$) evolution 
of the electron configurations is governed by the exact Hamiltonian 
propagation, 
$|\psi\rangle \to \exp( -  {\hat H}  \Delta \tau )  |\psi\rangle$, 
with a diffusion coefficient $D$ determined only by the free kinetic 
operator in ${\hat H}$. It is then possible to correct the approximate 
finite $\Delta \tau$ dynamic corresponding to the  fixed node 
Hamiltonian, by requiring that it satisfies exactly 
the  short time diffusion condition.  This condition  
can be simply written as 
\begin{equation} \label{diffusion}
[ \vec  x, [{\hat H} ,\vec x ] ] = D, 
\end{equation}
where $D=3 \hbar^2/m$ is the diffusion coefficient in three dimensions, 
and $\vec x$ is the electron position operator.

For lattice systems  with periodic boundary 
conditions (PBC), the position operator $\vec x$ is not well defined, 
as it cannot be matched with the boundary conditions.
This limitation can be easily solved by 
using  
 periodic spin position operators  
defined in the exponential form by 
\begin{eqnarray} \label{posmu}
{\hat X}_{\nu} &=& \exp\left[ i \sum_{\vec R} ( {\vec h}_{\nu} \cdot {\vec R})  \,{\hat S}^z_{\vec R} \right] 
\end{eqnarray} 
where ${\hat S}^z_{\vec R}$ is the $z$-component of the spin operator at site 
${\vec R}$, and $\nu$ labels the spatial coordinates, {\it e.g.}, 
${\vec h}_x= ( 2 \pi/l,0)$ and ${\vec h}_y = (0, 2 \pi/l)$ for 
a $L=l \times l$ square lattice. These operators are diagonal in the basis of 
configurations $\{ x \}$, 
$\langle x|{\hat X}_{\nu}|x'\rangle=X_{\nu}(x)\delta_{x,x'}$, 
as is the analogous position operator $\vec x$ 
in the continuous case. Remarkably  ${\hat X}_{\nu}$ 
is exactly equivalent to the well known Lieb-Schultz-Mattis operator, 
used to show a well known property on the low energy spectrum of spin-1/2 
Heisenberg Hamiltonians~\cite{lsm}. After simple inspection, a relation 
similar to the one (\ref{diffusion}) can be found also 
for the periodic spin position operators ${\hat X}_{\nu}$, by simply imposing 
the following equation:
\begin{equation} \label{relation}
\langle \psi_G | \left[ {\hat X}^{\dag}_{\nu}  ,
\left[  {\hat H} , {\hat X}_{\nu} \right] \right]
 |\psi_G \rangle  =
\langle \psi_G | \left[ {\hat X}^{\dag}_{\nu}  ,
\left[  {\hat H}^{\rm eff} , {\hat X}_{\nu} \right] \right]
 |\psi_G \rangle  
\end{equation}
For the lattice case, both the left hand side and the right hand side of this 
relation have non trivial expectation values.
Both of them can be simply calculated by a standard VMC 
method without particular difficulty, 
after  recasting them in a more conventional form for VMC calculations: 
\begin{widetext}
\begin{eqnarray} \label{relationvmc}
 \langle \psi_G | \left[ {\hat X}^{\dag}_{\nu}  ,
\left[  {\cal {\hat H}} , {\hat X}_{\nu} \right] \right]
 |\psi_G \rangle  &=&
-\sum_x \sum_{x^\prime}  \psi_G(x) \psi_G(x^\prime)
{\cal H}_{x,x^\prime} |X_{\nu}(x)-X_{\nu} (x^\prime) |^2 \nonumber   \\
&= &-\sum_{x}  |\psi_G(x)|^2 \left[ \sum_{x^\prime}  \psi_G(x^\prime)
{\cal H}_{x,x^\prime} |X_{\nu}(x)-X_{\nu} (x^\prime) |^2/\psi_G(x)  \right], 
\end{eqnarray}
\end{widetext}
a relation that is obviously valid 
both for ${\cal H}={\hat H}$ and for ${\cal H}= {\hat H}^{\rm eff}$. 
Here we assumed ${\cal H}_{x,x'}={\cal H}_{x',x}$. 
Therefore, similarly to the continuous scheme\cite{reynolds}, 
the value of the constant $K$ can be determined 
from Eq.~(\ref{relation}) with high statistical accuracy.  
Notice that if there is no sign problem, {\it i.e.}, $s_{x,x'}\le0$ 
for all configurations $\{x\}$, 
the constant $K$ turns out 
to be exactly one with vanishing statistical error, yielding again 
${\hat H}^{\rm eff}={\hat H}$.

After determining the constant $K$, the effective Hamiltonian 
${\hat H}^{\rm eff}$ [Eq.~(\ref{gammaham})] is defined, and the ground state 
$|\psi_0^{\rm eff}\rangle$ with its eigenvalue $E_0^{\rm eff}$ and the 
corresponding low energy excitations 
of ${\hat H}^{\rm eff}$ can be computed using the standard Green function 
quantum Monte Carlo method~\cite{calandra} without sign problem. 

To compute the expectation value of the energy
\begin{equation}
 E(\psi_0^{\rm eff})={ \langle \psi_0^{\rm eff}| {\hat H} | \psi_0^{\rm eff}  \rangle 
\over \langle \psi_0^{\rm eff} | \psi_0^{\rm eff} \rangle } 
\end{equation} 
using $| \psi_0^{\rm eff}\rangle$ as an approximate ground state for 
${\hat H}$, the method described in Ref.~\onlinecite{effective} can be 
applied, which very often improves sizably the upper bound estimate of the 
energy, {\it i.e.}, 
$E(\psi_0^{\rm eff}) \le E_0^{\rm eff}\le E(\psi_G)$, even in the 
standard FN case with $K=1$.~\cite{reynolds} 
As also remarked in Ref.~\onlinecite{effective}, contrary to the continuous 
case,  for lattice Hamiltonians the lowest variational energy 
$E(\psi_0^{\rm eff})$ does not correspond to $K=1$ in general.

In the following, we will indicate by FNE the improved FN effective 
Hamiltonian 
method [Eq.~(\ref{gammaham})] with the constant $K\le1$ determined by the 
condition (\ref{relation}), using for $|\psi_G\rangle$ the lowest energy 
variational wave function of the form described in Sec.~\ref{vwf}.

\subsection{ Calculation of dynamical correlation functions}  \label{secdyn}
The spin-one excitation spectrum of the effective Hamiltonian 
${\hat H}^{\rm eff}$ can be calculated by applying the forward walking 
technique~\cite{calandra} used to evaluate the imaginary time evolution of the 
following quantity 
\begin{equation} \label{sqimag}
S({\bf k},\tau) =
 { \langle   \psi_G |{\hat S}^z_{\bf k}   e^{ - \tau {\hat H}^{\rm eff} } {\hat S}^z_{-{\bf k}}  |\psi_0^{\rm eff}  \rangle  \over
  \langle   \psi_G | e^{ - \tau {\hat H}^{\rm eff} }  |\psi_0^{\rm eff}  \rangle  }
\end{equation}
where ${\hat S}^z_{\bf k} = { 1\over \sqrt{L}} \sum_{\bf r} e^{i {\bf k}\cdot{\bf r} }  {\hat S}^z_{\bf r}$ 
and $|\psi_0^{\rm eff}\rangle$ is the ground state of ${\hat H}^{\rm eff}$.
Note that the imaginary time propagation in Eq.~(\ref{sqimag}) can be 
evaluated without 
discretization errors in time $\tau$, as is common to many other 
Quantum Monte Carlo techniques,~\cite{evertz} and also pointed out in 
Ref.~\onlinecite{capriomethod} for the present Monte Carlo scheme. 
By simple inspection, the spin one excitation energy $E^{S=1}_{\bf k}$ of
${\hat H}^{\rm eff}$ for momentum ${\bf k}$ 
can be calculated by fitting the large imaginary time behavior of
$S({\bf k},\tau) \propto \exp\left[- E({\bf k})\tau \right]$. 
Here $E({\bf k})=E^{S=1}_{\bf k}-E^{\rm eff}_0$ and $E^{\rm eff}_0$ is the 
ground state energy of ${\hat H}^{\rm eff}$. 
Thus we fit $\log S({\bf k},\tau)$ with 
\begin{equation} \label{fit}
\log S({\bf k},\tau) = - \tau E({\bf k}) + A + B \log(\tau)
\end{equation}
for $\tau\agt \tau_{\rm c}$, where $A$, $B$, and $E({\bf k})$ are fitting 
parameters, and $\tau_{\rm c}$ is a suitable cutoff time, large enough so 
that the fitting form (\ref{fit}) can be used with good accuracy. 
In fact, the present fit is very stable in $\tau$, and a satisfactory 
convergence of $E({\bf k})$ is obtained even for relatively small 
$\tau_{\rm c} \simeq 2/J$.

\begin{figure}[hbt]
\includegraphics[width=5.cm,angle=-90]{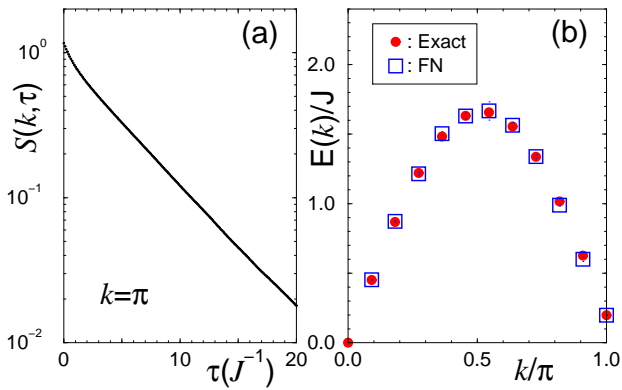}
\begin{center}
\caption{
(a): Imaginary time $\tau$ evolution of $S(k,\tau)$ at $k=\pi$ for the 1D 
spin-1/2 antiferromagnetic Heisenberg model ${\hat H}_{\rm 1D}$ on a $L=22$ 
ring. (b) Excitation energy $E(k)=E_k^{S=1}-E^{\rm eff}_0$ extracted from 
$S(k,\tau)$ with large $\tau$ shown in (a) (squares). 
The error bars are smaller than the size of the symbols. For comparison, 
the exact values are also presented by solid circles. 
}
\label{h1d}
\end{center}
\end{figure}

As a typical example, Fig.~\ref{h1d} (a) shows a semi log plot of 
$S(k,\tau)$ at $k=\pi$ for the 1D spin-1/2 antiferromagnetic 
Heisenberg model ${\hat H}_{\rm 1D}$. It is clearly seen that for a wide 
region of $\tau\agt 2/J$, $\log S(k,\tau)$ is linear so that we can safely 
estimate the excitation energy  $E({\bf k})$ with Eq.~(\ref{fit}).  
The error bars for  $E({\bf k})$ can be efficiently  evaluated by using the  
``bootstrap technique'',~\cite{boot} because the numerator and 
the denominator of Eq.~(\ref{sqimag}) are highly correlated. 
To demonstrate the accuracy and reliability of the 
method, the calculated $E({\bf k})$ for ${\hat H}_{\rm 1D}$ is shown in 
Fig.~\ref{h1d} (b). As seen in the figure, the agreement between 
our results and the exact values is excellent.

It is worthwhile to emphasize that,  within this method,  a single 
Monte Carlo simulation allows  to calculate all the lowest spin one 
excitation energies for various momenta with no extra effort. Moreover, 
the method is not restricted to the computation of the spin triplet spectrum 
alone, but can be  
easily generalized to 
arbitrary operators ${\hat{\cal O}}$ 
as long as they  
are  diagonal in terms of the chosen basis $\{ x \}$.
Although the high energy excitations are more difficult 
to calculate because the signal  decays much faster in time, 
 the most important low 
energy spectrum can be accurately determined.

%%%%%%%%%%%%%%%%%%%%%%%%%%%%%%%
%%%%%%%%%%%%%%%%%%%%%%%%%%%%%%%
\section{Numerical results} \label{secresults}
%%%%%%%%%%%%%%%%%%%%%%%%%%%%%%%
%%%%%%%%%%%%%%%%%%%%%%%%%%%%%%%

%%%%%%%%%%%%%%%%%%%%%%%%%%%%%%%%%%%%%%%%%%%%%%%%%%%%%%
\subsection{Projected BCS wave functions and  particle-hole symmetry}\label{pbcsphs}
%%%%%%%%%%%%%%%%%%%%%%%%%%%%%%%%%%%%%%%%%%%%%%%%%%%%%%

Before presenting our numerical results, here we will show 
an important relation between the projected BCS wave function $\pbcs$ and 
%%SSS  I cited Parola, we have written something similar before... 
%%which has been  already pointed out in Ref.~\onlinecite{parola},  and 
the particle-hole symmetry,~\cite{parola} 
which turns out to be crucial to differentiate spin systems defined on 
the conventional non-frustrated square lattice  
and the ones defined on the triangular lattice geometry. In both cases 
each site of the  
lattice is represented by a vector 
${\bf r}=r_1{\vec\tau}_1+r_2{\vec\tau}_2$ with $r_1$ and $r_2$ being 
integer (see, {\it e.g.}, Fig.~\ref{lattice}). 
Then a particle-hole transformation is defined with 
\begin{equation} \label{phole}
c^{\dag}_{{\bf r},\sigma} \to (-1)^{r_1+r_2}c_{{\bf r},\sigma}, 
\end{equation} 
or in the reciprocal lattice space with the reciprocal lattice vectors 
${\vec g}_1$ and ${\vec g}_2$, this transformation is equivalently defined 
with   
\begin{equation} \label{pholek}
c^{\dag}_{{\bf k},\sigma} \to c_{-{\bf k}+{\bf Q},\sigma}
\end{equation}
where ${\bf Q}=({\vec g}_1+{\vec g}_2)/2$.
As shown in App.~\ref{app:marshall}, whenever the BCS Hamiltonian 
${\hat H}_{\rm BCS}$ is invariant under the particle-hole transformation and 
$\Delta_k$ is real, {\it i.e.},
\begin{equation}\label{restriction}
\left\{ \begin{array}{l}
    \epsilon_{\bf k} = - \epsilon_{-{\bf k}+{\bf Q}} \\[0.1cm]
    \Delta_{\bf k} = -\Delta_{-{\bf k}+{\bf Q}} \\[0.1cm]
    \mu=0,
   \end{array}
  \right.
\end{equation}
the corresponding projected BCS wave function $\pbcs$ [Eq.~(\ref{wfq})] 
satisfies the so-called Marshall sign rule.~\cite{marshall} 
The Marshall sign rule is an exact property for the ground state of the 
spin-1/2 antifferomagnetic Heisenberg models on non frustrated lattices 
such as the square lattice, and indeed for these models the minimum 
variational 
energy is achieved when this rule is satisfied by the projected BCS wave 
function $\pbcs$.

%%%%%%%%%%%%%%%%%%%%%%%%%%%%%%%%%%%%%%%%%%%%%%%%%%%%%%
\subsection{One dimensional limit and spin fractionalization}\label{sub:1d}
%%%%%%%%%%%%%%%%%%%%%%%%%%%%%%%%%%%%%%%%%%%%%%%%%%%%%%

We shall first show the results for the uncoupled chain limit, {\it i.e.}, 
for the 1D spin-1/2 antiferromagentic Heisenberg model with nearest 
neighbor coupling: 
\begin{equation} \label{model1d}
{\hat H}_{\rm 1D}= J \sum\limits_{\langle i, j\rangle}  
\vec S_i \cdot \vec S_j. 
\end{equation}
Several previous studies~\cite{gros,sorella} have found that 
the ground state of ${\hat H}_{\rm 1D}$ can be described very accurately by 
the projected BCS wave function $\pbcs$ [Eq.~(\ref{wfq})] with only 
the nearest neighbor hopping $t_{i,j}=\delta_{i,j\pm1}$ 
and the first three neighbors for the gap function 
$\Delta_{i,j}=\Delta_{\l}  \delta_{i,j\pm l}$ $(l=1,2,3)$. 
Here the site-$i$ is represented by the vector 
${\vec r}_i=\,i\,{\vec\tau}_1$. Notice that the ground state of 
${\hat H}_{\rm 1D}$ satisfies the Marshall sign rule, and therefore from 
Eq.~(\ref{restriction}) with $Q=\pi$, $\mu$ and $\Delta_{\l}$ with $l$ even 
have to be identically zero.  It was also pointed out~\cite{sorella} that the 
inclusion of the third neighbor gap function  
$\Delta_3$ is crucial to improve the accuracy of the $\pbcs$. As shown in 
Fig.~\ref{plotmin}(a), the optimized 
parameters for $L=22$ are $\Delta_1=2.947\pm0.003$ and 
$\Delta_3=0.737\pm0.002$  with the notations  of 
Eq.~(\ref{defdeltak}), {\it i.e.}, 
$\Delta_k= 2 \Delta_1 \cos k+ 2 \Delta_3 \cos 3 k$. 
The  corresponding variational estimate of the total energy is 
$E/J=-9.78411\pm0.00005$ which is in excellent agreement 
with the exact value of $E/J=-9.78688$. 
For larger clusters the variational parameters change slightly and smoothly, 
and the same kind of accuracy is obtained even in the infinite size 
limit. 
By simple quadratic (linear) extrapolation  in $1/L$ for the energy 
($\Delta_l$) with data  up to $L=150$ sites,  
we found $E/J L=-0.442991(3)$ [$\Delta_1=3.41(3)$,
$\Delta_3=0.90(1)$] which compares very well with the well-known 
exact value $E/J L =-(\ln2-1/4)=-0.443147$.  
This result suggests that our variational method is particularly accurate 
in describing this 1D exactly solvable case, 
which is precisely our starting point for studying weakly coupled chains with 
$J'\ne0$.

\begin{figure}[hbt]
\includegraphics[width=9.cm,angle=0]{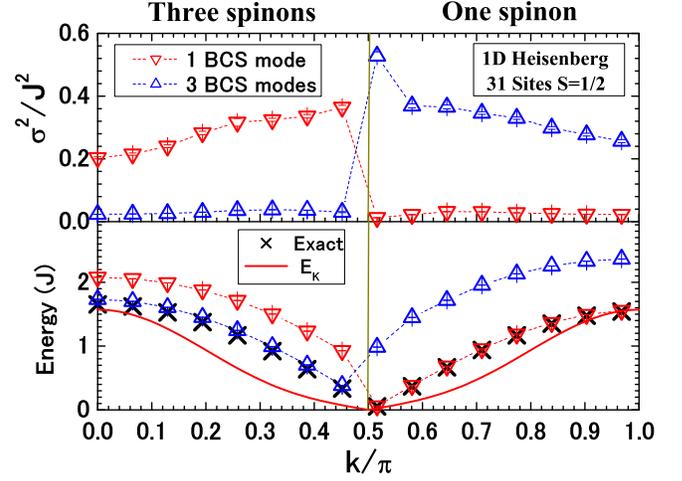}
\begin{center}
\caption{
Lowest spin-1/2 excitation energy $E(k)$ and $E(k^{(3)})$ (lower panel) and 
the variance $\sigma^2(|k \rangle)$ and $\sigma^2(|k^{(3)} \rangle)$ 
(upper panel) of the projected BCS states with a single BCS mode 
$|k \rangle$ and three BCS modes $|k^{(3)} \rangle$ (see text) 
as a function of momentum $k$. The model studied is the 1D spin-1/2 
antiferromagnetic Heisenberg model on a ring with L=31. 
For comparison, in the lower panel, the exact lowest excitation energy is 
plotted by crosses, and the BCS spectrum $E_{\bf k}$ 
denoted by dashed line is scaled by a factor to match the exact bandwidth.
 The remaining lines are guides to the eye. 
}
\label{bethe}
\end{center}
\end{figure}

In the rest of this subsection, we shall show that also the low-energy 
excited states can be described by projected BCS states. As we mentioned 
above, to satisfy the Marhsall sign rule, the optimized 
variational parameters in $\pbcs$ for the ground state meet the constraint 
relation (\ref{restriction}) with  $Q=\pi$. Therefore, at this 
momentum $k=Q$, the spin-1/2 BCS excitation spectrum $E_k$ 
[Eq.~(\ref{bcsspectrum})] shows gapless excitations at $k=\pm Q/2$, 
in perfect agreement with the exact spinon spectrum of 
the Bethe-ansatz solution.~\cite{faddeev} 
Since the elementary excitations of the
BCS Hamiltonian with energy $E_{\bf k}$ are described by the standard 
Bogoliubov modes,~\cite{text}
\begin{equation} \label{bog}
\left\{ \begin{array}{l}
\gamma^{\dag}_{k,\uparrow}= 
     u_k c^{\dag}_{k,\uparrow}-v_k c_{-k,\downarrow}\\[0.1cm]
\gamma_{-k,\downarrow}= 
     v_k c^{\dag}_{k,\uparrow} + u_k c_{-k,\downarrow},
   \end{array}
  \right.
\end{equation}
the simplest variational state for the spinon at momentum ${\bf k}$ is
\begin{equation} \label{elementary}
|{\bf k}\rangle= 
 {\cal P}_{\rm G} \gamma_{k,\downarrow}^\dag |{\rm BCS}\rangle.
\end{equation} 
To see whether this state $|{\bf k}\rangle$ corresponds to 
a spinon state, we consider a ring with $odd$ number of sites $L=31$ 
and {\it z}-component of the total spin $S_z^{\rm tot}=-1/2$. 
For this case it is known that a well defined  spinon exists only for half 
of the total Brillouin zone ($\pi/2 \le |{\bf k}| \le \pi $).~\cite{bethe} 
For this branch, as shown in Fig.~\ref{bethe}, the wave function 
$|{\bf k}\rangle$ represents very well a spinon
with momentum ${\bf k}$, 
as can be verified by the good accuracy in the energy and its small variance 
\begin{equation} \label{variance}
\sigma^2(|{\bf k}\rangle)=
{\frac{\langle {\bf k} | {\hat H}^2  |  {\bf k} \rangle}{\langle {\bf k}|{\bf k} \rangle}}  
- \left[{\frac{\langle {\bf k} | {\hat H}  |  {\bf k} \rangle}{\langle {\bf k}|{\bf k} \rangle}}\right]^2.
\end{equation} 
The variance $\sigma^2(|{\bf k}\rangle)$ is zero for an exact eigenstate, 
and is small for a very accurate variational state. 
As seen in Fig.~\ref{bethe}, this is clearly the 
case for the momenta $k \ge \pi/2$.

For the remaining branch of momenta ($|{\bf k}|\le\pi/2$), 
the excitations for ${\hat H}_{\rm 1D}$ are no longer elementary.~\cite{bethe} 
As shown in Fig.~\ref{bethe}, although the state 
$|{\bf k}\rangle$ is  
formally defined even for those momenta, 
 it represents a poor representation for the exact excited 
state. In fact 
 $E({\bf k})$  becomes quite off from the exact value, 
and  $\sigma^2(|{\bf k}\rangle)$ considerably increases when 
the momentum crosses  the ''spinon Fermi surface''. 
Remarkably, by  projecting  a non-elementary excitation of 
the BCS Hamiltonian, namely,  
\begin{equation}
|{\bf k}^{(3)}\rangle= 
{\cal P}_{\rm G}\gamma^{\dag}_{{\bf k}_{\rm F},\downarrow} 
\gamma^{\dag}_{-{\bf k}_{\rm F},\downarrow} 
\gamma^{\dag}_{{\bf k},\uparrow} |{\rm BCS} \rangle
\end{equation} 
with ${\bf k}_{\rm F}=  \pi (L+1)/2 L $, which mimics the correct 3-spinon 
eigenstate, we can achieve a good agreement for the spectrum even in the 
region outside the spinon Fermi surface (see Fig.~\ref{bethe}). 
Notice that, although the projection ${\cal P}_{\rm G}$ 
is crucial to gain a quantitative agreement for the spectrum, 
the  BCS  spectrum $E_{\bf k}$ already  gives  
a qualitatively correct feature of gapless  
 excitations  with finite spinon velocity at the 
right momentum ${\bf k}=\pi/2$ (see Fig.~\ref{bethe}).

Since the state $|{\bf k}\rangle$ [Eq.~(\ref{elementary})] is very accurate  
only in the momentum region $ \pi/2 \le |k| \le \pi$, we conclude that 
only half of the elementary excitations of the BCS Hamiltonian remain to be 
well defined excitations for the spin Hamiltonian ${\hat H}_{\rm 1D}$ after 
applying the Gutzwiller projection ${\cal P}_{\rm G}$. 
This effect is expected to hold  also in a 2D fractionalized phase. 
It turns out that the elementary BCS excitations which describe the correct 
spinons after applying the projection operator ${\cal P}_{\rm G}$ 
can be obtained by adiabatically switching off 
the gap function $\Delta_k$, a process 
 that defines naturally a Fermi surface, so that 
fermion quasiparticles can be created  (destroyed) 
in unoccupied (occupied) states only.

In 2D it is very difficult to confirm directly the above ``selection rules'' of
the Gutzwiller projection acting on the elementary BCS excitations. 
For instance, to study single spinon excitations, 
one might think of a 2D system on $(\l\times\l)$ lattice with $\l$ odd 
as a natural extention of the 1D system 
considered above. However this 2D system should contain many elementary 
excitations of spinons as it is easily understood by considering the 
$J^\prime \to 0$ limit, and thus it is not an ideal system for studying single 
spinon excitations.  
Nevertheless, it is very important to have reached a very  accurate 
1D limit, with the correct spin fractionalization, within the present 
variational approach, because this approach can be easily extended to higher 
dimensions.

Finally, it is worth mentioning  that,  according to the gauge theory by 
Wen,~\cite{wen,wentext}  the low energy excitations of several 2D 
spin liquids  described by this $\pbcs$ variational ansatz, 
can be understood at the mean field level without taking into account the 
Gutzwiller projection, simply because this projection becomes irrelevant 
for large distance correlations. 
This is the case for  a  "$Z_2$ gapless spin liquid". 
Without entering into too much details of this theory, we only  
mention here that a $Z_2$ spin liquid  
can be described by a  $\pbcs$ for which no SU(2) gauge transformation 
--- remaining in the restricted Hilbert space with no doubly occupation ---  
allows to eliminate 
the gap function in the corresponding BCS Hamiltonian. 
In such a case, only the gauge transformation 
$ c_{i,\sigma}  \to -c_{i,\sigma}$ leaves the BCS Hamiltonian unchanged, 
a transformation which defines the $Z_2$ group toghether with the identity 
operation. 
Most of the spin liquids which we will describe in this paper are $Z_2$ spin 
liquids in the triangular lattice geometry and should be stable according to 
the theory mentioned above.~\cite{wen,wentext}

%%%%%%%%%%%%%%%%%%%%%%%%%%%%%%%
\subsection{ Weakly coupled chains in the triangular geometry with $J'/J$ small} \label{sub:coldea}
%%%%%%%%%%%%%%%%%%%%%%%%%%%%%%%

In this subsection, as a first step towards the 2D limit, we shall consider 
the spin-1/2 antiferromagnetic Heisenberg model for weakly coupled chains 
in the triangular lattice geometry with $J'/J$ small (see Fig.~\ref{lattice}). 
Note that this region is appropriate for the material ${\rm Cs_2 Cu Cl_4}$, 
where $J'/J\simeq1/3$.~\cite{coldea}

\subsubsection{Variational results}

Motivated by the great success of the present variational approach in the 1D 
system discussed in the preceding section, the variational ansatz state which 
we shall consider here for this 2D model in the region $J'/J$ small 
is a similar projected BCS state $\pbcs$ described by
Eqs.~(\ref{bcshk})--(\ref{bcsspectrum}). Here both $t_{i,j}$ and 
$\Delta_{i,j}$ in the BCS Hamiltonian [Eq.~(\ref{bcshk})] parameterizing 
$\pbcs$ are  assumed translational invariant, and therefore they depend only 
on the relative vector 
${\vec \l} = \vec r_i -\vec r_j=l_1{\vec\tau}_1+l_2{\vec\tau}_2=(l_1,l_2)$ 
($l_1$ and $l_2$: integer) between the two sites $i$ and $j$, {\it i.e.}, 
$t_{\vec \l}$ and $\Delta_{\vec l}$. 
Also the ${\rm C_{2v}}$ point group symmetry is assumed for the variational 
parameters, {\it e.g.}, 
$\Delta_{\vec l}=\Delta_{{\cal R}_x{\vec l}}=\Delta_{{\cal R}_y{\vec l}}
=\Delta_{{\cal R}_x{\cal R}_y{\vec l}}$ 
for the $A_1$ symmetry, 
where  ${\cal R}_x$ (${\cal R}_y$) is the reflection operator about the 
$xz$ ($yz$) plane:
\begin{eqnarray}
{\cal R}_x {\vec l} & = & (l_1+l_2){\vec\tau_1}-l_2{\vec\tau_2}\nonumber\\
{\cal R}_y {\vec l} & = & -(l_1+l_2){\vec\tau_1}+l_2{\vec\tau_2}.\nonumber
\end{eqnarray}
To optimize the variational wave function, we use the SR optimization 
method described in Sec.~\ref{srmethod}, which enables us to determine the 
optimized variational parameters with very high accuracy.

As shown in Tab.~\ref{tablecoldea}, the optimized $t_{\vec \l}$ is found 
to be non zero only for the nearest neighbors along the chain direction 
${\vec \tau}_1$ (we set $t_{\vec\tau_1}=1$), and negligible otherwise, 
whereas the optimized $\Delta_{\vec l}$
is instead found to be sizable even among different  chains
(for instance, $\Delta_{(2,1)}$, shown in Tab.~\ref{tablecoldea}), 
displaying a true  two dimensional character. 
It is also found that the symmetry of the gap function $\Delta_{\vec\l}$ which 
minimizes the variational energy is the $A_1$ symmetry under the 
${\rm C_{2v}}$ point group. Finite-size corrections to the variational 
parameters scale as $1/ L$, and the  $18\times 18$ cluster is found large 
enough to be close to the thermodynamic limit, at least, for not too small 
values of $J'/J$.  
All the variational parameters for the  $18\times 18$ cluster as a 
function of  $J'/J(\le0.7)$ are tabulated in Tab.~\ref{tablecoldea} along 
with the variational energy. It should be noted that since the Marshall 
sign rule is no longer satisfied for this model, the constraint relation 
(\ref{restriction}) does not have to be met by the variational wave 
function. For example, as seen in Tab.~\ref{tablecoldea}, the optimized 
chemical potential $\mu$ turned out to be  different from zero.

\begin{table*}
\begin{ruledtabular}
 \begin{tabular}{cccccccc}
    $J'/J$  & 0.1 & 0.2 & 0.33 & 0.4  & 0.5 & 0.6 & 0.7\\   
  \hline
    $\mu$    & $-0.066(4)$ & $-0.085(5)$ &  $-0.109(4)$ & $-0.147(4)$ &  $-0.200(4)$ & $-0.232(4)$ & $-0.253(3)$  \\
  \hline
     $\Delta_{(1,0)}$ & 2.28(2) & 2.02(2) & 1.74(2)  & 1.61(3)  & 1.48(3)  & 1.36(2)  & 1.24(3)  \\
     $\Delta_{(0,1)}$ & 0.545(7) & 0.617(6) & 0.627(6) & 0.644(8) & 0.689(8) & 0.739(7) & 0.787(7)  \\
     $\Delta_{(1,1)}$ & 0.155(4) & 0.179(3) & 0.192(3) & 0.214(4) & 0.256(4) & 0.293(3) & 0.329(4)  \\
     $\Delta_{(-1,2)}$ & 0.007(5) & 0.099(6) & 0.147(6) & 0.160(6) & 0.162(8) & 0.153(4) & 0.143(6)  \\
     $\Delta_{(2,0)}$ & $-0.010(1)$ & $-0.006(2)$ & $-0.0012(8)$ & 0.002(2)& 0.013(1) & 0.025(2) & 0.037(2)  \\
     $\Delta_{(0,2)}$ & 0.195(3) & 0.159(5) & 0.070(3) & 0.058(4) & 0.062(4) & 0.068(5) & 0.066(4)   \\
     $\Delta_{(2,1)}$ & 0.346(4) & 0.381(4) & 0.370(4) & 0.363(5) & 0.360(5) & 0.361(6) & 0.356(6)   \\
     $\Delta_{(1,2)}$ & 0.041(2) & 0.085(3) & 0.098(3) & 0.102(4) & 0.102(5) & 0.097(6) & 0.095(4)   \\
     $\Delta_{(-2,3)}$ & 0.052(1) & 0.040(1) & 0.006(2) & 0.006(2) & 0.010(2) & 0.013(2) & 0.015(1)   \\
     $\Delta_{(3,0)}$ & 0.491(5) & 0.434(4) & 0.383(4) & 0.363(5) & 0.345(5) & 0.327(6) & 0.304(6)   \\
  \hline
    $K^{-1}$ & 1.00243(3) & 1.01014(7) &  1.0286(2) & 1.0434(2)  & 1.0700(3)  & 1.1030(3) &  1.1425(3)  \\
  \hline
    $E_{\rm VMC}/JL$  & $-0.44590(1)$  & $-0.44687(1)$ & $-0.44929(2)$  &  $-0.45118(2)$ & $-0.45474(2)$ & $-0.45932(2)$ & $-0.46514(2)$ \\
  \hline
    $E_{\rm FN}/JL$  & $-0.446074(1)$ & $-0.44723(1)$ & $-0.45005(1)$  & $-0.45223(1)$ & $-0.45628(1)$ & $-0.46156(1)$ & $-0.46823(1)$ \\
  \hline
    $E_{\rm FNE}/JL$  & $-0.446075(1)$ & $-0.447233(2)$ & $-0.45007(1)$  & $-0.45229(1)$  & $-0.45642(1)$ & $-0.46183(1)$ & $-0.46875(1)$ \\
 \end{tabular}
\end{ruledtabular}
\caption{
Optimized variational parameters of the wave function $\protect\pbcs$ 
for the spin-1/2 antiferromagnetic Heisenberg model on the anisotropic 
triangular lattice [Eq.~(\ref{model})] with various $J'/J$ and 
$L=18\times18$. 
$\Delta_{(n,m)}$ is the gap function $\Delta_{\vec r}$ for 
${\vec r} = n{\vec\tau}_1 +  m{\vec\tau}_2$. 
All other variational parameters are zero except for the chemical potential 
$\mu$ and the nearest neighbor hopping in the ${\vec\tau_1}$ direction, 
which is chosen to be one. The value of $K$, variational energy 
$E_{\rm VMC}=E(\protect\pbcss)$, FN (FNE) ground state energy 
$E_{\rm FN}=E_0^{\rm FN}$ ($E_{\rm FNE}=E_0^{\rm eff}$) with 
$|\psi_G\rangle=\protect\pbcs$ are also presented. 
The number in parentheses is the statistical error bar of the quantity 
corresponding to the last digit of the figure. 
}
\label{tablecoldea}
\end{table*}

\begin{figure}[hbt]
\includegraphics[width=9.cm,angle=0]{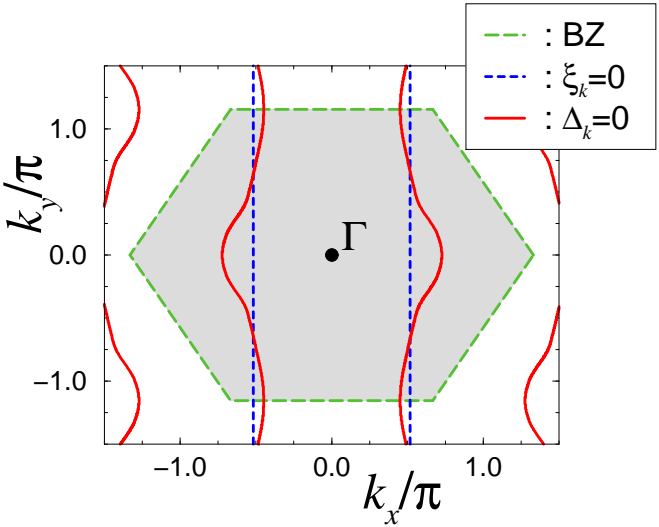}
\begin{center}
\caption{
Loci of ${\bf k}$-points where $\xi_{\bf k}=0$ (dashed lines) and 
$\Delta_{\bf k}=0$ (solid lines) determined by using the optimized 
variational parameters for $J'/J=0.33$. The boundary of the first Brillouin 
zone (BZ) for the triangular lattice is also denoted by long dashed lines.
}
\label{bz_jp033}
\end{center}
\end{figure}

\begin{figure}[hbt]
\includegraphics[width=8.cm,angle=0]{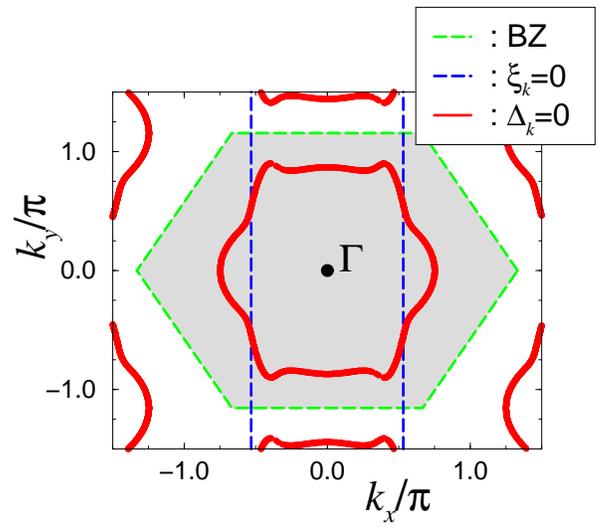}
\begin{center}
\caption{
The same as in Fig.~\ref{bz_jp033} but for $J'/J=0.5$.
}
\label{bz_jp05}
\end{center}
\end{figure}

To gain better insight on the physical nature and properties of these 
variational states, let us next evaluate the BCS excitation spectrum 
$E_{\bf k}$ given by Eq.~(\ref{bcsspectrum}) in the thermodynamic limit 
with the optimized variational parameters. 
In Fig.~\ref{bz_jp033} and Fig.~\ref{bz_jp05}, the contour lines for  
$\xi_{\bf k}=\epsilon_{\bf k}-\mu=0$ and $\Delta_{\bf k}=0$ are plotted 
together with the boundary of the first Brillouin zone (BZ) of the 
triangular lattice for $J'/J=0.33$ and 0.5, respectively.~\cite{note7} 
As can be seen clearly from these figures, $\xi_{\bf k}$ and 
$\Delta_{\bf k}$ vanish at the same momenta with incommensurate ``nodal'' 
points, and thus the corresponding BCS excitation spectrum 
$E_{\bf k}=\sqrt{\xi_{\bf k}^2+\Delta_{\bf k}^2}$ shows 
gapless excitations at these momenta.  It is also interesting to notice 
that with increasing $J'/J$ the shape of the contour line of the gap function 
$\Delta_{\bf k}=0$ changes form open lines to a  closed one in the BZ, 
acquiring a clear two dimensional characteristic. 
On the contrary, the minimum variational energy is always achieved with 
negligible values of the inter chain hopping $t_{\vec\l}$ even 
for the largest  $J^\prime/J$ considered.

%\begin{figure}[hbt]
%\includegraphics[width=7.cm,angle=0]{delta.eps}
%\begin{center}
%\caption{
%For $J'/J=0.33$.
%}
%\label{delta}
%\end{center}
%\end{figure}

Before considering the 2D limit, it is important to discuss in more details 
the properties of few coupled chains in the triangular lattice geometry 
displayed in Fig.~\ref{lattice}. 
If we consider only two chains in the ${\vec\tau}_1$ direction,  
this system corresponds to the well known zig-zag 
ladder with couplings $J$ and $J'$. 
It is now well established that the ground state of the zig-zag ladder 
is spontaneously dimerized with dimers along the rungs 
($\vec\tau_2$ direction) for $J^\prime/J\lesssim 4$, and that the low-lying 
spin excitations are gapped.~\cite{affleckwhite} 
For small values  of  $J^\prime/J $, it is impossible at 
present to detect the dimer order parameter numerically on finite size 
clusters 
simply because this quantity 
vanishes exponentially for $J^\prime/J \to  0$. 
However, it was also shown previously~\cite{chiral} that, 
for systems with finite number of chains, projected BCS 
wave functions can show   spontaneous dimerization 
provided there is a gap in the corresponding 
BCS excitation spectrum $E_{\bf k}$. This is manifestly the case for the 
zig-zag ladder system, as well as 
for any  system  with finite even number of chains, 
as the corresponding finite discretization for the momenta in the 
$y$-direction do not allow to fulfill simultaneously the two conditions 
$\Delta_{\bf k}=0$ and $\xi_{\bf k}=0$ and thus the presence of a gap in the 
BCS excitation spectrum is implied. 
Therefore, the projected BCS wave functions can naturally describe the 
crossover from a finite number of coupled chains,
dimerized and gapped,   to a gapless and fractionalized spin liquid 
in 2D, implying that the limit of infinite number of chains may be highly 
non trivial for a spin liquid wave function.

\begin{figure}[hbt]
\vskip 0.5cm
\includegraphics[width=7.5cm,angle=0]{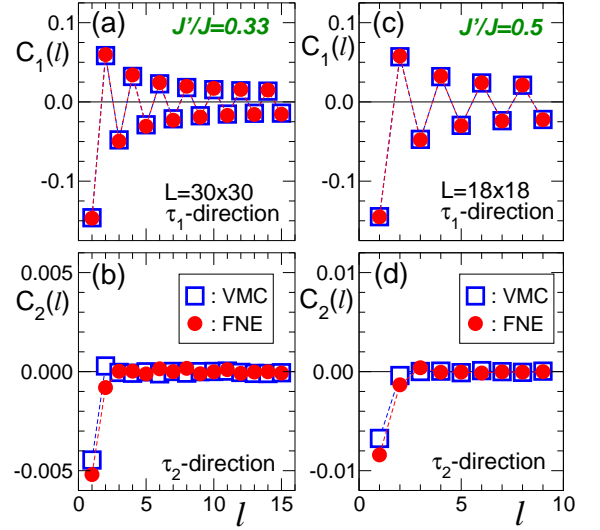}
\begin{center}
\caption{
Spin-spin correlation functions $C({\vec r})$ with 
${\vec r}=\l_1{\vec\tau_1}+\l_2{\vec\tau_2}$ 
for $J'/J=0.33$ [(a) and (b)] and $J'/J=0.5$ [(c) and (d)]
calculated using both variational Monte Carlo (VMC) and effective Hamiltonian 
(FNE) methods.  The cluster sizes used are $30\times30$ [(a) and (b)] and 
$18\times18$ [(c) and (d)]. $C_1(l)$ and $C_2(l)$ 
indicate $C(l \vec \tau_1)$ and $C(l \vec \tau_2)$, respectively. 
}
\label{sr_jp033}
\end{center}
\end{figure}

\begin{figure}[hbt]
\includegraphics[width=7.5cm,angle=-0]{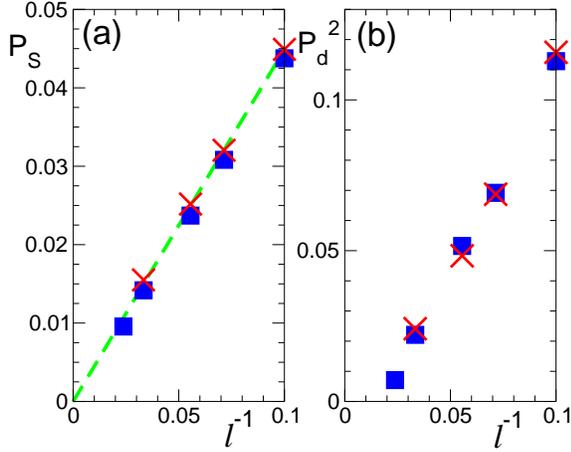}
\begin{center}
\caption{
System size dependence of (a) spin-spin correlation functions $C(\vec l)$ at 
the largest distance in the ${\vec\tau_1}$-direction ($P_{\rm s}$) and 
(b) the dimer order parameter squared $P_d^2$ (see the text) for clusters of 
$L=\l\times\l$.  The coupling studied is $J'/J=0.33$. The variational 
Monte Carlo and  effective Hamiltonian (FNE) 
results are denoted by squares and crosses, respectively. The straight dashed 
line in (a) is a guide to the eye. 
}
\label{Ps}
\end{center}
\end{figure}

In order to understand the properties of the present projected BCS wave 
function $\pbcs$, we now report several physical quantities. 
Fig.~\ref{sr_jp033} shows the spin-spin correlation functions 
\begin{equation}\label{spinspin}
C(\vec l) ={ \langle \psi  | {\hat S}^z_{\vec r}\ {\hat S}^z_{\vec r+\vec l}\ | \psi \rangle \over 
 \langle \psi | \psi \rangle }
\end{equation}
with $| \psi \rangle=\pbcs$ calculated in the $\vec\tau_1$ and $\vec\tau_2$ 
directions for typical couplings with $J'/J=0.33$ and $J'/J=0.5$. As seen in 
Fig.~\ref{sr_jp033}, even though the inter-chain spin correlations are very 
weak, the intra-chain spin correlations are appreciably large, at least, 
for the clusters studied. 
In order to examine whether magnetic long range order occurs, 
we have studied the system size dependence of 
$C(l_{\rm max}\vec\tau_1)\ (=P_s)$  at the maximum distance ($l_{\rm max}$) 
in the $\vec\tau_1$ direction compatible with periodic boundary conditions. 
There exists long range magnetic order only when 
$P_s$ is finite in the thermodynamic limit. 
Our results of $P_s$ are shown in Fig.~\ref{Ps} (a) for $J'/J=0.33$ and 
clusters up to $L=42\times42$. 
From the finite size scaling presented in Fig.~\ref{Ps} (a), 
it is clearly concluded that the projected BCS state $\pbcs$ is not 
magnetically ordered. This is apparently expected because the projected BCS 
wave function $\pbcs$ is a spin liquid state in 2D.

\begin{figure}[hbt]
\vskip 0.5cm
\includegraphics[height=7.cm,angle=-0]{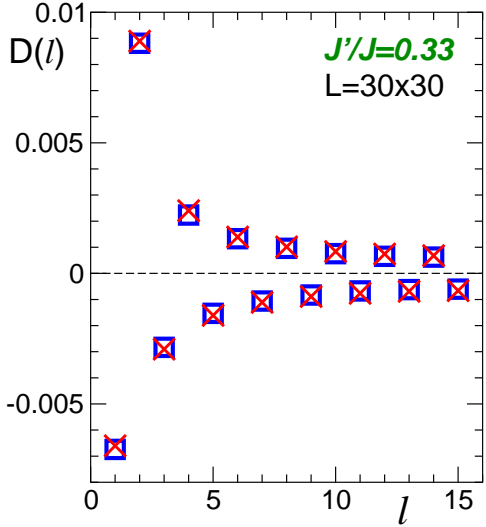}
\begin{center}
\caption{
Dimer-dimer correlation functions $D({\vec r})$ with 
${\vec r}=\l{\vec\tau_1}$  
for $J'/J=0.33$ calculated using both 
variational Monte Carlo (squares) and effective Hamiltonian (crosses) methods. 
}
\label{Odim}
\end{center}
\end{figure}

Another important quantity to study is the dimer-dimer correlation functions 
along the chain direction: 
\begin{equation}\label{def_dim}
D(\vec l) = {  \langle \psi | \left ( {\hat S}^z_{\vec r} 
{\hat S}^z_{\vec r+\vec\tau_1}\right) \left( {\hat S}^z_{\vec r+\vec\l} 
{\hat S}^z_{\vec r +\vec\l+\vec\tau_1}\right) | \psi \rangle \over    \langle \psi | \psi \rangle }-C(\vec\tau_1)^2. 
\end{equation}
As is well known, $D(\vec l)$ shows undamped oscillations at large 
distance $|\vec\l|\to\infty$ when there is a spontaneous 
broken translation symmetry characterized by a  dimerized spin-Pierls 
phase. 
Since for systems with finite number of chains projected BCS 
wave functions show dimer order,~\cite{chiral} 
it is not surprising to see 
large oscillations in $D(\vec l)$ as a function of the distance, as  
shown in Fig.~\ref{Odim} for $J'/J=0.33$. 
In order to rule out the possibility of a finite dimer order for the present 
projected BCS state in 2D, the system size dependence of the dimer 
order parameter  $P_d$ is studied in Fig.~\ref{Ps} (b) for 
clusters up to $L=42\times42$. 
The square of this order parameter ($P_d^2$) is related to 
the calculated largest distance dimer correlations through  
$$
P_d^2=18 \Big|[D(l/2 \vec\tau_1)
-D((l/2-1)\vec\tau_1)]\Big|
$$
for a cluster of $L=l\times l$ sites.~\cite{capriobrief} 
As seen in Fig.~\ref{Ps} (b), it is clear that 
$P_d \to0$ as $L\to\infty$. 
As expected, this projected BCS wave function $\pbcs$ does  not  
show spontaneous dimerization  
in 2D, and therefore it represents a genuine spin-liquid wave function.

\subsubsection{Effective Hamiltonian results}

One of the main advantage of our approach is that we can study 
the stability of the variational ansatz wave function using the effective 
Hamiltonian method described in Sec.~\ref{fnemethod}.~\cite{letter} 
Before showing our results, it should be noted first that  
one of the important quantities which measure the quality of 
our approximation in the effective Hamiltonian approach is the effective 
constant $K$, $K=1$ [determined as the result of Eq.~(\ref{relation})] 
meaning that there is no sign problem and the ground 
state of the effective Hamiltonian represents the exact ground state 
of $\hat H$. As shown in the Tab.~\ref{tablecoldea}, we found that the 
effective constant  $K$ is indeed very close to one for all the cases 
studied.  
This indicates that the number of off-diagonal matrix elements, which cause 
the sign problem and are taken into account only approximately in 
${\hat H}^{\rm eff}$,  represents only 
a very tiny fraction of the total number of matrix elements of $\hat H$. 
Therefore, ${\hat H}^{\rm eff}$ is expected to be rather close to ${\hat H}$, 
and indeed it coincides with ${\hat H}$ in the large 
anisotropic limit $J^\prime/J\to0$, where there is no sign problem.

Encouraged by this observation, 
we have calculated the spin-spin correlation functions $C(\vec l)$ and the 
dimer-dimer correlation functions $D(\vec l)$ using the FNE method, 
{\it i.e.}, $|\psi\rangle=|\psi_0^{\rm eff}\rangle$ in 
Eqs.~(\ref{spinspin}) and (\ref{def_dim}), 
with the optimized $\pbcs$ for $|\psi_G\rangle$ in ${\hat H}^{\rm eff}$. 
Here $|\psi_0^{\rm eff}\rangle$ is the exact ground state of 
${\hat H}^{\rm eff}$ calculated numerically. 
The typical results for $C(\vec l)$ and $D(\vec l)$ 
are presented in Fig.~\ref{sr_jp033} and Fig.~\ref{Odim}, respectively. 
By comparing the variational results with the FNE ones, it is evident 
that these correlation functions appear 
almost unchanged even at large distances, strongly suggesting that 
the projected BCS state $\pbcs$ is very stable and accurate. 
This should be contrasted to the isotropic case (cf. Fig.~\ref{sr_jp10}), 
which will be discussed in the next subsection.

A Systematic finite size scaling analysis of the order parameters 
$P_s=C(l_{\rm max}\vec\tau_1)$ and $P_d^2$ calculated using FNE is also 
reported in Fig.~\ref{Ps} for $J'/J=0.33$ and $L$ up to $30\times30$. 
It is clearly seen in Fig.~\ref{Ps} (b) that $P_d^2$ diminishes to zero 
in the limit $L\to \infty$. Even though the FNE results for $P_s$ shown in 
Fig.~\ref{Ps} (a) are almost the same as the ones for the variational 
calculations with $\pbcs$, it is still difficult to rule out completely 
the possibility of a very small non-zero magnetic order 
parameter $P_s \lesssim 0.001$ in the thermodynamic limit. 
Nevertheless, the fact that the spin-spin correlation functions 
calculated for 
$L \lesssim 1000$ using the variational wave function $\pbcs$ are almost 
identical to the ones calculated using the more accurate FNE approach 
(Fig.~\ref{sr_jp033}), strongly suggests that the magnetic long range order 
is not likely to occur even for the more accurate FNE ground state 
$|\psi_0^{\rm eff}\rangle$.

From these results, 
we conclude that the ground state of the spin-1/2 antiferromagnetic 
Heisenberg model on the triangular lattice with $J'/J\alt0.6$ -- 0.7 
(see also Sec.~\ref{conclusion}) is a spin liquid state with gapless spin 
excitations at the incommensurate momenta, described at least approximately 
by the present projected BCS wave function $\pbcs$.

\begin{figure}[hbt]
\vskip 0.5cm
\includegraphics[width=8.5cm,angle=-0]{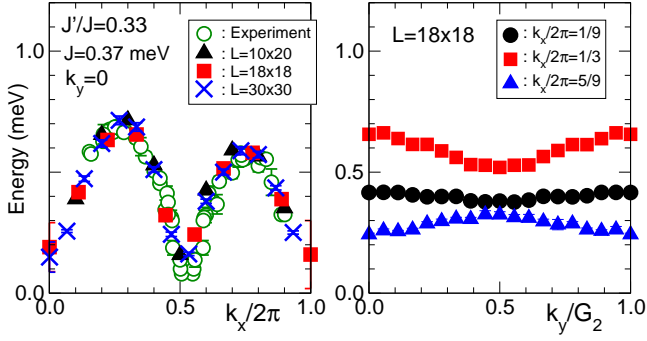}
\begin{center}
\caption{
Lowest triplet excitation energy as a function of momentum for $J'/J=0.33$ and 
different cluster sizes $L$ (indicated in the figures), 
calculated using the method introduced in Sec.~\ref{secdyn}. 
For comparison, the experimental values (open circles) measured by inelastic 
neutron scattering on ${\rm Cs_2 Cu Cl_4}$~\cite{coldea} are also plotted. 
Here $G_2=4\pi/\sqrt{3}$, and the experimentally estimated value of 
$J=0.37$ meV is used.~\cite{coldea} 
}
\label{sqw}
\end{center}
\end{figure}

Finally, let us discuss the implication of our numerical study on the 
experiments for ${\rm Cs_2 Cu Cl_4}$ where the estimated coupling 
is $J'/J\simeq1/3$.~\cite{coldea}  
Recent inelastic neutron scattering experiments for this material by 
Coldea {\it et al}~\cite{coldea} have revealed that the lowest spin 
excitation appears at an incommensurate momentum 
(see Fig.~\ref{sqw}), and the observed excitation spectrum consists of a broad 
incoherent continuum, at least, in the intermediate temperature region 
above the magnetic transition temperature $T_N$. 
As we have shown in the case of the 1D system (Sec.~\ref{sub:1d}), 
the BCS elementary excitations 
naturally define {\em true spinons} in our approach, and according to 
the gauge theory~\cite{wen,wenfermion} they 
should behave as free fermions at low enough 
energy, namely close to the nodal points of the incommensurate 
momenta where $\xi_k=\Delta_k=0$ (see Fig.~\ref{bz_jp033}).
Therefore, the low energy spin excitations observed experimentally can be 
explained by a two-spinon broad continuum. 
At present, we cannot calculate directly the dynamical spin 
correlation functions. However, using the technique described in 
Sec.~\ref{secdyn}, we can calculate the lowest spin-1 excitation energy at 
each momentum, and in Fig.~\ref{sqw} our results are compared 
with the experimental values estimated for ${\rm Cs_2 Cu Cl_4}$ by 
Coldea {\it et al}.~\cite{coldea} 
As seen in Fig.~\ref{sqw}, both results are in excellent agreement, 
considering that our 
calculated excitation spectra with $L$ up to $30\times30$ appear rather 
well converged to the thermodynamic limit. This remarkable agreement 
strongly supports the presence of a spin liquid state in 2D.

%%%%%%%%%%%%%%%%%%%%%%%%%%%%%%%
\subsection{
Isotropic triangular lattice with $J'= J$
} 
\label{secresults:sondhi}
%%%%%%%%%%%%%%%%%%%%%%%%%%%%%%%

It is well known~\cite{bernu} that for the spin-1/2 antiferromagnet 
Heisenberg model on the spatially isotropic triangular lattice ($J'=J$), 
a faithful variational ansatz is the so-called short range RVB state 
$|\psi_{\rm RVB}\rangle$ described by the following wave function: 
\begin{equation}\label{srrvb}
|\psi_{\rm RVB}\rangle = 
\sum_{\{C\}}\left[\prod_{{\scriptstyle i,j=1}
                  \atop {\scriptstyle (i<j)} }^L
|[ i, j]\rangle\right],
\end{equation} 
where the sum $\{C\}$ runs over all possible nearest neighbor  
dimer covering of the triangular lattice (with equal weights and therefore 
implying that all the spatial symmetries of the lattice are satisfied), 
whereas the  product is over 
the corresponding nearest neighbor singlet pairs of spins located on each 
dimer 
(defining singlet valence bond configurations for a given dimer covering) 
between sites $i$ and $j$, 
\begin{equation}
|[i,j]\rangle={\frac{1}{\sqrt{2}}}
\left( \uparrow_{i} \downarrow_{j}
-\downarrow_{i} \uparrow_{j}
\right), 
\end{equation} 
sorted  according to the lexicographic order ($i<j$). 

\begin{figure}[hbt]
\includegraphics[width=7.cm,angle=0]{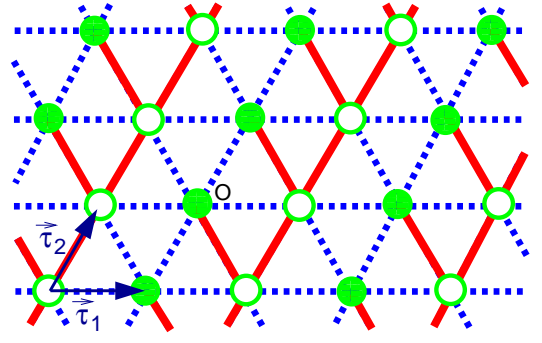}
\begin{center}
\caption{
Nearest neighbor gap functions, $\Delta({\vec\tau}_1)
=\Delta({\vec\tau}_2)=\Delta({\vec\tau}_2-{\vec\tau}_1)=\Delta$, 
which are consistent with the sign convention for the short range RVB state 
explained in App.~\ref{appendix:pfaffian}. The dashed 
(solid) bonds represent a positive (negative) sign 
[Eqs.~(\ref{phase1}) and (\ref{phase2})]. Notice that 
the unit cell is $(2\times1)$. The origin of the lattice $(0,0)$ 
[Eq.~(\ref{pairbroken})] is denoted by O. 
}
\label{lattice3}
\end{center}
\end{figure}

Indeed, very recent numerical calculations for the $6\times 6$ isotropic 
triangular antiferromagnet by the Lanczos method~\cite{liliana} 
have shown that the overlap between the short range RVB wave function 
$|\psi_{\rm RVB}\rangle$ and 
the exact ground state  $|\Psi_0\rangle$ is very large 
$|\langle\psi_{\rm RVB}|\Psi_0\rangle|^2=0.891$,~\cite{note10} 
and especially the average sign, 
\begin{equation}\label{sign_def}
\langle S\rangle= \sum_x  |\langle x|\Psi_0\rangle|^2  {\rm Sgn} 
\left[\langle x|\Psi_0\rangle \langle x|\psi_{\rm RVB}\rangle \right], 
\end{equation}  
is very close to the maximum value, namely,  $\langle S\rangle=0.971$. 
These results 
 clearly indicate that the phases of the exact ground state $|\Psi_0\rangle$ 
are very well described by the short range 
RVB ansatz state $|\psi_{\rm RVB}\rangle$. 
The values of the overlap $|\langle\psi_{\rm RVB}|\Psi_0\rangle|^2$ and 
the average sign $\langle S\rangle$ are even much better than the 
ones corresponding to classical N{\'e}el ordered wave functions considered 
previously, which also contain additional variational 
parameters.~\cite{caprio} 
It should be emphasized that an accurate description of the phases of 
the exact ground 
state $|\Psi_0\rangle$ by a simple variational wave function such as 
$|\psi_{\rm RVB}\rangle$ is the essential ingredient for reliable 
calculations based on the FN or FNE approach.

Although $|\psi_{\rm RVB}\rangle$ is a very good variational ansatz state 
for the ground state of the isotropic triangular antiferromagnet, 
the present representation of this state [Eq.~(\ref{srrvb})] is rather  
difficult to handle and is not convenient to improve the state 
$|\psi_{\rm RVB}\rangle$ systematically by including, for instance,  
long-range valence bonds,  because,  within  this 
representation, there is a ``sign problem''
even at the variational level~\cite{bernu}. 
In order to treat more conveniently the short range RVB 
wave function $|\psi_{\rm RVB}\rangle$ and its variants, we have derived 
in App.~\ref{appendix:pfaffian} an exact mapping between 
the short range RVB wave function $|\psi_{\rm RVB}\rangle$ 
and the projected BCS state $\pbcs$ on planar graphs,~\cite{kastlein} 
namely, for most interesting lattice geometries including 
the triangular lattice, the square lattice, and the Kagom\'e lattice.

As shown in Apps.~\ref{appendix:sondhi} and \ref{appendix:pfaffian}, 
the short range RVB state $|\psi_{\rm RVB}\rangle$ is equivalently 
described by the projected BCS wave function $\pbcs$ with a  special choice 
of the variational parameters: the only  non-zero parameters 
are the chemical potential $\mu$ and the nearest neighbor singlet gap 
functions $\Delta_{i,j}$ with the appropriate relative phases shown in 
Fig.~\ref{lattice3}, and the limit of $-\mu\gg|\Delta_{i,j}|$ is assumed so 
that the gap function $\Delta_{i,j}$ is proportional to the pairing function 
$f_{i,j}$ [Eqs.~(\ref{gap_pair})] considered in the 
App.~\ref{appendix:pfaffian}.

It is easily seen from Fig.~\ref{lattice3} that the corresponding 
BCS Hamiltonian ${\hat H}_{\rm BCS}$ is defined on a $(2\times 1)$ unit cell, 
and thus ${\hat H}_{\rm BCS}$ is not translation invariant. 
In fact, ${\hat H}_{\rm BCS}$ is invariant under an elementary translation 
${\cal T}_2:\ {\vec r}\to{\vec r}+\vec\tau_2$ in the $\vec\tau_2$ direction, 
whereas it is not under an elementary translation 
${\cal T}_1:\ {\vec r}\to{\vec r}+\vec\tau_1$ in the $\vec\tau_1$ direction. 
Let us now show briefly that the translation symmetry is recovered after the 
projection ${\cal P}_{\rm G}$, {\it i.e.}, $\pbcs$ is indeed  translation 
invariant. To  this end, one can combine 
the translation operation ${\cal T}_1$ with 
the ${\rm SU(2)}$ gauge transformation ${\cal U}$:
\begin{equation} \label{gauge}
 c^{\dag}_{\vec r,\sigma} \to - c^{\dag}_{\vec r,\sigma} 
\end{equation}
for $\vec r= m_1 \vec \tau_1 + m_2 \vec \tau_2 $ with $m_2$ {\em odd}. 
Under the composite application of the  transformations 
${\cal T}_1 \bigotimes {\cal U}$,  
the projected BCS wave function $\pbcs$ remains unchanged. Therefore, 
the $\pbcs$ is translation invariant because the ${\rm SU(2)}$ gauge 
transformation ${\cal U}$ acts as an identity in the physical Hilbert 
space with singly occupied sites due to the projection ${\cal P}_{\rm G}$.

\begin{figure}[hbt]
\vskip 0.5cm
\includegraphics[height=5.cm,angle=-0]{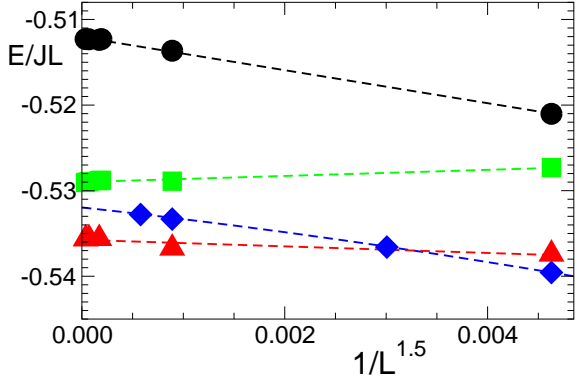}
\begin{center}
\caption{
Variational energy for the isotropic triangular lattice with $J'/J=1.0$ 
calculated using the short-range RVB state $|\Psi_{\rm RVB}\rangle$ (circles), 
the short-range RVB state with $\mu=0$ (squares), the classical N\'eel 
state $|\Psi_{\rm N{\acute e}el}\rangle$ 
(diamonds),~\cite{caprio} and the present projected BCS sate 
${\cal J}_{\rm s}\protect\pbcs$ with spin Jastrow factor 
${\cal J}_{\rm s}$ (triangles). 
}
\label{vmc_energy}
\end{center}
\end{figure}

Owing to this new, more convenient representation of the short range RVB state 
$|\psi_{\rm RVB}\rangle$ by the projected BCS wave function $\pbcs$, we can 
now calculate various physical quantities using the standard variational 
Monte Carlo method. For example, the variational 
energy $E(\psi_{\rm RVB})$ of the isotropic triangular antiferromagnet is 
plotted in Fig.~\ref{vmc_energy} for different system sizes. From the finite 
size scaling analysis of lattice sizes up to $L=30\times30$, we can easily 
provide a very accurate estimate of the variational energy 
in the thermodynamic limit: $E(\psi_{\rm RVB})/JL=-0.5123\pm0.0001$ for 
$L\to\infty$ (see also Tab.~\ref{tableopar}).

Another important advantage of  this projected BCS wave function 
representation is that it is easy to improve the variational ansatz state 
systematically. For instance, only by changing the chemical potential $\mu$ 
from a large value to zero, the variational energy is significantly improved 
as is shown in Fig.~\ref{vmc_energy} and in Tab.~\ref{tableopar}, where the 
variational state is denoted simply by $|\psi_{\rm RVB}\rangle$ with $\mu=0$.
Notice that in this case the $\pbcs$ is equivalent to a Gutzwiller 
projected free fermion state with nearest neighbor hoppings defined in a 
$(2\times 1)$ unit cell.~\cite{li,note11}

More generally, within the framework of the projected BCS wave function 
$\pbcs$, we can easily extend the variational ansatz state to include 
long-range singlet valence bonds simply by adding non zero gap functions 
$\Delta_{i,j}$'s to the BCS Hamiltonian ${\hat H}_{\rm BCS}$.~\cite{note9} 
In order for $\pbcs$ to preserve translation invariance, the 
pairing part of the BCS Hamiltonian is generalized in the 
following form: 
%\begin{widetext}
\begin{eqnarray}\label{pairbroken}
{\hat H}_{\rm pair}
&=& \sum_{\vec r} \Biggl[ {\sum_{{\vec t}_m}}^{\prime}
(-1)^{r_1 m_2}\Delta({\vec t}_m)  
\left(c^\dag_{{\vec r}\up} c^\dag_{{\vec r}+{\vec t}_m\dn}
-c^\dag_{{\vec r}\dn} c^{\dag}_{{\vec r}+{\vec t}_m\up} \right)\nonumber\\
&&~~~~~~~~+ {\rm H.c.}\Biggr], \
\end{eqnarray}
%\end{widetext}
where the first sum runs over all lattice vectors 
${\vec r}=r_1{\vec\tau}_1+r_2{\vec\tau}_2$ ($r_1,r_2$: integer), whereas 
the second sum $\sum_{{\vec t}_m}^{\prime}$ is for 
${\vec t}_m=m_1{\vec\tau}_1+m_2{\vec\tau}_2$ ($m_1,m_2$: integer) 
with $m_2>0$ or with $m_1\ge0$ and $m_2=0$ denoted by solid circles in 
Fig.~\ref{lattice2} (a). 
It is readily 
shown that ${\hat H}_{\rm pair}$ is invariant under 
${\cal T}_1 \bigotimes {\cal U}$ and ${\cal T}_2$.
In order to minimize the number of variational parameters, we have assumed 
here that $\Delta({\vec t}_m)=\Delta({\cal R}_y{\vec t}_m)$.
Because of the phases of the gap functions in ${\hat H}_{\rm pair}$ 
(see Fig.\ref{lattice3} for the nearest neighbor gap functions), 
this BCS Hamiltonian is not guaranteed to be reflection invariant  
around the $yz$-plain  ($R_y$). 
Nevertheless, as will be discussed later,
 our numerical calculations indicate empirically 
that the considered  projected BCS state 
$\pbcs$ becomes 
a fully symmetric spin liquid state only in the thermodynamic limit, 
a state being not only translation invariant but also reflection invariant. 
It is interesting to notice that even though this $\pbcs$ is not fully 
symmetric on  small lattice sizes such as a $6\times6$ cluster, a 
much better overlap $|\langle\pbcss|\Psi_0\rangle|^2$ as well as a much 
better average sign $\langle S \rangle$ are obtained for this $\pbcs$ 
compared to the ones for a translation invariant complex $\pbcs$ 
ansatz.~\cite{dagotto,palee} 
In App.~\ref{appendix:sondhi}, several peculiar and interesting properties 
of the corresponding unprojected BCS state $|{\rm BCS}\rangle$ are derived 
analytically. For example, the spectrum of the 
BCS Hamiltonian $E_{\bf k}$ has two branches 
[due to the ($2\times1$) unit cell] 
and a small excitation gap, which vanishes when the long range gap functions 
are turned off (provided $\mu= 0$). % $\mu \simeq 0$

\begin{table}
\begin{ruledtabular}
 \begin{tabular}{l|l|l}
 Method (wave function) &  ~~~~~~~~$E/JL $ & ~~~~~$m/m_0$ \\
\hline 
 VMC (RVB)   & $-0.5123\pm0.0001$ & 0.0 \\
 VMC (RVB with $\mu=0$) & $-0.5291\pm0.0001$        &   0.0 \\
 VMC (BCS+N{\'e}el)~\cite{mila} & $-0.532 \pm 0.001$ & 0.72 \\ 
 VMC (present study)   & $-0.5357\pm0.0001$ &      0.0  \\
 FN    & $-0.53989 \pm 0.00003$ & $0.325 \pm 0.006$    \\
 FNE   & $-0.54187  \pm 0.00006$ & $0.353 \pm 0.007$    \\
 SW    & $-0.538\pm0.002^*$ & 0.4774    \\
 GFMCSR~\cite{caprio} &  $-0.545\pm0.002^*$ & $0.41\pm 0.02$   \\
 \end{tabular}
\end{ruledtabular}
\caption{
The energy $E$ and the magnetic moment $m$ 
(divided by its maximum value $m_0=1/2$) estimated in the thermodynamic limit 
for the spin-1/2 antiferromagnetic Heisenberg model on the isotropic 
triangular lattice ($J=J'$). In the first four rows are 
the variational Monte Carlo (VMC) estimates for different wave functions: 
(from the top) the short-range RVB state $|\psi_{\rm RVB}\rangle$, 
$|\psi_{\rm RVB}\rangle$ with $\mu=0$ (see the text), 
the wave function studied recently by Weber, {\it et al}~\cite{mila} 
in which a classical N{\'e}el order and a singlet pairing (with symmetry 
different from ours) are included, and the present wave function 
${\cal J}_{\rm s}\protect\pbcs$. 
The fixed node (FN) and effective Hamiltonian (FNE) results are then 
provided. 
For comparison, the corresponding values estimated 
by the linear spin wave theory (SW) and the Green function Monte Carlo method 
with 
stochastic reconfiguration (GFMCSR)~\cite{caprio} are also presented. 
Note that the energies computed by the last two methods (denoted by $*$) 
are not a rigorous upper bound of the exact ground state energy.  
}
\label{tableopar}
\end{table}

\begin{figure}[hbt]
\includegraphics[width=8.cm,angle=0]{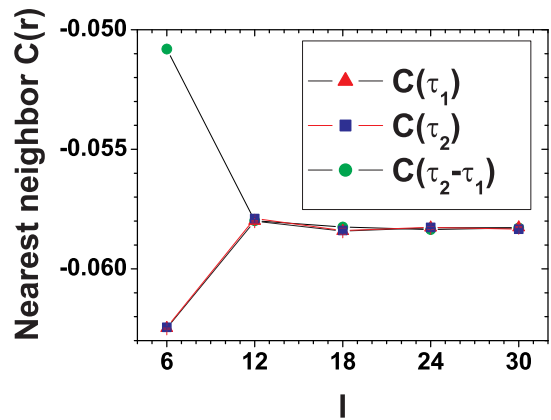}
\begin{center}
\caption{
System size dependence of the nearest neighbor spin-spin correlation functions 
$C(\vec r)$ [Eq.~(\ref{spinspin})] for different cluster sizes $L=l\times l$ 
up to $30\times30$ sites and for the three different directions of the 
triangular lattice 
($\vec r=\vec\tau_1$, $\vec\tau_2$, and $\vec\tau_2-\vec\tau_1$). 
Here $|\psi\rangle=\protect\pbcs$ with 
$\Delta(\vec\tau_1)=\Delta(\vec\tau_2)=\Delta(\vec\tau_2-\vec\tau_1)=1$ 
[see in Fig.~\ref{lattice3} and Eq.~(\ref{pairbroken})],  $t_{i,j}=0$, and 
$\mu=-1$.  
This state is a good but not optimal variational wave function for the 
spatially isotropic model ($J=J'$), and it is used only 
to show that wave functions of this type recover all spatial symmetries of 
the lattice in the thermodynamic limit. 
}
\label{notyet}
\end{center}
\end{figure}

As mentioned above, it is extremely important to show that
the projected BCS state $\pbcs$, constructed from $|{\rm BCS}\rangle$ with 
the paring interactions ${\hat H}_{\rm pair}$, 
preserve all reflection symmetries of the lattice in the thermodynamic limit, 
because otherwise some symmetry broken with finite  order parameter occurs, 
and this 
variational state is no longer a spin liquid. 
This symmetry property is easily proved for the short range RVB case. 
Within open boundary conditions, the short range RVB state 
$|\psi_{\rm RVB}\rangle$ [Eq.~(\ref{srrvb})] and the projected BCS state 
$\pbcs$ with the gap functions defined as in Fig.~\ref{lattice3}, 
$t_{i,j}=0$, and $\mu\to-\infty$ are exactly the same for any finite size 
clusters (see App.~\ref{appendix:pfaffian}). 
Therefore the symmetry of the state $\pbcs$ is implied by the one 
of the the short range RVB state $|\psi_{\rm RVB}\rangle$. 
For large clusters the boundary conditions cannot 
play a role, and hence the symmetry is approximately recovered even 
when periodic boundary conditions are used. 
We have checked numerically in Fig.~\ref{notyet} 
that the reflection symmetry is very well 
preserved in the thermodynamic limit also for the general case where the 
pairing function $f_{i,j}$ is not restricted only to nearest neighbor 
bonds. However in this case we could not rigorously prove our empirical 
observation of this symmetry invariance because the 
equivalence of RVB wave functions and $\pbcs$ no longer 
holds for long range pairing functions.

In order to further improve our variational ansatz wave function, we also 
introduce a simple spin Jastrow factor ${\cal J}_{\rm s}$ into the projected 
BCS wave function described above, {\it i.e.}, 
$\pbcs\to{\cal J}_{\rm S} \pbcs$, where 
\begin{equation} \label{jpbcs}
{\cal J}_{\rm S} = \exp\left[\sum_{{\scriptstyle i,j=1}\atop{\scriptstyle (i<j)}}^L 
v({\vec r}_i-{\vec r}_j) {\hat S}^z_i {\hat S}^z_j\right] 
\end{equation}
and $v(\vec r)$'s are variational parameters which are optimized using the SR 
method explained in Sec.~\ref{srmethod} (also see in App.~\ref{appmethod}). 
Since the most important contributions to the variational energy are from 
the short range $v(\vec r)$'s, in what follows we consider only terms with 
$|\vec r|<3$ and the $v(\vec r)$'s for larger distances are set identically 
to zero. 
As will be discussed later, the inclusion of the spin Jastrow factor 
${\cal J}_{\rm s}$ is also crucial for  stable FN and FNE calculations. 
Similarly, for the gap functions, only $\Delta({\vec t}_m)$'s with 
$|{\vec t}_m|<3$ are considered in the present variational wave function. 
After the SR optimization calculations, it is found that the optimized 
chemical potential $\mu$ is non zero and only nearest neighbor hopping terms 
with $t_{{\bf i},{\bf j}}=1$ 
for all the directions are relevant for the isotropic case ($J=J'$). 
In Fig.~\ref{vmc_energy} and in Tab.~\ref{tableopar}, 
the calculated variational energy is reported and compared with the results 
for other variational ansatz states. 
The improvement of the present variational 
wave function compared with previously studied states is remarkable. 
Our variational energy is even sizably better than a very recent estimate 
reported in Ref.~\onlinecite{mila}, in which a variational ansatz state 
studied includes both a classical N{\'e}el ordered magnetic moment and a 
singlet pairing with symmetry different from ours. 
To our knowledge, the present value for the energy per site in the 
thermodynamic limit, $-0.5357J\pm0.0001J$, is the lowest among the 
variational estimates reported so far.  
Of course, within the present ansatz, this value can be further improved 
by considering longer range terms in  $v(\vec r)$'s and 
$\Delta({\vec t}_m)$'s.

\begin{figure}[hbt]
\vskip 0.7cm
\includegraphics[width=7.5cm,angle=0]{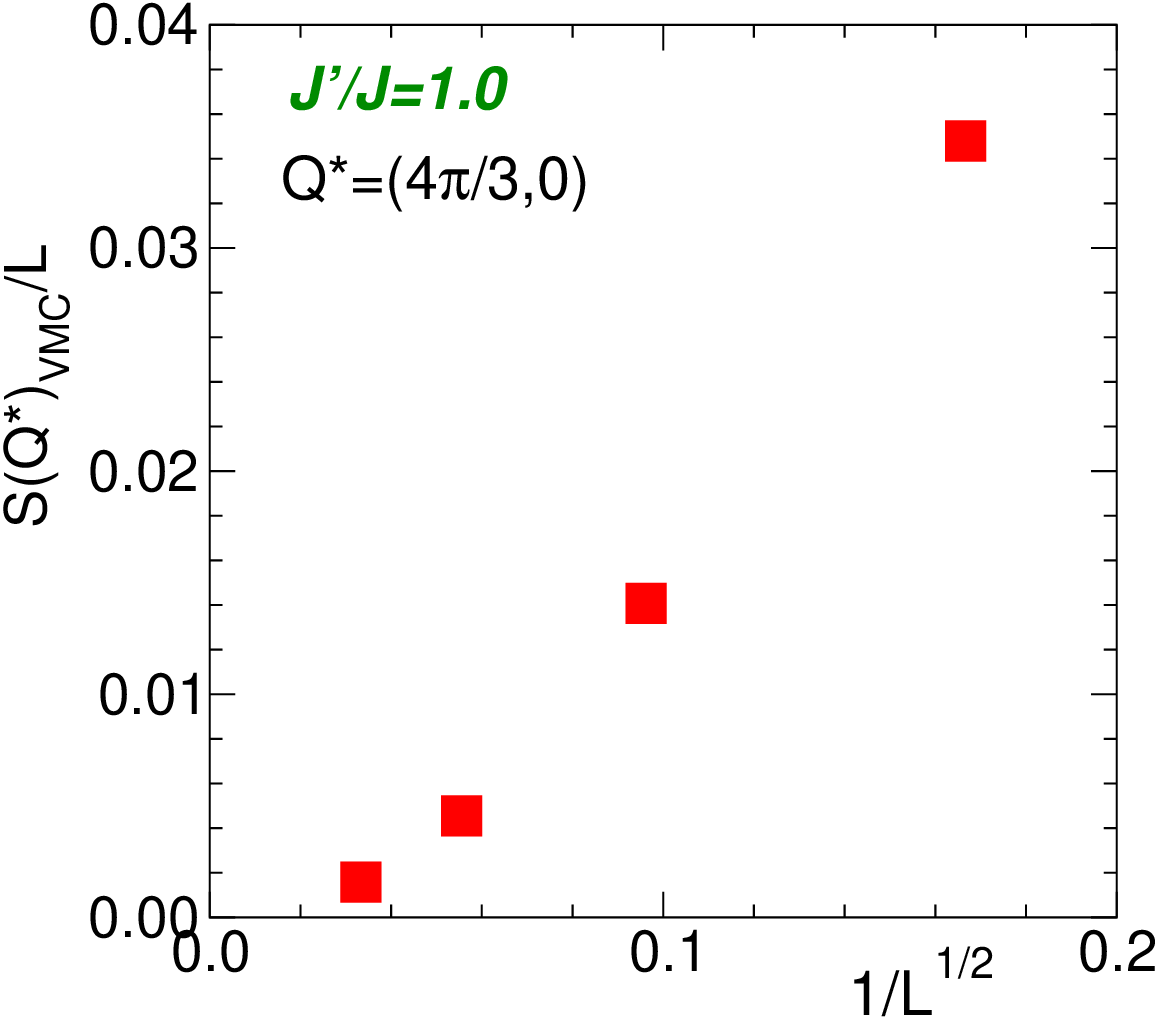}
\begin{center}
\caption{
System size dependence of spin structure factor $S({\vec q})$ at 
${\vec q}={\bf Q}^*=(4\pi/3,0)$ for the optimized variational state 
${\cal J}_{\rm s}\protect\pbcs$ (see the text)
for the spin-1/2 antiferromagnetic Heisenberg model on the isotropic 
triangular lattice of size $L$. Note that 
${\bf Q}^*$ corresponds to the momentum at which $S({\bf Q}^*)/L$ is 
finite for $L\to\infty$ when the state is classical N{\'e}el ordered with 
relative angle of $120^\circ$ between the nearest neighbor spins on the 
different sublattices. 
}
\label{sq_jp10_vmc}
\end{center}
\end{figure}

It is worth mentioning here that in general the BCS excitation spectrum 
$E_{\bf k}$ [Eq.~(\ref{spec2x1bcs})] has a finite gap when the longer range 
gap functions are included in the BCS Hamiltonian as in the present case, 
and therefore, as opposed to the spin liquid state discussed 
in Sec.~\ref{sub:coldea}, the spin liquid state described here by $\pbcs$ 
generally shows a finite gap in spin excitations. 
It should be also noted that, although the variational ansatz state with 
the spin Jastrow factor ${\cal J}_{\rm s}$ breaks the SU(2) spin rotation 
symmetry, non-singular ${\cal J}_{\rm s}$ does not imply long range magnetic 
order. This is clearly seen in Fig.~\ref{sq_jp10_vmc} where the spin 
structure factor 
$$
S({\vec q})=\sum_{\vec r}{\rm e}^{-i{\vec q}\cdot{\vec r}} C(\vec r)
$$
at ${\vec q}=(4\pi/3,0)$ is calculated for the optimized variational wave 
function mentioned above.

\begin{figure}[hbt]
\vskip 0.5cm
\includegraphics[width=7.2cm,angle=0]{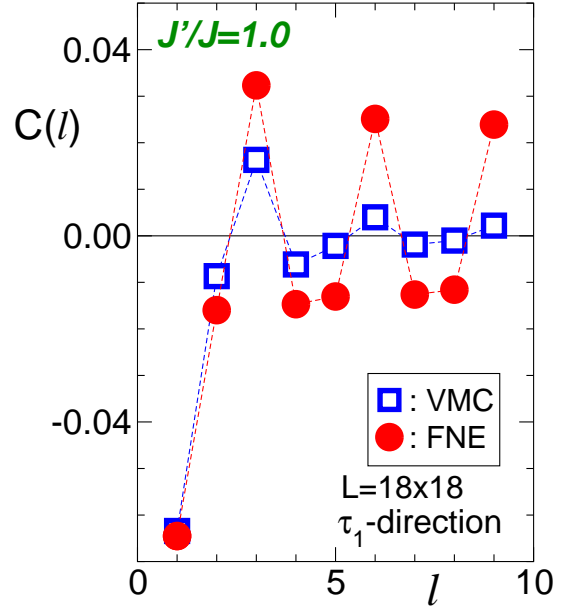}
\begin{center}
\caption{
Spin-spin correlation functions $C(\l{\vec\tau_1})$ in the 
${\vec\tau_1}$ direction for the spin-1/2 antiferromagnetic Heisenberg model 
on the isotropic triangular lattice with $L=18\times18$ calculated 
using both variational Monte Carlo (VMC) and effective Hamiltonian (FNE) 
methods. The variational wave function considered here is the projected BCS 
state ${\cal J}_{\rm s}\protect\pbcs$ with spin 
Jastrow factor ${\cal J}_{\rm s}$ (see the text), and this wave function is 
used as the guiding function $|\psi_G\rangle$ for the FNE calculation. 
The spin-spin correlation functions in the ${\vec\tau_2}$ and 
$\vec\tau_2-\vec\tau_1$ directions are essentially the same as the ones 
presented here. 
}
\label{sr_jp10}
\end{center}
\end{figure}

Although the projected BCS wave function considered here is a spin 
liquid state with a finite spin gap and without any type of long range 
magnetic order (Fig.~\ref{sq_jp10_vmc}), when the FN or FNE method is 
applied with $|\psi_G\rangle={\cal J}_{\rm s}\pbcs$, long range magnetic 
order appears with a finite magnetic moment in the $z$-direction 
(the effective Hamiltonian approach breaks the spin rotational invariance 
in this case and no detectable magnetic moment is observed in the other 
directions).
Indeed, as shown in 
Fig.~\ref{sr_jp10}, the spin-spin correlation functions $C(\vec r)$ 
along the $\vec\tau_1$ direction are drastically increased for 
the FNE ground state $|\psi_0^{\rm eff}\rangle$, 
compared with the ones for the spin liquid state ${\cal J}_{\rm s}\pbcs$. 
Furthermore, the oscillations observed in $C(\vec r)$ for the FNE ground 
state (Fig.~\ref{sr_jp10}) are consistent with the classical N{\'e}el 
order.  
These results strongly indicate that, in spite of its good variational energy, 
the spin liquid state is not stable against magnetic order for the 
isotropic case.

\begin{figure}[hbt]
\vskip 0.0cm
\includegraphics[width=5.5cm,angle=-90]{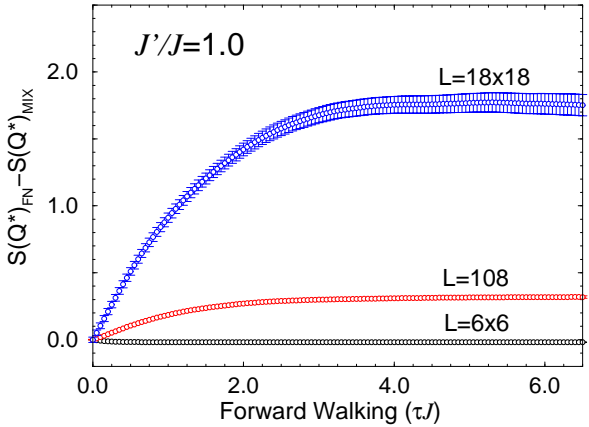}
\begin{center}
\caption{
Forward walking time evolution $\tau$ (unit $J^{-1}$) [Eq.~(\ref{forward2})] 
 of the spin structure factor $S({\vec q})$ 
at ${\vec q}={\bf Q}^*=(4\pi/3,0)$ for $J'/J=1.0$. Here our best variational 
wave function $\protect\pbcs$ with spin Jastrow factor ${\cal J}_{\rm s}$ 
(see the text) is used as the guiding function $|\psi_G\rangle$. 
For clarity, for each cluster, the FN results $S({\bf Q}^*)_{\rm FN}$ 
are referenced to the mixed-average estimate $S({\bf Q}^*)_{\rm MIX}$, 
which is set to zero. 
}
\label{forward}
\end{center}
\end{figure}

The FN and FNE calculations of the spin-spin correlation functions were 
carried out by applying the ''forward walking'' technique developed  
in Ref.~\onlinecite{calandra}, which will be described briefly in the 
following. With this technique, 
the expectation value of any operator ${\hat O}$ 
diagonal in configuration space $\{x\}$, can be 
computed for the ground state of the effective Hamiltonian.
This scheme removes completely the bias of the 
so-called mixed average estimate,~\cite{mixed}
$$ O_{\rm MIX} = 
{ \langle \psi_G  | {\hat O} |  \psi_0^{\rm eff} \rangle \over 
\langle \psi_G  |  \psi_0^{\rm eff} \rangle }, $$ 
by a large ``forward-walking'' time $\tau$ projection of the 
guiding function $|\psi_G\rangle$ in the above expression. 
More precisely, the expectation value 
of ${\hat O}$ for the state  $|\psi_0^{\rm eff}\rangle$ is computed 
through the equation 
\begin{equation} \label{forward2}
{ \langle\psi_0^{\rm eff} | {\hat O} | \psi_0^{\rm eff}\rangle \over 
\langle\psi_0^{\rm eff} |  \psi_0^{\rm eff} \rangle }
= \lim_{\tau\to\infty}
{ \langle \psi_G  |{\rm e}^{-\tau{\hat H}_{\rm eff}} {\hat O} |  \psi_0^{\rm eff} \rangle \over 
\langle \psi_G  | {\rm e}^{-\tau{\hat H}_{\rm eff}}|  \psi_0^{\rm eff} \rangle },
\end{equation}
where the simple relation 
$\lim_{\tau\to\infty}{\rm e}^{-\tau{\hat H}_{\rm eff}}|\psi_G\rangle\propto|\psi_0^{\rm eff} \rangle$ is used. 
This scheme provides for a quantitative estimate of the magnetic moment 
(order parameter), as shown in Fig.~\ref{forward}, where one can clearly see
that the values of the spin structure factor $S(\vec q)$ at the commensurate 
wave vector $\vec q={\bf Q}^*=(4/3 \pi,0)$ considerably increase with 
cluster sizes and 
with the forward-walking projection time $\tau$, the $\tau=0$ value 
corresponding to the much less accurate mixed average estimate.
In this figure, it is also apparent 
that a satisfactory convergence in $\tau$ [Eq.~(\ref{forward2})] is 
always obtained for the clusters studied.  
From the technical point of view, this calculation was made possible  
due to the quality of our variational  ansatz. For instance, 
without the  inclusion of the spin Jastrow 
factor ${\cal J}_{\rm s}$  a  
much longer forward-walking time $\tau$   
is necessary to achieve a reasonable convergence, with an almost 
prohibitive  computational 
effort to reduce the statistical errors  to an acceptable value.

\begin{figure}[hbt]
\vskip 0.9cm
\includegraphics[width=8.cm,angle=-0]{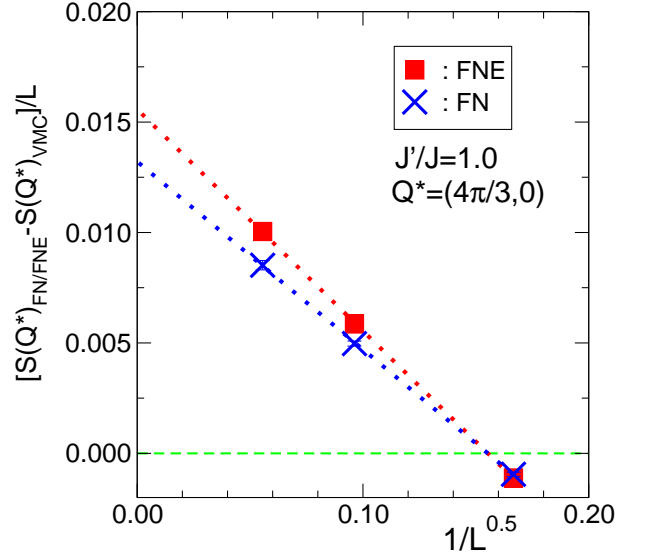}
\begin{center}
\caption{
System size dependence of the spin structure factor $S({\vec q})$ at 
${\vec q}={\bf Q}^*=(4\pi/3,0)$ for the spin-1/2 isotropic triangular 
antiferromagnet with $J'/J=1.0$ calculated using FN and FNE methods. 
Here our best variational wave function ${\cal J}_{\rm s}\protect\pbcs$ 
with spin Jastrow factor ${\cal J}_{\rm s}$ (see the text) is used as the 
guiding function 
$|\psi_G\rangle$. In order to reduce finite size effects, for each cluster, 
the variational result $S({\bf Q}^*)_{\rm VMC}/L$ for the spin liquid state 
${\cal J}_{\rm s}\protect\pbcs$ (which is zero for $L \to \infty$) is 
subtracted from the FN and FNE results $S({\bf Q}^*)_{\rm FN/FNE}/L$. 
}
\label{sq_jp10}
\end{center}
\end{figure}

Encouraged by these results, let us finally study the system size 
dependence of the spin structure factor $S(\vec q)$ at the commensurate 
wave vector $\vec q={\bf Q}^*$. 
The results calculated using both FN and FNE methods, $S({\bf Q}^*)_{\rm FN}$ 
and $S({\bf Q}^*)_{\rm FNE}$, 
are presented in Fig~.\ref{sq_jp10}. 
For each cluster, the FN and FNE Hamiltonians are constructed using the 
optimized variational state ${\cal J}_{\rm s}\pbcs$ as 
the guiding function $|\psi_G\rangle$. 
In this figure, to reduce the finite size effects for estimating the 
magnetic order parameter, the difference between the FN/FNE results 
$S({\bf Q}^*)_{\rm FN/FNE}$ and the corresponding variational estimates 
$S({\bf Q}^*)_{\rm VMC}$ is plotted. 
The difference should be extensive if there exists long range 
antiferromagnetic order, as the variational results 
$S({\bf Q}^*)_{\rm VMC}/L$ for the spin liquid state ${\cal J}_{\rm s}\pbcs$ 
decreases to zero in the limit $L\to\infty$ (Fig.~\ref{sq_jp10_vmc}).  
It is clearly observed in Fig~.\ref{sq_jp10} that the FN and FNE results for 
$S({\bf Q}^*)/L$ converge to finite values in the thermodynamic limit.

From these calculations we can estimate the magnetic moment quantitatively. 
To this purpose, it has to be taken into account that the long range magnetic 
order occurs in the $z$-direction because the FN Hamiltonian 
${\hat H}^{\rm FN}$ and the FNE Hamiltonian ${\hat H}^{\rm eff}$ 
do not commute with the total spin operator and display magnetic order only 
in the $z$-direction.~\cite{xypar}
Therefore, the magnetic structure factor is related to the magnetic order 
parameter (magnetic moment) $m$ up to a simple factor, namely, 
$S({\bf Q}^*)_{\rm FN/FNE}/L\to  m^2/2$.
Our estimates of $m$ are summarized in 
Tab.~\ref{tableopar} along with the ones of the  energy, $E_0^{\rm FN}$ and 
$E_0^{\rm eff}$, in the thermodynamic limit. For comparison, 
the estimates calculated by other methods and for different 
wave functions are also provided in Tab.~\ref{tableopar}. 
It is remarkable that even though a spin liquid wave function is used here 
as a guiding wave function, a finite magnetic moment $m$ is 
obtained with the FN and FNE approaches. These results clearly indicate that 
the spin liquid is eventually unstable in the isotropic model. 
It should be noted here that the magnetic moment estimated with approximate 
techniques such as the FN and FNE methods is considered as a lower bound 
because they are biased toward 
non magnetic order by the spin liquid guiding function.

It is also important to notice in Fig.~\ref{sq_jp10} and in 
Tab.~\ref{tableopar} that the FNE approach, which is a better 
variational method than the standard FN method, estimates a sizably larger 
value of $m$ compared with  the FN result, much closer to previous estimates 
based on different guiding  functions with explicit magnetic 
order.~\cite{caprio} 
In the previous work,~\cite{caprio} though the energy 
was not rigorously variational unlike in the present case,   
similar  corrections to the guiding function 
were implemented.~\cite{notesrvsfne} 
It is therefore very interesting that, by using two different variational wave 
functions with (overestimated) or without (strongly underestimated) magnetic 
order, both with  good variational energy, and by applying very similar 
techniques, a finite order parameter $m$ for the isotropic 
triangular lattice 
is obtained, rather independently of the guiding function used. 
This is a very important property of  
methods, such as FN, FNE, and the 
previously introduced Green function Monte Carlo method with 
stochastic reconfiguration,~\cite{caprio} 
which are all based on the numerical solution of an effective Hamiltonian.
 On the contrary, the simple variational 
approach does not lead to a reliable prediction for the magnetic moment $m$ 
simply because very different variational ansatz states with or without a 
finite magnetic moment may have very similar energy (see {\it e.g.}, 
Fig.~\ref{vmc_energy} and Tab.~\ref{tableopar}) but opposite long distance 
behavior. 
This is an additional confirmation that the present approach, 
a systematic correction to the variational ansatz, is very 
useful for understanding quantitative physical properties of 
strongly frustrated quantum systems, for which either numerically or 
analytically exact solutions are not known.

%%%%%%%%%%%%%%%%%%%%%%%%%%%%%%%%
\subsection{Nearly isotropic triangular lattice with $J'\alt J$: a spin liquid 
with a small spin gap} \label{secresults:liquid}
%%%%%%%%%%%%%%%%%%%%%%%%%%%%%%%%

In the previous subsection, we have shown a robust 
numerical evidence that a spin liquid state is not the ground state 
for the isotropic triangular antiferromagnet. 
However, it is natural to expect that quantum fluctuations become enhanced by 
increasing the spatial anisotropy $J/J'>1$, and that the magnetically ordered 
state eventually melts and a true spin liquid phase would emerge. 
In this subsection, we shall extend the spin liquid ansatz wave 
function discussed above away from the isotropic point, and study the 
stability of the spin liquid state with the FN or FNE method, as it was done 
successfully in the isotropic case.

\subsubsection{ Stability against magnetic ordering }

As reported in  previous studies,~\cite{kenzie_2} a simple semiclassical 
solution implies that the spin structure factor $S({\vec q})$ displays 
Bragg peaks at incommensurate momenta even slightly away from the isotropic 
case with $J'/J \ne  1$. 
It is confirmed by our variational approach, which provides 
a much better variational state compared to the classical solution, that 
these peaks appear in $S({\vec q})$ and their locations in the 
Brillouin zone smoothly evolve from the 
commensurate momenta ${\bf Q}^*=(4\pi/3,0)$ (and equivalent ones) to the 
incommensurate ones  within our accessible finite size clusters.

\begin{figure}[hbt]
\vskip 0.75cm
\includegraphics[width=7.5cm,angle=0]{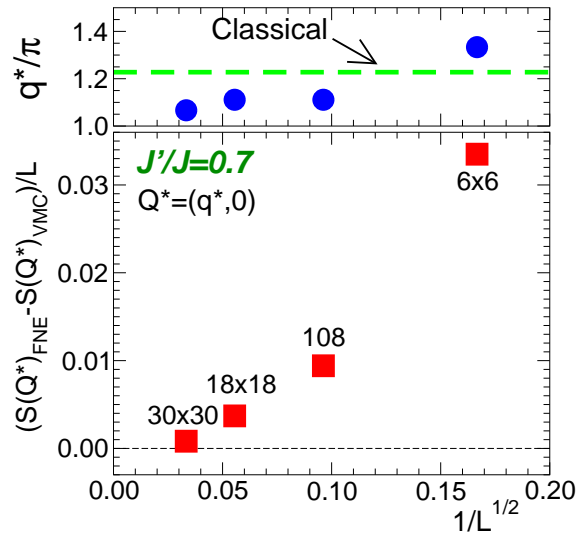}
\begin{center}
\caption{
System size dependence of the spin structure factor $S({\vec q})$ at 
${\vec q}={\bf Q}^*=(q^*,0)$ for $J'/J=0.7$ calculated using the FNE method. 
The incommensurate momenta $q^*$ as well as the corresponding value for the 
classical limit~\cite{kenzie_2} are plotted in the upper panel. 
For clarity, the variational estimates of the structure factor 
$S({\bf Q}^*)_{\rm VMC}$ are subtracted from the FNE results 
$S({\bf Q}^*)_{\rm FNE}$.  
The optimized projected BCS state $\protect\pbcs$ without 
spin Jastrow factor ${\cal J}_{\rm s}$ (see the text) is used for 
the variational calculations. The same wave function $\protect\pbcs$ is used 
as the guiding 
function $|\psi_G\rangle$ for the FNE calculations. The system sizes used 
are indicated in the figure. 
}
\label{sq_jp07}
\end{center}
\end{figure}

At the variational level with the same type of spin liquid wave functions 
considered for the isotropic case (see also Tab.~\ref{tablesondhi2}), 
the spin structure factor is finite 
for these incommensurate peaks, and therefore, as opposed to the classical 
solution, no long range magnetic order is implied.  
To study this property within the FN or FNE approach, we have to note 
that it is very difficult to perform a finite 
size scaling analysis when incommensurate correlations are studied, 
a situation which is further complicated by 
the proximity to a possible phase transition from a magnetic to a non 
magnetic ground state, because the value of $m$, as we have shown, 
is rather small already in the isotropic case. 
In order to carry out a reliable finite size scaling, in the following, 
we consider a sizable anisotropy within the hypothesis to be far enough from 
the critical point (which is unaccessible within the  
finite size clusters studied, $L \lesssim 1000$).
Indeed, for $J'/J=0.7$, we are able to successfully carry out the 
analysis of the FNE calculations, and the results are 
presented in Fig.\ref{sq_jp07}. Here the guiding wave function 
$\pbcs$ does not require the spin Jastrow factor ${\cal J}_{\rm s}$ to be 
accurate enough, and it consists of 
$\Delta({\vec t}_m)$ with $|{\vec t}_m|<3$ [Eq.~(\ref{pairbroken})] as 
well as the hopping integrals and the chemical potential. All these parameters 
are optimized using the SR minimization method. 
As shown in Tab.~\ref{tablesondhi2}, it is found that  
the optimized chemical potential is non zero, and only the 
nearest neighbor hopping in the $\vec\tau_1$ direction is relevant 
and can be set to unity ($t_{\vec\tau_1}=1$).

\begin{table}
%\begin{ruledtabular}
\renewcommand{\arraystretch}{1.2}
\begin{tabular}{l| l|l}
\hline \hline
  $~J'/J$          & ~~~~~0.7       & ~~~~~0.8  \\
\hline
  ~~~~$\mu$   &  ~~0.304(3)  & ~~0.243(6) \\
\hline
  $\Delta(1,0)$   &  ~~2.01(3)  & ~~2.33(4)       \\
  $\Delta(0,1)$   &  ~~1.08(2)  & ~~1.42(2)      \\
  $\Delta(1,1)$   &  ~~0.205(3) & ~~0.215(3)      \\
  $\Delta(-1,2)$   &  ~~0.002(3)& ~~0.001(3)       \\
  $\Delta(2,0)$   & ~~$-0.36(1)$& ~~$-0.32(2)$        \\
  $\Delta(0,2)$   &  ~~0.01(1)  & ~~$-0.02(3)$       \\
  $\Delta(2,1)$   &  ~~0.011(2) &  ~~0.063(2)     \\
  $\Delta(1,2)$   &  ~~0.002(2) &  ~~0.001(2)     \\
  $\Delta(-2,3)$  & ~~0.001(1)  &  ~~0.0008(9)     \\
\hline
  ~~~~$K^{-1}$   &  ~~1.1886(3)  &  ~~1.2242(3) \\
\hline
  $E_{\rm VMC}/JL$   &  $-0.46467(3)$  &  $-0.47840(3)$ \\
\hline
  ~$E_{\rm FN}/JL$   &  $-0.47051(2)$  &  $-0.48521(2)$   \\
\hline
  $E_{\rm FNE}/JL$   &  $-0.47171(3)$  &  $-0.48691(4)$  \\
\hline \hline
\end{tabular}
%\end{ruledtabular}
\caption{The optimized variational parameters of the projected BCS wave 
function for $J'/J=0.7$ and 0.8 with $L=18\times18$. 
$\Delta(m_1,m_2)$ is the singlet gap function $\Delta({\vec t}_m)$
for 
${\vec t}_m = m_1{\vec\tau}_1 +  m_2{\vec\tau}_2$ in Eq.~(\ref{pairbroken}), 
and $\mu$ is the chemical potential. The remaining variational parameters are 
zero except for the nearest neighbor hopping in the ${\vec\tau_1}$ direction, 
which is chosen to be one. 
The value of $K$, variational energy $E_{\rm VMC}=E(\protect\pbcss)$, 
FN (FNE) ground state energy $E_{\rm FN}=E_0^{\rm FN}$ 
($E_{\rm FNE}=E_0^{\rm eff}$) with $|\psi_G\rangle=\protect\pbcs$ are also 
presented.
The number in parentheses is the statistical error bar corresponding to the 
last digit of the figure. 
}
\label{tablesondhi2}
\end{table}

The excitation spectrum $E_{\bf k}$ of the corresponding BCS 
Hamiltonian are analytically calculated in App.~\ref{appendix:sondhi}, 
from which several very interesting features are observed.
If the gap functions $\Delta({\vec t}_m)$ are restricted to the nearest 
neighbors, the spectrum of the BCS Hamiltonian is generically gapless 
(provided $\mu= 0$). 
Once the gap functions are non-zero for longer bonds, a tiny 
energy gain is obtained, and correspondingly a finite excitation 
gap in $E_{\bf k}$ is opened. However this gap is very small. 
In fact, it is found that the low energy 
BCS spectrum $E_{\bf k}$ on finite size is almost indistinguishable from  
a gapless one, and that the lowest excitation gap in $E_{\bf k}$ is estimated 
as small as $\sim0.3\%$ of $2W$ for $J'/J=0.7$ where $W$ is the maximum 
excitation energy of $E_{\bf k}$.~\cite{note6}  
The reason for the appearance of a finite excitation gap in $E_{\bf k}$ 
is simply understood because the BCS Hamiltonian with broken translation 
symmetry has two sites per unit cell, and thus the BCS spectrum is in 
general gapped. 
It should be emphasized here that, as explained in the 
previous  subsection, after applying the projection ${\cal P}_{\rm G}$ onto 
the GS $|{\rm BCS}\rangle$ of this BCS Hamiltonian, the translation symmetry 
of the projected BCS state $\pbcs$ is recovered. 
Therefore, a quite new, 
remarkable possibility to form a spin liquid state with a finite spin 
gap~\cite{note3} and with a single spin per unit cell can be easily 
established here within the present variational framework 
{\em without breaking the translation symmetry}.

Let us now discuss the static magnetic properties for $J'/J=0.7$.
As seen in Fig.\ref{sq_jp07}, the finite size scaling analysis of the FNE 
results for the spin structure factor clearly indicates a vanishing magnetic 
order parameter. We have also found that the FNE spin structure factors as 
well as the FNE spin-spin correlation functions do not differ significantly 
from the ones computed for the variational spin liquid state $\pbcs$ 
explained above (Tab.~\ref{tablesondhi2}). 
This should be contrasted to the isotropic case shown in 
Sec.~\ref{secresults:sondhi}, where significant differences were observed. 
These results strongly suggest that the spin liquid ansatz state described 
here by the projected BCS wave function $\pbcs$ is stable within the present 
approach. Therefore, another spin liquid region different from the one 
discussed in Sec.\ref{sub:coldea} might exist in the anisotropic 
triangular lattice with $J'/J$ closer to one, the state of which can be 
described, at least approximately, by this proposed spin liquid wave 
function.

\subsubsection{ Stability against spontaneous dimerization }

\begin{figure}[hbt]
\vskip 0.75cm
\includegraphics[width=6.8cm,angle=0]{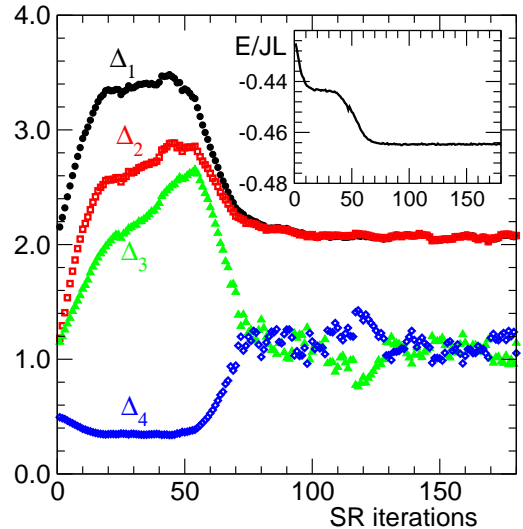}
\begin{center}
\caption{
Monte Carlo evolution of the variational parameters defined in 
Eq.~(\protect\ref{defdelta}) as a function of SR iterations 
for the spin-1/2 antiferromagnetic Heisenberg model on the anisotropic 
triangular lattice with $J'/J=0.7$ and 
$L=18\times18$. Here the SR minimization method explained in 
Sec.~\ref{srmethod} is used with $\delta t=0.25/J$ [Eq.~(\ref{iterforce})]. 
Inset: the corresponding energy evolution as a function of SR iterations. 
At each SR iteration, the energy is computed for the wave function with 
the fixed variational parameters given at the corresponding iteration 
in the main figure. 
}
\label{hist}
\end{center}
\end{figure}

Since the BCS Hamiltonian is defined with a $(2\times1)$ unit 
cell [Eq.~(\ref{pairbroken})], it is natural to ask ourselves  
if a valence bond $solid$ (not $liquid$) would be stabilized, which 
should also exhibit a finite spin excitation gap at the expense of a 
broken translation symmetry. 
In order to examine whether the valence bond solid is energetically more 
favored than the spin liquid state for the present anisotropic 
antiferromagnetic Heisenberg model, 
it is sufficient to study the translation symmetry of the optimized 
projected state because the valence bond solid necessarily breaks this 
symmetry. 
This possibility can be easily considered within our approach 
by using a BCS Hamiltonian that is not invariant under the 
transformation  ${\cal T}_1 \bigotimes {\cal U}$. 
In Eq.~(\ref{pairbroken}) it is assumed that the gap  function 
$\Delta_{{\bf i},{\bf j}}$ connecting sites ${\bf i}$ and ${\bf j}$ 
depends only on  ${\bf j}-{\bf i}$ up to the sign. 
To check the possible instability 
toward the valence bond solid, we have released these constraints for 
$\Delta_{{\bf i},{\bf j}}$ but with remaining in the same $(2\times 1)$ 
unit cell; for example, $\Delta_{0,\vec\tau_1}$ can be different from 
$\Delta_{\vec\tau_1,2\vec\tau_1}$, but 
$\Delta_{0,\vec\tau_1}=\Delta_{2\vec\tau_1,3\vec\tau_1}$. 
We have then optimized independently the first four 
$\Delta_{{\bf i},{\bf j}}$'s at the shortest distances:
\begin{equation}\label{defdelta}
\left\{\begin{array}{l}
\Delta_1 = \Delta_{0,\vec \tau_1}, \\[0.1cm]
\Delta_2 = \Delta_{\tau_1,\vec   2 \tau_1}, \\[0.1cm]
\Delta_3 = \Delta_{0,\vec \tau_2},~ {\rm and}   \\[0.1cm]
\Delta_4 = -\Delta_{\tau_1,\tau_1+\vec \tau_2}, 
   \end{array}
  \right.
\end{equation}
whereas all the other ones are kept fixed at the values obtained 
assuming the translation invariant ansatz (shown in Tab.~\ref{tablesondhi2}). 
Whenever  the optimized parameters satisfy $\Delta_1=\Delta_2$ and 
$\Delta_3=\Delta_4$ within the statistical errors, the projected BCS state 
is a  spin liquid. Otherwise the optimized state is a valence bond solid 
simply because this state is no longer translation invariant. 
In order to optimize these variational parameters, 
we have used the SR minimization 
method described in Sec.~\ref{srmethod}. Our results are presented in 
Fig.~\ref{hist}, where the Monte Carlo evolution of these four parameters are 
plotted for $J'/J=0.7$ and 
$L=18\times18$. It is clearly seen in Fig.~\ref{hist} that, although the 
initial values of the parameters 
are far off form the symmetric condition ($\Delta_1=\Delta_2$ and 
$\Delta_3=\Delta_4$), after a few hundred  SR iterations these parameters 
converge to the symmetric values (Tab.~\ref{tablesondhi2}). 
The inset of Fig.~\ref{hist} shows the 
Monte Carlo evolution of the corresponding energy as a function of the SR 
iterations (each iteration corresponds to a small variational Monte 
Carlo simulation with fixed variational parameters). 
After the first few SR iterations the energy appears  to be   
trapped in  a metastable state with a broken symmetry solution, but then 
after one hundred SR iterations the energy eventually converges to 
a lower value corresponding to the spin liquid  fully symmetric 
solution. These results, therefore, strongly indicate 
that the optimized state is translation invariant, and spontaneous 
dimerization is very unlikely to occur in this model because it is not 
stabilized even when the variational ansatz wave function allows to stabilize 
this kind of order.

\begin{figure}[hbt]
\vskip 0.75cm
\includegraphics[width=7.5cm,angle=-0]{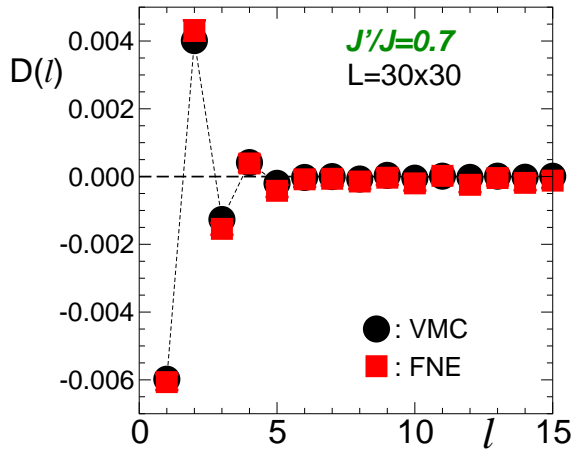}
\begin{center}
\caption{
Dimer-dimer correlation functions $D({\vec r})$ with ${\vec r}=\l{\vec\tau_1}$ 
for the spin-1/2 triangular antiferromagnet with $J'/J=0.7$ and $L=30\times30$
calculated using both variational Monte Carlo (VMC) and FNE methods. 
The variational ansatz state considered here is the projected BCS state 
$\protect\pbcs$ (see text), and this state is used as the guiding function 
$|\psi_G\rangle$ for the FNE calculations. 
}
\label{dim_jp07}
\end{center}
\end{figure}

In principle, as shown in Ref.~\onlinecite{chiral} for 1D 
systems, also a translation invariant state can 
give rise to a spontaneously dimerized order after applying 
the projection operator ${\cal P}_{\rm G}$ on this state. 
In order to rule out this possible order, we have also calculated 
explicitly the dimer-dimer correlation functions $D(\vec r)$ defined 
in Eq.~(\ref{def_dim}) for the variational wave function described 
above (see also Tab.~\ref{tablesondhi2}).
The results 
are compared with the ones computed using the FNE method. 
A typical example of the results is presented in Fig.~\ref{dim_jp07} for 
$J'/J=0.7$ and $L=30\times30$.~\cite{note1} 
As seen in Fig.~\ref{dim_jp07}, 
the two results do not show significant differences, 
confirming  that   the spin liquid 
variational ansatz appears very stable in this case.  
It should be emphasized once again that the  
situation is very different from the isotropic case  
where a magnetic instability was clearly detected with 
the FNE method (Sec.~\ref{secresults:sondhi}).

%\begin{figure}[hbt]
%\includegraphics[width=7.cm,angle=0]{sondhi_0.eps}
%\begin{center}
%\caption{
%Phases used for the states for $J'/J\sim1.0$. 
%}
%%\label{phases}
%\end{center}
%\end{figure}

The numerical calculations presented here strongly suggest that 
a  new type of spin liquid discussed here is an appropriate ground state 
description of the spin-1/2 antiferromagnetic Heisenberg model on the 
anisotropic 
triangular lattice, at least in the region 
around $J'/J\approx0.7$--0.8 (also see Sec.~\ref{conclusion}).  
In the present work, we have not attempted 
to determine the critical value of $J'$ above which 
the incommensurate magnetically ordered state is stable,
which was previously suggested in, 
{\it e.g.}, Ref.~\onlinecite{kenzie_2}.
Indeed, there exists another more interesting phase boundary
between the two possible 
spin liquid states, the one presented here and the one considered in 
Sec.~\ref{sub:coldea}, which will be discussed in the next section.

%SSSfig

%%%%%%%%%%%%%%%%%%%%%%%%%%%%%%%
%%%%%%%%%%%%%%%%%%%%%%%%%%%%%%%
\section{Conclusions and final remarks}\label{conclusion}
%%%%%%%%%%%%%%%%%%%%%%%%%%%%%%%
%%%%%%%%%%%%%%%%%%%%%%%%%%%%%%%

\begin{figure}[hbt]
\vspace{1.cm}
\includegraphics[height=7.cm,angle=0]{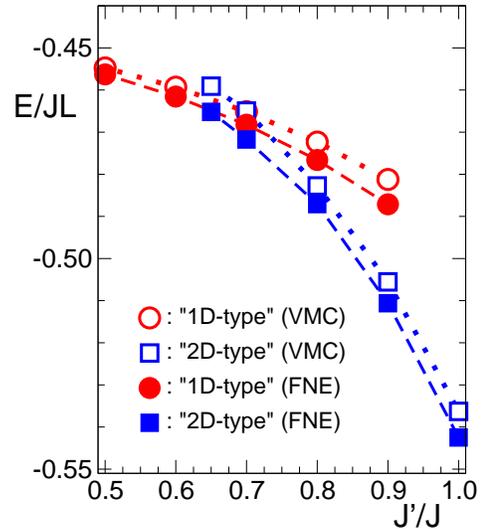}
\begin{center}
\caption{ 
Energy per site for the spin-1/2 antiferromagnetic Heisenberg model 
on the spatially anisotropic triangular lattice calculated  
using the variational Monte Carlo (VMC) and the effective Hamiltonian (FNE) 
methods. Here two types of spin liquid wave functions are considered, 
the one described in Sec.~\ref{sub:coldea} (denoted by ``1D-type'') and 
the other described in Sec.~\ref{secresults:sondhi} (denoted by ``2D-type''). 
These wave functions are used as the guiding functions for the FNE 
calculations. 
}
\label{phaseenergy}
\end{center}
\end{figure}

In this paper, using various quantum Monte Carlo techniques, we have studied 
the ground state phase diagram as well as the low-lying spin excitations for 
the spin-1/2 antiferromagnetic Heisenberg model on the triangular lattice 
as a function of the spatially anisotropic coupling $J'/J$ 
(Fig.~\ref{lattice}). We have found numerical evidence for the presence of 
two different spin liquid states. The first spin liquid  
(``algebraic spin liquid'') is stable for $J'/J\alt0.65$ 
(see Fig.~\ref{phaseenergy}), and is characterized by gapless spin 
excitations, thus very similar to 1D spin liquids. 
Conversely, the other spin liquid is rather a new type of spin liquid state, 
stable for $0.65\alt J'/J\alt0.8$, and should show a small 
spin excitation gap. 
Starting from the isotropic limit $J'=J$ where the ground state is 
magnetically ordered (classical N{\'e}el ordered), quantum fluctuations 
increase strongly with decreasing the coupling $J'/J$ down to zero. 
Therefore, the stability of a spin liquid in this region of the phase diagram 
is quite clear.

The critical coupling $J'_{\rm C}$ discriminating these two spin liquid 
phases can be determined in principle by comparing the corresponding energy, 
and our best estimates of the energy as a function of $J'/J$ are 
summarized in Fig.~\ref{phaseenergy}. From these results, it is concluded 
that the critical coupling $J'_{\rm C}/J$ is about 0.65. 
Because of the limitation of currently available cluster sizes, our approach 
can not either determine precisely the critical coupling or describe 
accurately the nature of the transition. 
At the variational level, a first order transition occurs at a critical 
point where the energy curves of the two spin liquid phases (as a function of 
$J'/J$), shown in 
Fig.~\ref{phaseenergy}, intersect with different slopes. 
In the same figure, it is also clear that the two slopes become very close 
within the FNE calculations, suggesting that the transition would eventually 
turn to a conventional second order transition, when the quantum 
fluctuations are more accurately  taken into account.

Likewise, we have not tried to determine the critical coupling and 
to study the nature of the transition between the spin liquid phase and 
the magnetically ordered phase with incommensurate magnetic order. 
The latter phase should appear somewhere around 
$J'/J \gtrsim 0.8$.~\cite{kenzie_2} 
However, it is very difficult to perform a finite size scaling analysis to 
determine the magnetically ordered incommensurate phase simply because 
too large clusters  are necessary for an accurate estimate of the magnetic 
moment.

Our numerical results were obtained  using the quantum 
variational Monte Carlo method as well as the lattice FN and FNE methods, 
which are essentially the Green function quantum Monte Carlo method with 
the  fixed node approximation. The FNE (``effective Hamiltonian approach'') 
is a further improved version of the standard FN method and developed here 
in Sec.~\ref{fnemethod}. 
Although our results might be affected by this approximation in general, 
we have shown that the present methods provide very sensible results for 
the isotropic 
triangular antiferromagnet with $J'=J$ (Sec.~\ref{secresults:sondhi}), and the 
numerically exact results for the strongly anisotropic limit of the 1D 
uncoupled chains with $J'=0$ (Sec.~\ref{sub:1d}). Therefore, 
we have obtained rather reliable results for both limits of the phase diagram, 
and the same numerical tools have been applied also to the still 
controversial region of the phase diagram for $ 0 < J'/J<1$.

The quality of our approximation in the FN and FNE methods 
depends mostly on the choice of the guiding 
function $|\psi_G\rangle$, for which we used the best variational ansatz 
state,  optimized using the SR minimization method (Sec.~\ref{srmethod}). 
The optimization of $|\psi_G\rangle$ is a crucial ingredient of our 
approach, because the approximate ground state which we consider, 
{\it i.e.} the numerically exact 
ground state $|\psi_0^{\rm eff}\rangle$ of the effective Hamiltonian 
${\hat H}^{\rm eff}$, can be 
computed with  the restriction to have the same signs of $|\psi_G\rangle$. 
Indeed, on small clusters for which numerically exact 
diagonalization of the systems can be done, the variational ansatz state 
described by an projected BCS wave function $\pbcs$ provides 
a very good average sign $\langle S\rangle$ [Eq.~(\ref{sign_def})]  
for both frustrated and non frustrated systems as discussed in 
Sec.~\ref{secresults:sondhi} and 
also in Refs.~\onlinecite{caprio},~\onlinecite{letter}, 
and~\onlinecite{arrachea}.

Within this approach, which was shown quite reliable, we have 
found a surprisingly stable spin liquid (``algebraic spin liquid'') 
phase in the regime of large 
anisotropic couplings $J/J'\alt0.65$ (Sec.~\ref{sub:coldea}). 
Our numerical calculations also indicate that 
this spin liquid state shows gapless, fractionalized spin 
excitations 
(Fig.~\ref{bz_jp033} and Fig.~\ref{bz_jp05}).~\cite{wen,wenfermion} 
Therefore, we predict that this 2D algebraic spin liquid state should 
show peculiar low energy properties similar to the 1D systems. 
Although our conclusion is based on numerical calculations for rather large 
clusters (up to $42\times42$ sites), 
to be fair, we cannot rule out a very weak 
instability toward symmetry-broken ordered states, as predicted in 
the $J'/J \to 0$ limit by using the susceptibility criterion based on a 
random phase approximation (RPA).~\cite{rpa} 
Even in that  study, the instability occurs in 
an irrelevantly small temperature region, as also pointed out by the 
authors.~\cite{rpa} 
Moreover, it is not clear how reliable the RPA calculation is for finite 
values of $J'/J$.

We have also found another type of spin liquid phase in the region $J'/J$ 
close to the isotropic limit, {\it i.e.}, for $0.65\alt J'/J\alt0.8$ 
(Sec.~\ref{secresults:liquid}). 
This rather new type of spin liquid state is characterized by a small  
spin excitation gap, and is described by the projected BCS state 
$\pbcs$ with a  gap function defined by Eq.~(\ref{pairbroken}) with 
a $(2\times1)$ unit cell.  This spin liquid state is an extension of the 
conventional short range RVB state $|\psi_{\rm RVB}\rangle$ 
[Eq.~(\ref{srrvb})], which has been considered before in the context of 
the isotropic triangular lattice~\cite{bernu} and is indeed a very good 
representation of the exact ground state of the isotropic 
triangular antiferromagnet for small clusters 
(Sec.~\ref{secresults:sondhi} and Ref.~\onlinecite{liliana}).

To extend the short range RVB state, we have constructed an exact mapping 
between the short range RVB state and the projected BCS state $\pbcs$ 
(Sec.~\ref{secresults:sondhi} and App.~\ref{appendix:pfaffian}). In doing so, 
the unit cell of the BCS Hamiltonian, defining the  
$|{\rm BCS}\rangle$ state, is expanded to a ($2 \times 1$) unit cell. 
This mapping is  crucial in the present study because within this approach
the short range RVB state can be easily extended  and improved  
systematically by including hopping integrals, a 
finite chemical potential, and most importantly long range gap functions 
in the BCS Hamiltonian, with no particular numerical effort.  
An additional advantage of using the $\pbcs$ representation is 
that it is easy to gain qualitative insight of the low-lying spin excitations 
from the corresponding BCS excitation spectrum 
$E_{\bf k}$.~\cite{wen,wenfermion} 
For instance, $E_{\bf k}$ of the BCS Hamiltonian, 
corresponding to the short range RVB state, shows a finite excitation gap 
simply because $-\mu\gg|\Delta_{i,j}|$ 
(Sec.~\ref{secresults:sondhi} and App.~\ref{appendix:sondhi}), and therefore 
a finite spin gap is expected in the short range RVB state. 
This is indeed the correct property of the short range RVB state, because  
the presence 
of a finite spin gap and a very short correlation length have been 
established before.~\cite{sondhi}

It is worth mentioning a further remarkable property of this new type 
of spin liquid state described by this projected BCS wave function. 
Without the projection ${\cal P}_{\rm G}$, the BCS state 
$|{\rm BCS}\rangle$ breaks translation and reflection symmetries because 
the BCS Hamiltonian is defined with a $(2\times1)$ unit cell. 
As a consequence, the BCS excitation spectrum $E_{\bf k}$ has a finite gap 
in general. However, as discussed in Sec.~\ref{secresults:sondhi}, 
the translation symmetry is recovered after applying the 
projection operator onto $|{\rm BCS}\rangle$. Therefore, within the projected 
BCS wave functions, we have discovered a peculiar way to open a finite spin 
gap without breaking the translation symmetry of the state. 
It is also important to emphasize that the reflection symmetry of the 
projected state is also restored in the thermodynamic limit, as we have 
systematically tested 
numerically (Fig.~\ref{notyet}), which however cannot be proved 
rigorously except 
for the limiting case of the symmetric short range RVB state.

Our finding of stable spin liquid phases in the spin-1/2 anisotropic 
triangular antiferromagnet is in good agreement with recent 
studies of the half-filled Hubbard model on the triangular lattice with 
spatial anisotropic hopping based on the Gutzwiller 
approximation~\cite{kenzie_1} and the variational Monte Carlo 
simulations.~\cite{nandini} 
However, it should be remarked here that the gap function in $\pbcs$ 
which we found energetically favorable for the nearly isotropic case is 
rather different from the ones considered in the previous 
studies.~\cite{kenzie_1,nandini}
Our spin liquid state is defined by a BCS Hamiltonian with a non translation 
invariant gap function, whereas the conventional ansatz state such as the 
ones considered in Refs.~\onlinecite{kenzie_1} and \onlinecite{nandini} is 
defined with a homogeneous gap function, which we found much less 
accurate close to the isotropic point (see Fig.~\ref{phaseenergy}). 
Although both the Hubbard model in the limit of large 
on-site repulsion $U$ and the Heisenberg model considered here  
should be the same, we just note that,  
with a variational method sensitive only to the energy, it is much 
easier to find good  variational  
wave functions  for a low energy effective Heisenberg 
model rather than for the corresponding Hubbard model, because the latter 
contains also the large energy scale $U$.

In order to have better insight on the low-lying spin excitations 
of the spin liquid phases, we have directly calculated the spin one 
excitation dispersion with the method described in Sec.~\ref{secdyn}. 
This quantity is particularly important since it can be compared directly 
with inelastic neutron 
scattering experiments. The detailed measurements on ${\rm Cs_2 Cu Cl_4}$ 
by Coldea {\it et al}~\cite{coldea} are indeed available, for which a spin 
liquid like behavior has been observed.~\cite{coldea} 
It has been also reported that this material can be described by the model 
Hamiltonian Eq.~(\ref{model}) with $J'/J\approx1/3$.~\cite{coldea} 
As shown in Fig.~\ref{sqw}, our calculations were found to be in 
excellent agreement with the experiments with no fitting 
parameters. It is also clear from Fig.~\ref{sqw} that the dispersion is 
2D characteristic 
because for uncoupled chains ($J'=0$) the dispersion should be symmetric 
around the momenta $(\pi/2,0)$ and $(\pi,0)$. 
This excellent agreement strongly supports 
the exitance of this type of 2D spin liquid state. 
The successful comparison also indicates that our numerical technique can 
compute accurately the excitation dispersion for the non-trivial 
strongly anisotropic antiferromagnetic Heisenberg model on the triangular 
lattice.

\begin{figure}[hbt]
vskip 0.5cm
\includegraphics[width=8.5cm,angle=-0]{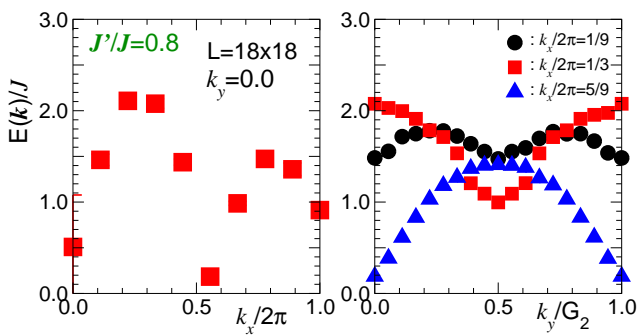}
\begin{center}
\vskip -2.0cm
\caption{
Lowest triplet spin excitations as a function of momentum for the spin-1/2 
antiferromagnetic Heisenberg model on the triangular lattice with $J'/J=0.8$ 
and $L=18\times18$, calculated using the method introduced in 
Sec.~\ref{secdyn}. Here $G_2=4\pi/\sqrt{3}$
}
\label{sqw_jp08}
\end{center}
\end{figure}

Encouraged by these results, we have also calculated the low-lying 
spin one dispersion for the other spin liquid phase which appears in the 
region of $0.65\alt J'/J \alt 0.8$. A typical results for $J'/J=0.8$ 
is reported in Fig.~\ref{sqw_jp08}.  
This coupling regime is relevant 
for the organic material {$\kappa$-(ET)$_2$Cu$_2$(CN)$_3$} for which a spin 
liquid like behavior has been also observed.~\cite{kanoda} 
Since there is no experimental data available for the 
spin excitations of this material, our results shown in Fig.~\ref{sqw_jp08} 
provide a theoretical prediction for the spin excitation dispersion, which 
should be compared with the future neutron scattering experiments.  
It is interesting to notice that the dispersions for $J'/J=0.8$ and 
for $J'/J=0.33$ are very similar along the $x$-direction, whereas the 
difference becomes evident in other momentum regions, for example, along 
the $y$-direction. This feature should be also checked experimentally 
in the feature.

In both spin liquid phases described by our variational ansatz, the spin 
excitation gap in the thermodynamic limit cannot be resolved by directly 
computing the excitation dispersions with the method introduced in 
Sec.~\ref{secdyn} as the present available system sizes are not 
large enough. According to the argument presented in the previous section, 
for $J'/J\simeq 0.8$ a finite spin excitation gap is implied by the 
corresponding gap in the BCS excitation spectrum $E_{\bf k}$ of 
the $\pbcs$ ansatz. However, this gap should be very small because 
i) it is close to a transition 
to an ordered state appearing for larger $J'/J$ and ii) it is due to the 
rather small long range part of the gap functions $\Delta_{i,j}$'s.  
In fact, we have found that the lowest excitation gap in $E_{\bf k}$ is 
as small as $\sim0.2\%$ of $2W$ for $J'/J=0.8$ (where $W$ is the maximum 
excitation energy in $E_{\bf k}$).~\cite{note6}     
This small gap might  also explain an 
apparent finite spin susceptibility observed on 
{$\kappa$-(ET)$_2$Cu$_2$(CN)$_3$},~\cite{kanoda} which should eventually 
vanish exponentially at very low temperatures with an activated behavior.

It is also very interesting to note that the organic material 
{$\kappa$-(ET)$_2$Cu$_2$(CN)$_3$}, which shows a spin liquid like behavior 
under ambient pressure,\cite{kanoda} has been very recently 
found to become a superconductor under small applied pressure (about 
4--6 $\times10^{-1}$ GPa).~\cite{kanoda2,kanodanature} 
If we assume that the origin of the superconductivity is intimately related 
to the spin liquid like nature of the phase observed next to the 
superconducting phase, the present study implies a rather unique, unexpected 
pairing symmetry of the superconductor with a non translation 
invariant pairing amplitude. This is a measurable effect because 
the projected BCS state $\pbcs$ discussed in Sec.~\ref{secresults:liquid} 
is no longer translation invariant once the projection constraint 
${\cal P}_{\rm G}$ is released or mobile carriers are introduced into the 
homogeneous spin liquid state. Therefore, we expect that this unconventional 
pairing formation can be probably observed by, {\it e.g.}, scanning tunneling 
microscopy experiments.~\cite{davis}

\begin{figure}[hbt]
\includegraphics[width=8.5cm,angle=0]{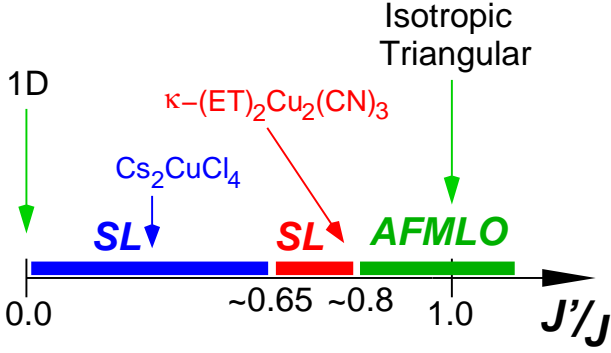}
\begin{center}
\caption{
A schematic phase diagram of the spin-1/2 antiferromagnetic Heisenberg model 
on the triangular lattice. $J$ ($J'$) is the nearest neighbor 
antiferromagnetic coupling between the spins in the chain (in different 
chains) (see Fig.~\ref{lattice}). 
SL and AFMLO stand for spin liquid and antiferromagnetic long range order 
(incommensurate spin order for $J'/J$ away from the isotropic case), 
respectively. Materials possibly 
described by this class of models are also indicated. 
}
\label{phase}
\end{center}
\end{figure}

Based on our numerical calculations and their comparison with experiments, 
we conclude that the spin liquid is a generic phase of quantum matter, and 
that,  for its stability,  it is not 
important to be very close to a phase transition, unlike the recently 
proposed scenario in which a spin liquid state appears only at a transition 
$point$ between a magnetically ordered phase and a spontaneously 
dimerized phase.~\cite{fisher} 
Within the projected BCS wave function approach, spontaneous 
dimerization can be correctly described in quasi 1D systems, but 
not in 2D unless the translation symmetry of the variational state is 
explicitly broken.~\cite{chiral} 
Even allowing this possible symmetry breaking, by adding appropriate 
symmetry breaking terms to   
the BCS Hamiltonian, we have not found a stable dimerized phase in the 
present 2D system. Instead, we have found rather stable spin liquid phases. 
Finally, our phase diagram of the spin-1/2 
antiferromagnetic Heisenberg model on the triangular lattice is summarized 
in Fig.~\ref{phase}, where possibly related materials are also indicated.  
In particular, it would be very important 
to distinguish experimentally the two different spin liquid phases 
predicted here for different values of the coupling $J^\prime/J$.

%%%%%%%%%%%%%%%%%%%%%%%%%%%%%%%
%%%%%%%%%%%%%%%%%%%%%%%%%%%%%%%
\begin{acknowledgments}
%%%%%%%%%%%%%%%%%%%%%%%%%%%%%%%
%%%%%%%%%%%%%%%%%%%%%%%%%%%%%%%

We thank F. Becca and L. Arrachea for providing us Exact 
Diagonalization results for the $6\times6$ clusters, and especially 
R. Coldea for sending us his 
neutron scattering data on ${\rm Cs_2 Cu Cl_4}$. 
One of us (S.S.) acknowledges Kavli Institute for Theoretical Physics (KITP) 
in Santa Barbara for its kind 
hospitality at the early stage of this work. 
This work has been partly supported by INFM and MIUR (COFIN 2004).

\end{acknowledgments}

%%%%%%%%%%%%%%%%%%%%%%%%%%%%%%%
%%%%%%%%%%%%%%%%%%%%%%%%%%%%%%%
\appendix 
%%%%%%%%%%%%%%%%%%%%%%%%%%%%%%%
%%%%%%%%%%%%%%%%%%%%%%%%%%%%%%%

%%%%%%%%%%%%%%%%%%%%%%%%%%%%%%%
\section{ Details of the numerical calculations} \label{appmethod}
%%%%%%%%%%%%%%%%%%%%%%%%%%%%%%%

In this appendix, we will explain how to define appropriately and calculate 
efficiently the  quantity $O_k(x)$  
[Eq.~(\ref{opx})] for each variational parameter $\alpha_k$ of the wave 
function $|\Psi_{\{\alpha_k\}}\rangle$. 
Since the derivation is rather 
general and not restricted to the 
particular form of the wave function used,  
we will here indicate the  
determinantal  part of the wave function by 
$|{{\bf\Phi}_0\rangle}$, instead of $|{\rm BCS}\rangle$, and thus 
$|\Psi_{\{\alpha_k\}}\rangle={\cal P}_{\rm G}|{{\bf\Phi}_0\rangle}.$ 
Since it is generally found much more efficient, we will 
consider a wave function with definite number $N$ of electrons, for which 
there are also important  simplifications  for deriving  
computationally convenient expressions for the quantities $O_k(x)$. 
 
To this purpose, we will use a particle-hole transformation introduced 
by Yokoyama and Shiba,~\cite{shiba} which is defined by the following 
canonical transformation:
\begin{equation} \label{spindownph}
\left\{
 \begin{array}{l}
c^\dag_{i\up}=\ccd_i \\[0.2cm]
c^\dag_{i\dn}=(-1)^{i_1+i_2} \cc_{i+L}.
 \end{array}
\right.
\end{equation}
Here $c^\dag_{i\sigma}$ ($c_{i\sigma}$) is an electron creation 
(annihilation) operator at 
site $\vec r_i=i_1{\vec\tau_1}+i_2{\vec\tau_2}$ with spin 
$\sigma(=\uparrow,\downarrow)$, whereas $\ccd_i$ and $\cc_i$ represent 
the new canonical operators corresponding to the original spin 
up (spin down) electrons for $i \le L$  ($L < i \le 2 L$). 
$L$ is the total number of sites.

After this particle-hole transformation (\ref{spindownph}), it is clear that 
the BCS Hamiltonian [Eq.~(\ref{bcsh})] can be written as 
\begin{equation}\label{hamg}
\tl{\cal H}_{\rm BCS} = \sum_{I,J=1}^{2L}\ccd_I
\left({\bar H}^{(\rm HF)}\right)_{I,J}\cc_J,
\end{equation}
where $\left({\bar H}^{(\rm HF)}\right)_{I,J}$ are appropriate $2L\times 2L$ 
matrix elements, 
which can be straightforwardly computed from Eq.~(\ref{bcsh}). 
Note that this matrix can be easily evaluated also for 
rather general Hartree-Fock, BCS Hamiltonians containing, {\it e.g.}, non 
translation invariant terms, or several types of orders.  
The total number of particles $\np$ after the transformation (\ref{spindownph})
is related to the total spin along the $z$-direction 
$S_z=(N_\uparrow-N_\downarrow)/2$ in the original 
representation through the following equation: 
$$
\np\equiv \sum\limits_{I=1}^{2L} C^{\dag}_I C_I
    =\sum\limits_{i=1}^{L} 
    \left(c^{\dag}_{i,\uparrow} c_{i,\uparrow} + c_{i,\downarrow} c^{\dag}_{i,\downarrow}\right)=  L+2 S_z.
$$ 
Since $S_z$ is conserved in the original BCS Hamiltonian, the transformed 
Hamiltonian (\ref{hamg}) has a definite number $\np$ of particles, as 
anticipated. 
With this particle-hole transformation, it is possible 
to study the spin excitations of the BCS Hamiltonian with $S_z \ne 0$, 
as well as the ground state which belongs to the singlet $S_z=0$ sector. 
Therefore, in the following, we will consider the general case with 
unrestricted $\np$.

By using an appropriate unitary transformation, 
\begin{eqnarray}\label{eigeninv}
\left\{
\begin{array}{c}
\displaystyle{\cc_I=\sum_{\alpha=1}^{2L}\left(\bar{U}\right)_{I\alpha}{\hat\gamma}_\alpha}\\[0.4cm]
\displaystyle{{\hat\gamma}^\dag_\alpha=\sum_{i=I}^{2L}\left(\bar{U}\right)_{I\alpha}}
\ccd_I,
\end{array}
\right.
\end{eqnarray}
$\tl{\cal H}_{\rm BCS}$ is diagonalized with a new set of  quasiparticle 
operators 
$\{{\hat\gamma_\alpha^\dag},{\hat\gamma_\alpha}\}$: 
\begin{equation}
\tl{\cal H}_{\rm BCS} = \sum_{\alpha=1}^{2L}\eps_\alpha{\hat\gamma_\alpha^\dag}
{\hat\gamma_\alpha}
\end{equation}
where, for convenience, the eigenvalues are sorted in ascending order
$\eps_1\le\eps_2\le\cdots\le\eps_{2L}$.

A natural choice of the Slater determinant part $|{\bf\Phi}_0\rangle$ of the 
variational wave function is the ground state of 
${\tl{\cal H}}_{\rm BCS}$ with $\np$ particles: 
\begin{equation}
|{\bf\Phi}_0\rangle={\hat\gamma}^\dag_1 {\hat\gamma}^\dag_2\cdots 
{\hat\gamma}^\dag_{\np}|\tl 0\rangle
\end{equation}
where $|{\tl 0}\rangle$ is the vacuum  
($\cc_I|{\tl 0}\rangle=0$) of the Hilbert space after the particle-hole 
transformation. 
Thus the Slater determinant for a  
particle-hole transformed configuration 
$| x\rangle=\ccd_{I_1}\ccd_{I_2}\cdots \ccd_{I_{\np}}|\tl 0\rangle$ is 
\begin{equation} \label{detslater}
\langle  x|{\bf\Phi}_0\rangle=\det\left[{\bar S}\right],
\end{equation}
where ${\bar S}$ is a $\np\times\np$ matrix, whose elements are 
\begin{equation}
\left({\bar S}\right)_{l, \alpha} 
= \langle{\tl 0}|\cc_{I_{l}}{\hat\gamma}^\dag_\alpha|{\tl 0}\rangle
= \left(\bar U\right)_{I_{l}\alpha},
\end{equation}
$I_l$'s ($1\le I_l \le 2L$) are the ``positions'' of the $\np$ particles 
($l=1,2,\cdots\np$)  
 corresponding to the configuration $| x \rangle$, and 
$\alpha=\{1,2,\cdots,{\np}\}$.

Now  let us 
consider small changes for  $\alpha_k\to\alpha_k'=\alpha_k+\delta\alpha_k$. 
Since ${\tl{\cal H}}_{\rm BCS}$ depends linearly on $\{\alpha_k\}$, the 
perturbed system is described by 
\begin{equation}
{\tl{\cal H}}'_{\rm BCS}={\tl{\cal H}}_{\rm BCS}
+\sum_{k=1}^{p}\delta\alpha_k\cdot{\hat V}_k
\end{equation}
where 
%\begin{equation}
${\hat V}_k=\sum_{I,J=1}^{2L}\left({\bar V}^{(k)}\right)_{IJ}\ccd_I\cc_J$ 
%\end{equation}
is a suitable operator proportional to the chosen variational parameter 
$\alpha_k$. 
For instance, when the chemical potential is changed 
$\mu \to \mu + \delta \mu$, namely a term  
$-\delta \mu\sum_{i=1}^L\sum_{\sigma}c^\dag_{i,\sigma}c_{i,\sigma}$ is added 
to ${\hat H}_{\rm BCS}$, 
the corresponding matrix element $\left({\bar V}^{(k)}\right)_{IJ}$ is 
$-\delta_{I,J}$ for $I\le L$ 
and $\delta_{I,J}$ for $I> L$, due to the particle-hole 
transformation (\ref{spindownph}). Here $\delta_{i,j}$ is the Kronecker 
$\delta$-function. 
When a small change in a pairing term $\left[\Delta_{i,j}c^\dag_{i,\uparrow}c^\dag_{j,\downarrow}+{\rm h.c.}\right]$ is considered, the corresponding matrix element 
$\left({\bar V}^{(k)}\right)_{IJ}$ is 
$(-1)^{j_1+j_2}\left[\delta_{I,i}\delta_{J,j+L}+\delta_{I,j+L}\delta_{J,i}\right]$.

With the basis set formed by the quasiparticle operators 
$\{{\hat\gamma_\alpha^\dag},{\hat\gamma_\alpha}\}$, 
the matrix ${\bar V}^{(k)}$ is transformed as 
$   {\bar V}^{(k)} \to {\bar U}^{\dag} {\bar V}^{(k)} {\bar U}$, and thus the 
first order correction to the state $|{\bf\Phi}_0\rangle$ is 
easily computed as follows:
\begin{equation}
|{\bf\Phi}'_0\rangle= \left[1+\sum_{k=1}^p
\delta \alpha_k \sum_{\eta,\nu=1}^{2L}\left({\bar Q}^{(k)}\right)_{\eta\nu} 
\gamma^{\dag}_\eta \gamma_\nu\right] |{\bf\Phi}_0\rangle+ {\cal O}(\delta\alpha_k^2)
\end{equation}
where
%\begin{widetext}
\begin{equation} \label{defq}
 \left({\bar Q}^{(k)}\right)_{\eta\nu} =\left\{
\begin{array}{cl}
\displaystyle{
{\frac{\left[{\bar U}^{\dag} {\bar V}^{(k)}  {\bar U} \right]_{\eta\nu}}{\eps_\nu-\eps_\eta}}  } 
&:{\rm for}~ \eta>\np  {\rm ~and~} \nu\le  \np \\[0.4cm]
  0 &:{\rm otherwise}
\end{array}
\right. 
\end{equation}
%\end{widetext}
This expression is well defined as long as 
 $|{\bf\Phi}_0\rangle$ is non degenerate, so that 
the denominator in the definition of ${\bar Q}^{(k)}$ is always non zero.
This condition is rather generally  satisfied for the BCS Hamiltonian. 
We can now readily express the state $|{\bf\Phi}'_0\rangle$ 
in terms of the $\{\ccd,\cc\}$ basis set,
\begin{equation}\label{wf1o}
|{\bf\Phi}'_0\rangle=\left[1 
+\sum_{k=1}^p\delta \alpha_k \sum_{I,J=1}^{2L} \left({\bar M}^{(k)}\right)_{IJ}
 \ccd_I \cc_J 
\right]|{\bf\Phi}_0\rangle 
+ {\cal O}(\delta\alpha_k^2)
\end{equation}
where  ${\bar M}^{(k)}={\bar U} {\bar Q}^{(k)} {\bar U}^\dag$. 
At each iteration of the SR minimization procedure described in 
Sec~\ref{srmethod}, ${\bar M}^{(k)}$ 
has to be computed, because the operators 
$\{{\hat\gamma_\alpha^\dag},{\hat\gamma_\alpha}\}$ change every time  
a new set of $\{\alpha_k\}$ is calculated. 
This can be done using four matrix-matrix multiplications.

From Eq.~(\ref{wf1o}), it is now easy to evaluate 
the perturbed Slater determinant $\langle x|{\bf\Phi}'_0\rangle$, and 
thus an explicit expression for  $O_k(x)$ defined by Eqs.~(\ref{opx}) and 
(\ref{opx2}): 
\begin{eqnarray} \label{finalokx} 
O_k(x) &=& \sum_{I=1}^{2L}\sum_{J=1}^{2L}\left({\bar M}^{(k)}\right)_{IJ}
{\frac{\langle x|\ccd_I \cc_J|{\bf\Phi}_0\rangle}
{\langle x|{\bf\Phi}_0\rangle}}\nonumber \\
&=&\sum_{l=1}^{\np}\sum_{J=1}^{2L}\left({\bar M}^{(k)}\right)_{I_{\l}J}
{\cal G}_{J\l}.
\end{eqnarray}
Here  ${\cal G}_{J\l}$ is a local single-particle ``Green's function'', 
\begin{eqnarray}
{\cal G}_{J\l} &\equiv& {\frac{\langle x|\ccd_{I_{\l}}\cc_J|\bf\Phi_0\rangle}
{\langle x|\bf\Phi_0\rangle}}\nonumber \\
&=&\sum_{\alpha=1}^{\np} \left(\bar U\right)_{J\alpha}.
\left({\bar S}^{-1}\right)_{\alpha\l},
\end{eqnarray}
which is computed and updated during the variational Monte Carlo 
iterations.  
Since the matrix ${\bar M}^{(k)}$ does not depend on the 
configuration $|x\rangle$  and ${\cal G}_{J\l}$ is 
always known, only about $ L^2$ 
operations are required to evaluate $O_k(x)$ for each $k$ and for 
each sampled configuration $|x\rangle$.

Finally, we note that when the variational wave function 
$|\Psi_{\{\alpha_k\}}\rangle$ contains the spin Jastrow factor 
$
{\cal J}_S=\exp\left[\sum_{i,j=1\ (i<j)}^L v_{ij}{\hat S}_i^z{\hat S}_j^z\right],
$
{\it i.e.}, 
$|\Psi_{\{\alpha_k\}}\rangle={\cal P}_{\rm G}{\cal J}_S|{\bf\Phi}_0\rangle$, 
the calculation of $O_k(x)$ corresponding to the variational parameter 
$v_{ij}$ is much simpler and straightforward, because $O_k(x)$ is simply 
$\langle x|{\hat S}_i^z{\hat S}_j^z|x\rangle$ in this case.

%%%%%%%%%%%%%%%%%%%%%%%%%%%%%%%%%%%%%%%%%%%%%
\section{Marshall sign rule and projected BCS wave functions}\label{app:marshall}
%%%%%%%%%%%%%%%%%%%%%%%%%%%%%%%%%%%%%%%%%%%%%

In this appendix, we will show that a BCS wave function 
projected onto the subspace of singly occupied sites satisfies 
the Marshall sign rule provided the corresponding BCS Hamiltonian 
\begin{equation} \label{bcsgeneral}
{\hat H}_{\rm BCS} =\sum_{j,l} \left[  t_{j,l} 
\left(\sum_{\sigma} c^{\dag}_{j\sigma} c_{l\sigma} 
\right)
+ \left( \Delta_{j,l} 
c^{\dag}_{j\uparrow} c^{\dag}_{l\downarrow} + {\rm h.c.} \right)\right] 
\end{equation} 
satisfies the following conditions:
\begin{equation}\label{relationapp}
 t_{j,l}= \Delta_{j,l}= 0: {\rm ~for~{\it j}~and~{\it l}~on~the~same~sublattice.}
%&&~~~~~i.e.,~|j_x-l_x|+|j_y-l_y|~{\rm even} \label{relationapp}
\end{equation} 
Here $c^\dag_{i\sigma}$ ($c_{i\sigma}$) is an electron 
creation (annihilation) operator at 
site ${\bf r}_i=(i_x,i_y)$ with spin $\sigma(=\uparrow,\downarrow)$, 
and $t_{j,k}$ and $\Delta_{j,k}$ are assumed symmetric under 
$j \leftrightarrow  k$ interchange and real. 
It is  also assumed that the ground state of the BCS Hamiltonian 
$|{\rm BCS}\rangle$ is  unique, namely with a finite size gap to the first 
excitation,  a condition which can be generally met for non trivial 
values of $\Delta_{i,j}$ and $t_{i,j}$.

In what follows, we will show that the projected BCS wave function 
$\pbcs ={\cal P}_{\rm G} |{\rm BCS}\rangle$ constructed from the BCS 
state above satisfies the Marshall sign rule, namely,
\begin{equation} \label{marshall}
{\rm Sgn} \left[ \langle x|{\rm BCS} \rangle \right] =  (-1)^{N_{\rm A}},
\end{equation}
 where $N_{\rm A}$ is the number of
down spins on one of the two sublattices, and 
$|x\rangle$ is an electron configuration 
with no doubly occupied sites, which can be written as
$$\langle x | =  \langle 0 | \prod\limits_{i=1}^L  c_{i\sigma_i} $$
with the $c_{i\sigma_i}$ factors being ordered 
from left to right according to the increasing index $i$ 
($\sigma_i$ is the spin of the electron on site $i$). 
Here it is assumed that the number of sites ($L$) is even.

First, it is convenient to perform the following particle-hole
transformation for the down spins only:  
\begin{eqnarray} \label{phdown}
c_{l\uparrow} &=& d_{l\uparrow}\nonumber\\
c_{l\downarrow} &=& (-1)^{l_x+l_y} d^\dag_{l\downarrow}.
\end{eqnarray}
With this transformation, the BCS Hamiltonian is transformed to a 
standard one-body Hamiltonian commuting with the total number of 
particles, {\it i.e.}, 
\begin{eqnarray} \label{bcstransf}
{\hat H}_{\rm BCS}&=& \sum_{j,l} \left[ t_{j,l} \left(\sum_{\sigma} 
d^{\dag}_{j\sigma} d_{l\sigma}\right)\right.   \nonumber \\
&+& 
 \left. \Delta_{j,l} (-1)^{l_x+l_y} \left(\sum_{\sigma,\sigma^\prime} 
 d^{\dag}_{j\sigma} d_{l\sigma^\prime} 2 i  S^y_{\sigma\sigma^\prime} 
\right)\right],
\end{eqnarray}
where ''$i$'' denotes the imaginary unit, $S^y$ is the $y$-component 
of the spin matrix, and condition (\ref{relationapp}) is used.
On the other hand, after this transformation~(\ref{phdown}), the basis of no 
doubly occupied sites turns  into
\begin{eqnarray} \label{mapping}
|\uparrow\rangle  &\to&  |\uparrow\downarrow\rangle  \nonumber \\
|\downarrow\rangle &\to& |0\rangle, \nonumber  
\end{eqnarray}
namely, the sites are either doubly occupied or empty in the new 
representation, and all configurations $\{\hat x \}$ are now 
defined by the $L/2$ positions of the doubly occupied sites:
$$\langle \hat x |=\langle   0 | d_{{\bf R}_1\downarrow} d_{{\bf R}_1\uparrow} 
d_{{\bf R}_2\downarrow} d_{{\bf R}_2\uparrow}
\cdots  d_{{\bf R}_{L/2}\downarrow} d_{{\bf R}_{L/2}\uparrow},$$ 
where ${\bf R}_l\in\{{\bf r}_1,{\bf r}_2,\cdots,{\bf r}_L\}$. 
Because of the additional phase in Eq.~(\ref{phdown}), $\langle x|$ is related 
to $\langle{\hat x}|$ by 
$ \langle  x | = (-1)^{N_A} \langle \hat x | $. Therefore, 
using the definition of the Marshall sign (\ref{marshall}), 
a state with the Marshall sign rule in the original basis 
$\{ x \}$  is equivalent to a bosonic state in the particle-hole 
transformed basis $\{ \hat x \}$: for  each 
configuration $|\hat x\rangle $, the wave function 
should be always positive (or always negative).

In order to show this bosonic rule for the projected BCS state considered, 
it should be noticed that the new  basis $\{ \hat x \}$ 
is invariant under a global rotation of the spins which transforms  
the spin matrix $S^y_{\sigma\sigma^\prime}$ into 
$S^z_{\sigma\sigma^\prime}= \sigma \delta_{\sigma\sigma^\prime}/2 $: 
% in Eq.(\ref{bcstransf}):
\begin{equation}\label{su2t}
\left(
   \begin{array}{c}
    d_{j\uparrow}\\[0.1cm]
    d_{j\downarrow}
   \end{array}
\right)
={\frac{1}{\sqrt{2}}}\left(
   \begin{array}{cc}
    1 & i\\[0.1cm]
    -i & -1
   \end{array}
\right)
\left(
   \begin{array}{c}
    a_{j\uparrow}\\[0.1cm]
    a_{j\downarrow}
   \end{array}
\right).
\end{equation}
Indeed, $d_{j\uparrow}d_{j\downarrow}$ is transformed into $
a_{j\downarrow}a_{j\uparrow}$. It is also easily shown that 
this transformation~(\ref{su2t}) factorizes the Hamiltonian (\ref{bcstransf}) 
as 
\begin{equation} \label{transformed} 
{\hat H}_{\rm BCS} = \sum_{\sigma}\left[
\sum_{i,j}\left({\bar H}_{\sigma}\right)_{ij}a^\dag_{i\sigma}a_{i\sigma}
\right],
\end{equation}
where ${\bar H}_{\uparrow}={\bar H}_{\downarrow}^*$, ${\bar H}_\uparrow$  
and ${\bar H}_{\downarrow}$ being appropriate one-body Hamiltonian 
matrices whose eigenstates are $\phi_{\alpha}({\bf r})$ and 
$\phi^*_{\alpha} ({\bf r}) $, 
respectively ($\alpha=1,2\cdots, L$). 
Therefore, the ground state $| {\rm BCS} \rangle$ 
 of ${\hat H}_{\rm BCS}$ once computed in the basis 
of doubly occupied and empty sites $\{ \hat x \}$ reads
\begin{equation}
\langle \hat x |  {\rm BCS} \rangle = \left|{\rm Det} \left({\bar S}\right)\right|^2>0, 
\end{equation}
where ${\bar S}$ is a $(L\times L)$ matrix with 
$\left( {\bar S} \right)_{l,\alpha}=\phi_\alpha ({\bf R}_l)$.
This proves the statement given at the beginning of this appendix.

Finally, we note that the condition~(\ref{relationapp}) is satisfied 
whenever the BCS 
Hamiltonian~(\ref{bcsgeneral}) is invariant under a particle-hole 
transformation:
\begin{equation} \label{phsim}
c^{\dag}_{i\sigma} \to (-1)^{i_x+i_y} c_{i{\sigma}}, 
\end{equation}
provided $t_{j,k}=t_{k,j}$ and $\Delta_{j,k}$ is real. 
Eq.~(\ref{restriction}) in Sec.~\ref{pbcsphs} follows readily from this 
particle-hole symmetric condition.

%%%%%%%%%%%%%%%%%%%%%%%%%%%%%%%
\section{ BCS wave function with a broken translation symmetry} \label{appendix:sondhi}
%%%%%%%%%%%%%%%%%%%%%%%%%%%%%%%

In this appendix, several important properties are discussed on the 
ground state $|{\rm BCS}\rangle$ of the BCS Hamiltonian 
which breaks primitive lattice translation symmetry by extending the unit 
cell to $(2\times1)$. 
The resulting state $|{\rm BCS}\rangle$ is used to build a projected BCS 
states $\pbcs={\cal P}_{\rm G}|{\rm BCS}\rangle$ for the isotropic 
and nearly isotropic triangular systems discussed in Sec.~\ref{secresults:sondhi} and 
Sec.~\ref{secresults:liquid}, respectively.

\begin{figure}[hbt]
\includegraphics[width=8.cm,angle=0]{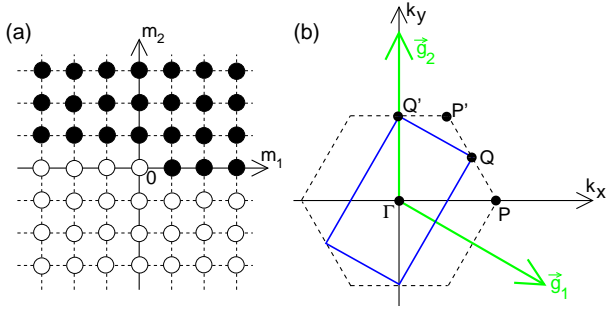}
\begin{center}
\caption{
(a): Region of $(m_1,m_2)$ (denoted by solid circles) considered 
for the sum over ${\vec t}_m=m_1{\vec\tau}_1+m_2{\vec\tau}_2$. 
Each circle corresponds to a pair of integers $(m_1,m_2)$. 
(b): The first Brillouin zone (hexagon drown by dashed lines) for 
the triangular lattice shown in Fig.~\ref{lattice}. The corresponding 
reciprocal lattice vectors are ${\vec g}_1=2\pi(1,-{\frac{1}{\sqrt{3}}})$ 
and ${\vec g}_2=2\pi(0,{\frac{2}{\sqrt{3}}})$. The reduced 
Brillouin zone (tilted rectangle) is also given by solid lines. 
Several symmetric points are $\Gamma$: (0,0), P: $({\frac{4}{3}}\pi,0)$, 
Q: $(\pi,{\frac{\pi}{\sqrt{3}}})$, 
P': $({\frac{2}{3}}\pi,{\frac{2}{\sqrt{3}}}\pi)$, and 
Q': $(0,{\frac{2}{\sqrt{3}}}\pi)$. 
%A lattice constant is set to be one.
}
\label{lattice2}
\end{center}
\end{figure}

Let us start with defining the BCS Hamiltonian ${\hat{H}}_{\rm BCS}$ 
introduced in Sec.~\ref{vwf} in a slightly different way. A part of 
${\hat{H}}_{\rm BCS}$ which is invariant under any primitive lattice 
translations on the triangular lattice (Fig.~\ref{lattice}) can be generally 
described by the following Hamiltonian:   
%\begin{widetext}
\begin{eqnarray} 
{\hat H}^{(0)}_{\rm BCS}
&=& -\sum_{{\vec r},\sigma} \left[ {\sum_{{\vec t}_m}}^{\prime}
t_{{\vec t}_m}  
\left(c^\dag_{{\vec r}\sigma} c_{{\vec r}+{\vec t}_m\sigma}
+c^{\dag}_{{\vec r}+{\vec t}_m\sigma} c_{{\vec r}\sigma}\right)\right]
\nonumber\\
&+&\sum_{\vec r} \left[ {\sum_{{\vec t}_m}}^{\prime}
\Delta_{{\vec t}_m}  
\left(c^\dag_{{\vec r}\up} c^\dag_{{\vec r}+{\vec t}_m\dn}
-c^\dag_{{\vec r}\dn} c^{\dag}_{{\vec r}+{\vec t}_m\up} \right) 
+ {\rm H.c.}\right]\nonumber\\
&-&\mu \sum_{{\vec r},\sigma} c^\dag_{{\vec r}\sigma} c_{{\vec r}\sigma},
\end{eqnarray}  
%\end{widetext}
where the first sum $\sum_{\vec r}$ runs over all lattice vectors 
${\vec r}=r_1{\vec\tau}_1+r_2{\vec\tau}_2$ ($r_1,r_2$: integer), whereas 
the second sum $\sum_{{\vec t}_m}^{\prime}$ is for 
${\vec t}_m=m_1{\vec\tau}_1+m_2{\vec\tau}_2$ ($m_1,m_2$: integer) 
with $m_2>0$ or with $m_1>0$ and $m_2=0$ as denoted by solid circles in 
Fig.~\ref{lattice2} (a). 
%%%%%%%%%%%%%%%%%%%%%%%%%%%%%%%%%%%%%%%%%%%%%%%%%%%%%%%%%%%%%%%%
%%%%%%%%%%%%%%%%%%%%%%%%%%%%%%%%%%%%%%%%%%%%%%%%%%%%%%%%%%%%%%%%
Now let us add to ${\hat H}^{(0)}_{\rm BCS}$ an additional pairing term 
${\hat V}_{\rm pair}$ which breaks the underlying lattice translation 
symmetry:  
\begin{eqnarray} 
{\hat V}_{\rm pair}
&=& \sum_{\vec r} \Biggl[ {\sum_{{\vec t}_m}}^{\prime}
(-1)^{r_1}{\bar{\Delta}}_{{\vec t}_m}  
\left(c^\dag_{{\vec r}\up} c^\dag_{{\vec r}+{\vec t}_m\dn}
-c^\dag_{{\vec r}\dn} c^{\dag}_{{\vec r}+{\vec t}_m\up} \right) \nonumber\\
&&\quad\quad\quad+ {\rm H.c.}\Biggr]. 
\end{eqnarray}  
Note that the unit cell of the total BCS Hamiltonian 
${\hat{H}}_{\rm BCS}={\hat H}^{(0)}_{\rm BCS}+{\hat V}_{\rm pair}$ is now 
extended to $(2\times1)$. 
In the following, $t_{{\vec t}_m}$ , $\Delta_{{\vec t}_m}$, 
${\bar{\Delta}}_{{\vec t}_m}$, and $\mu$ are all assumed to be real.

Let us first explore the excitation spectrum of ${\hat{H}}_{\rm BCS}$. 
For this purpose, it is convenient to Fourier transform the BCS Hamiltonian 
to the momentum space with the reciprocal lattice 
vectors, ${\vec g}_1=2\pi(1,-{\frac{1}{\sqrt{3}}})$ and 
${\vec g}_2=2\pi(0,{\frac{2}{\sqrt{3}}})$ [see Fig.~\ref{lattice2} (b)]. 
After the Fourier transformation with  
$c^\dag_{{\vec r}\sigma}={\frac{1}{\sqrt{L}}}\sum_{\vec k}
{\rm e}^{-i{\vec k}\cdot{\vec r}} c^\dag_{{\vec k}\sigma}$, 
${\hat H}_{\rm BCS}$ can be conveniently described by the following 
$(4\times4)$ matrix form: 
\begin{widetext}
\begin{eqnarray}\label{bcs2x1}
{\hat H}_{\rm BCS}={{\sum_{\vec k}}^{\prime}}
\left( \begin{array}{cccc}
        c^\dag_{{\vec k}\up}, & c^\dag_{{\vec k}+{\vec Q}\up}, 
        & c_{-{\vec k}\dn}, & c_{-({\vec k}+{\vec Q})\dn} 
       \end{array}
\right) %\nonumber \\
\left( \begin{array}{cccc}
       \xi_1({\vec k}) & 0 & \Delta_{11}({\vec k}) 
       & \Delta_{12}({\vec k}) \\[0.1cm]
       0 & \xi_2({\vec k}) & {\Delta_{12}({\vec k})}^* 
       & \Delta_{22}({\vec k}) \\[0.1cm]
         \Delta_{11}({\vec k}) & \Delta_{12}({\vec k}) & -\xi_1({\vec k})
       & 0 \\[0.1cm]
         {\Delta_{12}({\vec k})}^* & \Delta_{22}({\vec k}) 
         & 0 & -\xi_2({\vec k})
       \end{array}
\right)
\left( \begin{array}{c}
        c_{{\vec k}\up} \\[0.1cm]  c_{{\vec k}+{\vec Q}\up} \\[0.1cm]
        c^\dag_{-{\vec k}\dn}\\[0.1cm] c^\dag_{-({\vec k}+{\vec Q})\dn} 
       \end{array}
\right), 
\label{mat_broken}
\end{eqnarray}
\end{widetext}
where 
\begin{equation}
\left\{
 \begin{array}{l}
  \displaystyle{
  \xi_1({\vec k})=-2{\sum_{{\vec t}_m}}^{\prime}t_{{\vec t}_m}
      \cos\left({\vec k}\cdot{\vec t}_m\right) - \mu}  \\[0.2cm]
  \displaystyle{
  \xi_2({\vec k})=-2{\sum_{{\vec t}_m}}^{\prime}t_{{\vec t}_m}
       \cos\left[\left({\vec k}+{\vec Q}\right)\cdot{\vec t}_m\right] - \mu},
 \end{array}
\right.
\end{equation}
and 
\begin{equation}
\left\{
 \begin{array}{l}
   \displaystyle{
   \Delta_{11}({\vec k})=2{\sum_{{\vec t}_m}}^{\prime}\Delta_{{\vec t}_m}
         \cos\left({\vec k}\cdot{\vec t}_m\right)}\\[0.2cm]
   \displaystyle{
   \Delta_{22}({\vec k})=2{\sum_{{\vec t}_m}}^{\prime}\Delta_{{\vec t}_m}
         \cos\left[\left({\vec k}+{\vec Q}\right)\cdot{\vec t}_m\right]}\\[0.2cm]
   \displaystyle{
   \Delta_{12}({\vec k})={\sum_{{\vec t}_m}}^{\prime}
         {\bar{\Delta}}_{{\vec t}_m}\left(
          {\rm e}^{i\left({\vec k}+{\vec Q}\right)\cdot{\vec t}_m}
         +{\rm e}^{-i{\vec k}\cdot{\vec t}_m}
         \right)}
 \end{array}
\right.
\end{equation}
with ${\vec Q}={\vec g}_1/2$ so that ${\vec Q}\cdot{\vec r}=\pi r_1$, 
The primed sum ${\sum_{\vec k}}^{\prime}$ runs over the reduced Brillouin 
zone shown in Fig.~\ref{lattice2} (b), or equivalently, {\it e.g.}, for 
parallelogram lattice of $L=L_1\times L_2$ ($L_1$, $L_2$: even) with 
${\vec k}={\frac{k_1}{L_1}}{\vec g}_1+{\frac{k_2}{L_2}}{\vec g}_2,$ 
the primed sum is taken over $k_1=0,1,\cdots,L_1/2-1$ and 
$k_2=0,1,\cdots,L_2-1$.

The excitation spectrum of the BCS Bogoliubov mode for this BCS Hamiltonian 
is now easily calculated by diagonalizing the $(4\times4)$ Hamiltonian 
matrix given in Eq.~(\ref{mat_broken}). 
It turned out that the excitation spectrum $E_{\vec k}$ is doubly degenerate 
at each momentum ${\vec k}$ and is given by $E_{\vec k}=E_{\vec k}^{(\pm)}$, 
where
\begin{equation}\label{spec2x1bcs}
E_{\vec k}^{(\pm)}=\left[Q_0({\vec k}) \pm Q_1({\vec k})\right]^{1/2} 
\end{equation} 
%and $-E_{\vec k}^{(\pm)}$, 
and
\begin{widetext}
\begin{eqnarray}
Q_0({\vec k}) &=& {\frac{1}{2}}\left[ 
\xi_1({\vec k})^2 + \xi_2({\vec k})^2 + \Delta_{11}({\vec k})^2 
+ \Delta_{22}({\vec k})^2 +  2\left|\Delta_{12}({\vec k})\right|^2 \right]\nonumber\\
Q_1({\vec k}) &=& {\frac{1}{2}}
  \left[ \left( \xi_1({\vec k})^2 - \xi_2({\vec k})^2 
            + \Delta_{11}({\vec k})^2 - \Delta_{22}({\vec k})^2 \right)^2 
 + 4\left|\Delta_{12}({\vec k})\right|^2\left\{\left(\xi_1({\vec k})-\xi_2({\vec k})\right)^2
+\left(\Delta_{11}({\vec k})+\Delta_{22}({\vec k})\right)^2
\right\}\right]^{1/2}. \nonumber
\end{eqnarray}
\end{widetext}
From this expression, it is clear that generally 
$E_{\vec k}^{(\pm)}$ 
has a finite excitation gap except for some special cases.

So far no assumption has been made for the values of real gap functions 
$\Delta_{{\vec t}_m}$ and ${\bar{\Delta}}_{{\vec t}_m}$. As will be discussed 
in App.~\ref{appendix:pfaffian}, in order to be consistent with the phase 
rule [Eqs.~(\ref{phase1}) and (\ref{phase2})] and for ${\hat H}_{\rm BCS}$ to be invariant under the transformation 
${\cal T}_1 \bigotimes {\cal U}$ (Sec.~\ref{secresults:sondhi}), 
the gap functions $\Delta_{{\vec t}_m}$ and ${\bar{\Delta}}_{{\vec t}_m}$ 
are subject to the following conditions: 
\begin{equation}
\left\{
\begin{array}{rl}
\Delta_{{\vec t}_m}  = 0, & ~~ {\rm for~m_2~odd} \\ [0.1cm]
{\bar{\Delta}}_{{\vec t}_m} = 0, & ~~ {\rm for~m_2~even}.
\end{array}
\right. 
\end{equation}
With this condition, $\Delta({\vec t}_m)$ given in Eq.~(\ref{pairbroken}) 
is equivalent to $\Delta_{{\vec t}_m}$ and ${\bar{\Delta}}_{{\vec t}_m}$ 
defined here.

Let us now consider one special but important case for which 
all hopping terms $t_{{\vec t}_m}$ are zero and 
$\Delta_{11}({\vec k})=-\Delta_{22}({\vec k})$. 
In this case, $Q_1({\vec k})=0$, and 
therefore the excitation spectrum for the Bogoliubov mode is simply 
\begin{equation}
E_{\vec k} = \left[{\mu}^2+\Delta_{11}({\vec k})^2+
\left|\Delta_{12}({\vec k})\right|^2\right]^{1/2}.  
\end{equation}
After tedious but straight forward calculations, the ground state 
$|{\rm BCS}\rangle$ of the BCS Hamiltonian ${\hat H}_{\rm BCS}$ for this 
special case can be calculated analytically: 
\begin{widetext}
\begin{equation}
\begin{array}{l}
\displaystyle{
|{\rm BCS}\rangle= \exp\left[{\sum_{\vec k}}^{\prime}
\left\{f_{11}({\vec k})c^\dag_{{\vec k}\up} c^\dag_{-{\vec k}\dn}
+f_{22}({\vec k})c^\dag_{{\vec k}+{\vec Q}\up} c^\dag_{-({\vec k}+{\vec Q})\dn}
+f_{12}({\vec k})\left(c^\dag_{{\vec k}\up} c^\dag_{-({\vec k}+{\vec Q})\dn}-
c^\dag_{{\vec k}\dn} c^\dag_{-({\vec k}+{\vec Q})\uparrow}  \right)
\right\}
\right]|0\rangle
}
\end{array}
\end{equation}
\end{widetext}
where
\begin{eqnarray}\label{gap_pair}
f_{11}({\vec k})&=&-f_{22}({\vec k})=-{\frac{\Delta_{11}({\vec k})}{E_{\vec k}-\mu}}, \nonumber\\
f_{12}({\vec k})&=&-{\frac{\Delta_{12}({\vec k})}{E_{\vec k}-\mu}},
\end{eqnarray}
and ${\Delta_{12}({\vec k})}^*=\Delta_{12}(-{\vec k})$ is used. 
This state should be compared with the more conventional one given 
by Eq.~(\ref{wfbcs}) where there is only one site per unit cell.  
Finally, let us consider two more specific cases separately. 
{\it Case}~(1): $\mu=0$ and only nearest 
neighbor gap functions are finite, {\it i.e.}, 
$\Delta_{{\vec\tau}_1}={\bar\Delta}_{{\vec\tau}_2}={\bar{\Delta}}_{{\vec\tau}_2-{\vec\tau}_1}=\Delta$ 
(see Fig.~\ref{lattice3}). Because 
$\Delta_{11}({\vec k}) = -\Delta_{22}({\vec k}) =
2\Delta\cos({\vec k}\cdot{\vec\tau}_1)$ and 
$\Delta_{12}({\vec k})=2\Delta\left[\cos({\vec k}\cdot{\vec\tau}_2)
-i\sin\left({\vec k}\cdot({\vec\tau}_2-{\vec\tau}_1)\right)\right]$,  
$E_{\vec k}$ has gapless excitations 
at ${\vec k}={\vec g}_1/4\pm{\vec g}_2/4$ and $-{\vec g}_1/4\pm{\vec g}_2/4$. 
{\it Case}~(2): $\mu\to-\infty$. Since 
$f_{11}({\vec k})\propto-\Delta_{11}({\vec k})$ and  
$f_{12}({\vec k})\propto-\Delta_{12}({\vec k})$ in this limit, {\it i.e.}, 
a pairing function is proportional to a gap function, 
a projected BCS wave function $\pbcs$ built from this BCS state 
becomes 
\begin{widetext}
\begin{equation}\label{pbcs-svb}
\displaystyle{
\lim_{\mu\to-\infty}\pbcs 
={\cal P}_{\rm G}\left[\sum_{\vec r}
\left\{
{\sum_{{\vec t}_m}}^{\prime} 
\left( \Delta_{{\vec t}_m} + (-1)^{r_1}{\bar{\Delta}}_{{\vec t}_m}\right)
\left( c^\dag_{{\vec r}\up} c^\dag_{{\vec r}+{\vec t}_m\dn}
-c^\dag_{{\vec r}\dn} c^{\dag}_{{\vec r}+{\vec t}_m\up}  \right)
\right\}
\right]^{L/2}|0\rangle. }
\end{equation}
\end{widetext}
From this result, it it clear that the projected BCS state discussed here 
indeed includes a short range RVB wave function $|\psi_{\rm RVB}\rangle$ 
since if $\Delta_{{\vec\tau}_1}={\bar\Delta}_{{\vec\tau}_2}={\bar{\Delta}}_{{\vec\tau}_2-{\vec\tau}_1}$ (all other gap functions are zero) 
is chosen, the above wave function (\ref{pbcs-svb}) consists of the pairing 
functions with exactly the same phase as in Eqs.~(\ref{phase1}) and 
(\ref{phase2}). 
More details of this relation will be found in the next appendix.

%%%%%%%%%%%%%%%%%%%%%%%%%%%%%%%%%%%%%%%%%%%%%%%%%%%
\section{ Pfaffian and Projected BCS wave functions} \label{appendix:pfaffian}
%%%%%%%%%%%%%%%%%%%%%%%%%%%%%%%%%%%%%%%%%%%%%%%%%%%

We start from the definition of a Pfaffian in terms of an antisymmetric 
 $2N\times 2N$ matrix $\bar f$ where $f_{i,j}=-f_{j,i}$. 
The Pfaffian of matrix ${\bar f}$ is
%\begin{widetext}
\begin{equation} \label{pfaf}
 P[{\bar f}]=\sum\limits_{{\scriptstyle (i_1<j_1),(i_2<j_2),\cdots, (i_N < j_N)}
                  \atop {\scriptstyle~{\rm and}~ i_1 <i_2 \cdots < i_N} }  
  (-1)^p   \prod\limits_{k=1}^N    f_{i_k,j_k},    
\end{equation}
%\end{widetext}
where the sum runs  over all  possible covering of indices 
$\{(i_1,j_1),(i_2,j_2),\cdots,(i_N,j_N)\}$ 
such that $i_k<j_k$ and $i_1 < i_2< i_3<\cdots <i_N$, being $p$ the parity of 
the permutation of the $2 N $ indices: 
\begin{equation}
\left(
\begin{array}{ccccccc}
1 & 2 & 3 & 4& \cdots & 2N-1 & 2N \\
i_1 & j_1 & i_2 & j_2 & \cdots & i_N & j_N
\end{array}
\right)\nonumber
\end{equation}
The most important relation known for the Pfaffian 
is that $\left(P[{\bar f}]\right)^2={\rm Det}[{\bar f}]$, which however 
will not be used in the following. 

Now let us suppose that the indices $1,2,3,...,2N$ label the positions 
of a lattice (not necessarily one dimensional). Then each covering of the 
indices in the Pfaffian is interpreted as a particular dimer configuration 
in which for example a spin singlet pair located at sites  $(i_k,j_k)$ is 
assigned.  
  
We can now define two spin wave functions in terms of these dimer coverings. 
 The first wave function is   expanded in the 
well known valence bond basis:~\cite{liang}
\begin{widetext}
\begin{equation}  \label{rvb}
  |{\rm RVB}\rangle =\sum\limits_
{{\scriptstyle (i_1<j_1),(i_2<j_2),\cdots,(i_N < j_N)}
                  \atop {\scriptstyle~{\rm and}~ i_1 <i_2 \cdots < i_N} }
 (-1)^p   \left[ \prod\limits_{k=1}^N   f_{i_k,j_k}  (S^{-}_{j_k}-S^{-}_{i_k} ) \right] 
 |F \rangle 
\end{equation}
\end{widetext}
where $S^-_i$ is the spin-$1/2$ lowering operator at site $i$, 
$|F\rangle $ is the ferromagnetic state defined by
\begin{equation} \label{ferro} 
|F\rangle =c^{\dag}_{1,\uparrow}, c^{\dag}_{2,\uparrow}, \cdots 
c^{\dag}_{2N,\uparrow} |0\rangle, 
\end{equation}
and $|0\rangle $ is the electron vacuum. Notice that for convenience we have 
included in the definition of the wave function the same permutation 
sign $(-1)^p$ appearing in the Pfaffian (\ref{pfaf}). 
The second wave function is the projected BCS state:
\begin{equation} \label{pbcs}
\pbcs = {\cal P}_{\rm G} \exp\left[ {\sum_{\scriptstyle{i,j=1}
                  \atop {\scriptstyle{(i<j)} }}^{2N}
f_{i,j} 
 \left( c^{\dag}_{i,\uparrow} c^{\dag}_{j,\downarrow}  -  c^{\dag}_{i,\downarrow} c^{\dag}_{j,\uparrow}  \right) 
} \right]  |0\rangle 
\end{equation}
where ${\cal P}_{\rm G}$ is the Gutzwiller projection operator onto singly 
occupied sites. In  both wave functions, we have used the upper triangular 
part of the 
antisymmetric matrix  ${\bar f}$, whereas the tedious permutation sign 
$(-1)^p$ is present only in the wave function $|\rm{RVB}\rangle$.

We shall now show that the two wave functions $\pbcs$ and 
$|{\rm RVB}\rangle$ are actually the same {\em for any ${\bar f}$ and 
on any lattice with all possible boundary conditions}.
\vskip 0.5 truecm 
\noindent {\bf Proof.}  
Because the projection ${\cal P}_{\rm G}$ forbids double occupancy and 
thus singlet bonds sharing the common site, the projected BCS wave 
function $\pbcs$ can be expanded for all dimer coverings as follows: 
\begin{widetext}
\begin{equation}
\pbcs =  {\frac{1}{N!}}{\cal P}_{\rm G}  \left[ \sum_{\scriptstyle{i,j=1}
                  \atop {\scriptstyle{(i<j)} }}^{2N} f_{i,j}
  (c^{\dag}_{i,\uparrow} c^{\dag}_{j,\downarrow}  -  c^{\dag}_{i,\downarrow} c^{\dag}_{j,\uparrow} )   \right]^N  |0\rangle 
 =  \sum\limits_{{\scriptstyle (i_1<j_1),(i_2<j_2),\cdots, (i_N < j_N)}
                  \atop {\scriptstyle~{\rm and}~ i_1 <i_2 \cdots < i_N} }
   \prod\limits_{k=1}^N  \left[ f_{i_k,j_k} 
  (c^{\dag}_{i_k,\uparrow} c^{\dag}_{j_k,\downarrow}  -  c^{\dag}_{i_k,\downarrow} c^{\dag}_{j_k,\uparrow} )   \right] |0\rangle, 
\end{equation}
\end{widetext}
where in the last formula the constant $1/N!$ cancels out the factor due to 
the ordering of the indices $i_1<i_2,\cdots < i_N$. 
On the other hand, by substituting the definition of the ferromagnetic state 
in the wave function $|{\rm RVB}\rangle$ 
and making the necessary $p^*$ fermion permutations  for each 
dimer covering,  we arrive at the following expression: 
\begin{widetext}
\begin{equation}  \label{rvbnew}
  |{\rm RVB}\rangle =
\sum\limits_{{\scriptstyle (i_1<j_1),(i_2<j_2),\cdots,(i_N < j_N)}
                  \atop {\scriptstyle~{\rm and}~ i_1 <i_2 \cdots < i_N} }
 (-1)^p  (-1)^{p^*} 
  \left[ \prod\limits_{k=1}^N   f_{i_k,j_k}  (S^{-}_{j_k}-S^{-}_{i_k} )   
  c^{\dag}_{i_k,\uparrow} c^{\dag}_{j_k,\uparrow}
) \right] 
 |0 \rangle.
\end{equation}
\end{widetext}
By simple inspection, it is readily realized that the 
fermion sign $(-1)^{p^*}$ is exactly the one  $(-1)^p$ defining the  
permutation sign of the covering of indices $\{(i_k,j_k)\}$, and that 
 $  (S^{-}_{j_k}-S^{-}_{i_k} )
  c^{\dag}_{i_k,\uparrow} c^{\dag}_{j_k,\uparrow}|0\rangle = 
  c^{\dag}_{i_k,\uparrow} c^{\dag}_{j_k,\downarrow}  -  c^{\dag}_{i_k,\downarrow} c^{\dag}_{j_k,\uparrow} |0\rangle$.  
Thus it is proved that two wave functions $|{\rm RVB}\rangle$ and 
$\pbcs$ are exactly the same.

In the remaining of this appendix, we will show that even for the triangular 
lattice geometry, the so-called short range RVB wave 
function $|\psi_{\rm RVB}\rangle$ [Eq.~(\ref{srrvb})] can be described by 
the projected BCS wave function (\ref{pbcs}) with a particular choice of 
the matrix $\bar f$.

\subsection{Consequences of the theorem: short range RVB wave function}

The short range RVB wave function $|\psi_{\rm RVB}\rangle$ [Eq.~(\ref{srrvb})] 
is defined with the same weight for all nearest neighbor 
valence bonds. Therefore, if we can find the matrix $\bar f$ in 
Eq.~(\ref{rvb}) in such a way that the sign of the permutation for each dimer 
covering is exactly canceled, {\it i.e.},
\begin{equation} \label{condition}
(-1)^p \prod\limits_{k=1}^N   f_{i_k,j_k}  =1, 
\end{equation}
then a relation is established between RVB wave functions and projected 
BCS wave functions. 
The condition (\ref{condition}) is highly non trivial and difficult to satisfy 
as the number of dimer coverings is exponentially large and the entries 
of the matrix are few in comparison. 
Nevertheless, this problem for the case of nearest neighbor dimer covering was 
solved in all planar graphs.~\cite{kas}

By applying these old results,~\cite{kas} we can generalize the 
Read-Chakraborty relation between the short range RVB and projected BCS 
wave functions on the square lattice~\cite{readn} for the triangular lattice 
case. Namely, by using the known matrix ${\bar f}$ reported in 
Ref.~\onlinecite{kas}, it is possible to satisfy the 
condition (\ref{condition}) even for the triangular case.~\cite{sondhi} 
Here we use open boundary conditions for a lattice of $l\times l$ sites 
($l$ being multiple of $6$) where site $i$ is ordered lexicographically: 
$i=l m_2 +m_1= (m_1,m_2)$.   
The Cartesian coordinates of the triangular lattice (Fig.~\ref{lattice}) 
are thus 
$${\vec r}_i=m_1{\vec\tau}_1+m_2{\vec\tau}_2=(m_1+m_2/2, \sqrt{3}\ m_2/2 ).$$ 
The pairing functions $f_{k,j}$ which meet the condition (\ref{condition}) 
reads: 
%\begin{eqnarray*} 
%f=1~  &{\rm ~if~}& j=(1,0)  \\ 
%f=1~  &{\rm~ if~}& j=(0,1)  \\ 
%f=-1 &{\rm~ if~}& j=(1,1)           {\rm ~~ the~ diagonal~ bond}  \\ 
%\end{eqnarray*}
%if $k =(0,0)$
\begin{equation} \label{phase1}
f_{k,j}=\left\{
\begin{array}{cl}
1 &  ~\mbox{for $j=(1,0)$}\\ [0.1cm]   
1 &  ~\mbox{for $j=(0,1)$}  \\ [0.1cm]
1 & ~\mbox{for $j=(-1,1)$} %, the diagonal bond}  
\end{array}
\right.
\end{equation}
for $k =(0,0)$, whereas 
%\begin{eqnarray*} 
%f=1~   &{\rm~ if~} & j=(2,0) \\
%f=-1  &{\rm~ if~} & j=(1,1)  \\
%f=1~   &{\rm~ if~} & j=(2,1)  {\rm ~~the~ other~ diagonal~ bond} \\
%\end{eqnarray*}
%if $k =(1,0)$.
\begin{equation} \label{phase2}
f_{k,j}=\left\{
\begin{array}{cl}
1 &  ~\mbox{for $j=(2,0)$}\\ [0.1cm]   
-1 &  ~\mbox{for $j=(1,1)$}  \\ [0.1cm]
-1 & ~\mbox{for $j=(0,1)$} %, the other diagonal bond}  
\end{array}
\right.
\end{equation}
for $k =(1,0)$. 
All the other values of $f_{k,j}$ are obtained by translation of 
$2{\vec\tau_1}=(2,0)$  and/or ${\vec\tau}_2=(1,0)$. Therefore, the unit 
cell of $f_{k,j}$ is $(2\times1)$ on the triangular lattice 
(see Fig.~\ref{lattice3}).

\subsection{ Complex representation~\cite{sondhi}}% (Sondhi).

We can multiply by the imaginary unit $i$ all the $c^{\dag}_{j,\sigma}$ 
for the odd ${\vec\tau}_1$ components of $j$  since the resulting wave 
function is equivalent to the original one [Eq.~(\ref{pbcs})] in the presence 
of the projection operator ${\cal P}_{\rm G}$ (apart from an overall phase). 
The obtained new complex matrix $\bar f$ consists of:
\begin{equation*}
f_{k,j}=\left\{
\begin{array}{cl}
i &  ~\mbox{for $j=(1,0)$}\\ [0.1cm]   
1 &  ~\mbox{for $j=(0,1)$}  \\ [0.1cm]
i & ~\mbox{for $j=(-1,1)$, the diagonal bond}  
\end{array}
\right.
\end{equation*}
for $k =(0,0)$, whereas 
\begin{equation*}
f_{k,j}=\left\{
\begin{array}{cl}
i &  ~\mbox{for $j=(2,0)$}\\ [0.1cm]   
1 &  ~\mbox{for $j=(1,1)$}  \\ [0.1cm]
-i & ~\mbox{for $j=(0,1)$, the other diagonal bond}  
\end{array}
\right.
\end{equation*}
for $k =(1,0)$. All the other values of $f_{k,j}$ are obtained by 
translation of $2{\vec\tau_1}=(2,0)$  and/or ${\vec\tau}_2=(1,0)$.
%\begin{eqnarray*} 
%f=i~   &{\rm~ if~}&  j=(1,0) \\
%f=1~   &{\rm~ if~}&  j=(0,1) \\
%f=-i  &{\rm~ if~}&  j=(1,1)          {\rm ~~ the~ diagonal~ bond } \\
%\end{eqnarray*}
%if $k =(0,0)$   
%\begin{eqnarray*}  
%f=i   &{\rm~ if~}& j=(2,0)   \\
%f=1   &{\rm~ if~}& j=(1,1)   \\
%f=i   &{\rm~ if~}& j=(2,1)          {\rm ~~ the~ other~ diagonal~ bond}  \\
%\end{eqnarray*}
%if $k =(1,0)$

Notice that if the diagonal bonds are eliminated, we end up with again a 
$(1\times1)$ unit cell with $f_{\bf k}\propto \cos(k_x)-i \cos(k_y)$ 
in Fourier space, thus recovering the Read-Chakraborthy result for the 
square lattice with ${\vec r}_i=(m_1,m_2)$.~\cite{readn}

It is important to emphasize that for the triangular lattice geometry 
a $(2 \times 1)$ unit cell is inevitable even with the complex 
representation (as the diagonal bonds acquire different signs). 
Thus, 
we conclude that the translation symmetry in $f_{k,j}$ has to be broken 
for a good projected BCS wave function in the triangular case. 
It is interesting that two sites per a unit cell is enough, 
in contrast to the classical N{\'e}el order state with $120^{\circ}$ 
degrees of mutual spin orientations which contains instead three sites per 
unit cell.

\subsection{Periodic boundary conditions}

For the short range RVB state with periodic boundary conditions, it is known 
that condition (\ref{condition}) can not be satisfied by a single $\pbcs$, 
but by four, all obtained with the possible choices of periodic and 
antiperiodic boundary conditions in $\vec\tau_1$ and $\vec\tau_2$ 
directions.~\cite{kas} 
The proof can be applied for each of such $\pbcs$, showing that these 
four projected BCS wave functions have to be used to exactly match 
the short range RVB wave function with periodic boundary conditions.

%%%%%%%%%%%%%%%%%%%%%%%%%%%%%%%
%%%%%%%%%%%%%%%%%%%%%%%%%%%%%%%


\begin{thebibliography}{99}
\bibitem{note5} All results presented here in this paper are calculated with 
periodic boundary conditions (PBC) unless otherwise stated.
\bibitem{note4} This condition can be checked for small size systems.
\bibitem{note7} In Fig.~\ref{bz_jp033} and Fig.~\ref{bz_jp05}, the variational 
parameters optimized for $L=18\times18$ (Tab.~\ref{tablecoldea}) are used.
\bibitem{note10} Assuming that $|\Psi_{\rm RVB}\rangle$ and 
$|\Psi_0\rangle$ here are properly normalized.
\bibitem{xypar} This property has been verified by numerically 
computing both diagonal and off diagonal spin-spin correlation functions. 
Namely, the mixed average estimate of the magnetic order parameter in the 
$xy$-plane was computed and found to remain negligibly small.
\bibitem{notesrvsfne} FNE provides results similar to GFMCSR,~\cite{caprio} 
when the same guiding function is used. However, FNE is much more efficient, 
as far as the statistical errors are concerned, 
and provides a variational upper bound of the energy.
\bibitem{note3}
The way to open a finite spin excitation gap discussed here is similar 
to the one 
realized in the quantum dimer model on the triangular lattice,~\cite{sondhi} 
where in order to treat the dimer covering of the lattice, an algebraic 
method based on the Pfaffian was introduced. The corresponding matrix 
is also defined in a $(2\times 1)$ unit cell 
(see also App.~\ref{appendix:pfaffian}).
\bibitem{note1} The variational parameters are optimized for clusters 
up to $L=18\times 18$, and the ones for $L=18\times 18$ are used for 
the $30\times 30$ cluster calculations.
\end{thebibliography}
\end{document}